\documentclass{jfm}
\usepackage{graphicx}
\usepackage{epstopdf, epsfig}
\usepackage{amsmath}
\usepackage{mathrsfs}

\usepackage{siunitx}
\newcommand{\st}[1]{\textcolor{black}{#1}}
\newcommand{\jl}[1]{\textcolor{black}{#1}}

\shorttitle{Unsteady surfactant-contaminated superhydrophobic channels}
\shortauthor{S. D. Tomlinson and others}

\title{Unsteady evolution of slip and drag in surfactant-contaminated superhydrophobic channels}

\author{Samuel D. Tomlinson\aff{1}
\corresp{\email{samuel.tomlinson@manchester.ac.uk}},
  Fr\'{e}d\'{e}ric Gibou\aff{2},
  Paolo Luzzatto-Fegiz\aff{2},
  Fernando Temprano-Coleto\aff{3},
  Oliver E. Jensen\aff{1}
  \and Julien R. Landel\aff{1,\,4}}
\begin{document}

\affiliation{
\aff{1}Department of Mathematics, University of Manchester, Oxford Road, Manchester M13 9PL, UK
\aff{2}Department of Mechanical Engineering, University of California, Santa Barbara, CA 93106, USA
\aff{3}Andlinger Center for Energy and the Environment, Princeton University, Princeton, NJ 08544, USA
\aff{4}Universite Claude Bernard Lyon 1, Laboratoire de Mecanique des Fluides et d'Acoustique (LMFA), UMR5509, CNRS, Ecole Centrale de Lyon, INSA Lyon, 69622
Villeurbanne, France
}

\maketitle

\begin{abstract} 
%
\jl{Recognising that surfactants can impede the drag reduction resulting from superhydrophobic surfaces (SHSs), we investigate the impact of spatio--temporal fluctuations in surfactant concentration on the drag-reduction properties of SHSs. 
We model the unsteady transport of soluble surfactant in a channel flow bounded by two SHSs.}
The flow is laminar, pressure-driven, and the SHSs are periodic in the streamwise and spanwise directions. 
We assume that the channel length is much longer than the streamwise period, the streamwise period is much longer than the channel height and spanwise period, and bulk diffusion is sufficiently strong for cross-channel concentration gradients to be small.
By combining long-wave and homogenisation theories, we derive an unsteady advection--diffusion equation for surfactant-flux transport over the length of the channel, which is coupled to a quasi-steady advection--diffusion equation for surfactant transport over individual plastrons.
As diffusion over the length of the channel is typically small, the surfactant flux is governed by a nonlinear wave equation. 
\jl{In the fundamental case of the transport of a bolus of surfactant, we predict its propagation speed and describe its nonlinear evolution via interaction with the SHS.}
The propagation speed can fall below the average streamwise velocity as the surfactant adsorbs and rigidifies the plastrons. 
Smaller concentrations of surfactant are advected faster than larger ones, so that wave-steepening effects can lead to shock formation in the surfactant-flux distribution.
\jl{Our asymptotic results reveal how unsteady surfactant transport can affect the spatio--temporal evolution of the slip velocity, drag reduction and effective slip length in SHS channels.} 
%
\end{abstract}


\section{Introduction} \label{sec:introduction}

Surfactants are chemical compounds that are advected and diffuse throughout a fluid, where they adsorb onto liquid--liquid or liquid--gas interfaces \citep{manikantan2020surfactant}.
They have been shown to 
impair the effective slip length and drag reduction in superhydrophobic microchannels \citep{peaudecerf2017traces}. \jl{Surfactants that have been adsorbed onto the liquid--gas interfaces of SHSs are advected downstream by the flow and accumulate at stagnation points (i.e. liquid--solid contact lines), generating an adverse Marangoni force that may negate any drag-reducing effects in laminar \citep{kim2012pressure,bolognesi2014evidence,peaudecerf2017traces, song2018effect} or turbulent \citep{tomlinson2023model} flows.}
Superhydrophobic surfaces (SHSs) use chemically-coated microscopic structures to suspend a fluid over a series of gas pockets \citep{lee2016superhydrophobic}. 
The combination of no-slip structures and shear-free liquid--gas interfaces generates the drag reduction \jl{in surface} flows. 
Hence, SHSs have been considered for applications in biofluidics \citep{darmanin2015superhydrophobic}, heat transfer \citep{lam2015analysis} and marine hydrodynamics \citep{xu2020superhydrophobic}, both in laminar and turbulent flows. 
Field studies have shown that surfactant is present in the ocean and that the surfactant concentration can vary significantly in space and time \citep{pereira2018reduced, frossard2019properties}.
Traces of surfactant have been measured in rivers, estuaries and fog \citep{lewis1991chronic, facchini2000surface}. They are also present in most industrial and laboratory environments \citep{manikantan2020surfactant}.
\jl{In all these natural, industrial and laboratory environments, the surfactant concentration can fluctuate in space and time. 
Motivated by this observation, we study an idealised scenario representative of some of these complicated engineering applications. 
From a fundamental view, we seek to understand how variations in time and space in surfactant transport can affect the drag-reducing properties of SHSs in a canonical channel flow. 
Our simplified scenario considers the unsteady transport of surfactant in a laminar pressure-driven channel flow bounded between streamwise- and spanwise-periodic SHSs. 
We model how surfactant is advected and diffuses over length scales and time scales that are large compared to the dimensions of the SHS texture.}

Experimental studies first suggested that naturally-occurring surfactants could affect channel flows bounded by SHSs comprising spanwise ridges \citep{kim2012pressure}, as well as finite-length streamwise ridges \citep{bolognesi2014evidence}. 
They found that the flow rate and wall shear stress closely resembled a channel with solid walls, and thus their SHSs offered only a modest drag reduction \citep{kim2012pressure, bolognesi2014evidence}.
%
\cite{schaffel2016local} showed that experimentally-measured slip lengths on SHSs consisting of pillars were much smaller than predicted by surfactant-free simulations; this was true whether surfactant was explicitly added or not, suggesting that naturally-occurring surfactants played a key role. 
As noted earlier, a requirement for surfactant effects to manifest on SHSs is the presence of stagnation points perpendicular to the flow, at which point surfactant can accumulate to generate a surface tension gradient.
\citet{song2018effect} showed that surface tension gradients emerged in their experiments for finite streamwise ridges, increasing the drag compared to those configurations with concentric ridges that lack stagnation points in the flow.

\st{To investigate the effect of weak surfactant concentrations, \cite{peaudecerf2017traces} introduced simulations inclusive of surfactant dynamics; they showed that a plastron could be immobilised by concentrations below levels commonly occurring in the environment and in engineered systems.} 
The simulations of \cite{peaudecerf2017traces} predicted that surfactant impairment would decrease as the streamwise plastron interface length increased. This was confirmed by their experiments \citep{peaudecerf2017traces}, which showed that if the driving pressure was suddenly removed, a reverse flow was established at the interface, decaying with time as $1/t$ at intermediate times. 
This time scaling was predicted by a similarity solution driven by surfactant relaxation, assuming advection-dominated flow.  In contrast to these plastron-scale findings, 
there is presently no theory that includes the combined effects of solubility, advection and diffusion, that describes inhomogeneous surfactant transport across 
multiple 
plastrons, or that can model the effects of unsteady surfactant concentration at the inflow.
%
%
%

%

Steady scaling theories were constructed for a pressure-driven channel flow with two-dimensional gratings \citep[approximating long, spanwise-oriented gratings][]{landel2020theory}, as well as for long gratings with finite spanwise extent, assuming spatially periodic flow \citep{temprano2023single}. Both theories are in agreement with the slip velocity and drag predicted in full numerical simulations. 
\citet{temprano2023single} further validated their theory by performing experiments with SHS gratings of various lengths, finding that surfactant effects decrease with the square of the interface length.
\citet{landel2020theory} and \citet{temprano2023single} assume that the surfactant concentration is small (as may be expected when surfactant is not explicitly added) and that the shear stress 
is approximately uniform at the liquid-gas interface. 
They do not consider the stagnant cap regime, first reported for air bubbles rising in surfactant-contaminated water \citep{bond1928lxxxii, frumkin1947effect}. 
%

\st{To provide a more comprehensive theory that accommodates non-uniform shear stresses at the liquid--gas interface, \citet{tomlinson2023laminar} assumed that bulk diffusion was strong enough to suppress cross-channel concentration gradients, allowing systematic asymptotic approximations to be developed.
They considered gratings of finite spanwise extent and small surfactant concentrations (allowing linearisation of physicochemical relations). 
The surfactant flux was assumed to be uniform and was prescribed along the length of the channel.} 
\st{Several dimensionless groups were identified by \citet{temprano2023single} and \citet{tomlinson2023laminar} that influence the drag in superhydrophobic channels.
}
\citet{temprano2023single} showed that surfactant impairment in their simulations and experiments was well predicted by a single dimensionless group, when the surfactant properties, SHS dimensions and flow velocities are constrained within physically realizable ranges. 
Using these physical constraints, scaling analysis identified the dimensionless group as the ratio between the streamwise length of the interface and a surfactant-determined lengthscale, labelled ``mobilization length''. 
\st{Without these physical constraints on surfactant or flow properties, \citet{tomlinson2023laminar} found several other relevant dimensionless groups by calculating asymptotic solutions for the concentration field and drag across the whole parameter space; these depend on a velocity scale generated by interfacial Marangoni effects, the surfactant diffusivity and the flow rate.}
Another dimensionless group found by \citet{tomlinson2023laminar} can be used to predict whether the surfactant concentration field is in the stagnant-cap regime.
\st{This dimensionless group was also identified in the numerical simulations performed by \citet{sundin2022slip} in a two-dimensional channel with liquid-infused surfaces (LIS) when the applied shear stress is high.
\citet{sundin2022slip} found that LISs may be more susceptible to surfactant effects than SHSs and derived a scaling theory for LISs when the applied shear stress is small.}
\st{The stagnant cap regime in a two-dimensional shear flow with insoluble surfactant has been investigated with a linear \citep{baier2021influence,baier2022shear} and non-linear \citep{mayer2022superhydrophobic} equation of state.
Assuming a linear equation of state, our unsteady model allows us to investigate how these stagnant cap distributions evolve as the bulk and interfacial surfactant concentration field varies in time.  
}

Both \citet{temprano2023single} and \citet{tomlinson2023laminar} assumed that the velocity and bulk concentration fields are steady and spatially periodic.
That is, they did not allow surfactant to enter the channel with a non-uniform distribution that varies in space and time over multiple periods, \jl{as arises in various environments} \cite[][]{frossard2019properties}.
\jl{Below, we use multiscale homogenisation techniques \cite[such as those outlined in][]{bottaro2019flow} and a long-wave approximation to study the time- and space-varying effects of surfactant over the whole SHS. These theoretical techniques provide mathematical and physical understanding, without the need for expensive numerical simulations which would have to numerically resolve the small details over each texture period.}
We show how a time-dependent one-dimensional asymptotic theory, derived from the three-dimensional Stokes and surfactant transport equations, can be adapted to describe the unsteady evolution of slip and drag in a laminar pressure-driven channel flow with streamwise- and spanwise-periodic grooves, allowing for time-dependent distributions of surfactant flux at the channel inlet.  
The problem exhibits multiple length and time scales, which we exploit to derive and solve a quasi-steady advection--diffusion problem for surfactant concentration over moderate length scales (i.e. the streamwise period of the SHS) and an unsteady advection--diffusion problem for surfactant flux over long length scales (i.e. the streamwise length of the channel), whilst assuming that bulk diffusion is strong enough for cross-channel concentration gradients to be small \citep{tomlinson2023laminar}. 
\st{The surfactant concentration transport equations are nonlinear and of mixed hyperbolic-parabolic type; the unsteady evolution of the surfactant flux over the length of the channel is predominantly hyperbolic, allowing the formation of shocks.}
The problem possesses a number of distinct asymptotic regimes, which we exploit to reveal how the shocks forming in the space- and time-dependent surfactant-flux distribution affect the slip length and drag reduction. 
The slip length and drag reduction are key quantities of interest for practical applications that can be shown to satisfy their own unsteady partial differential equations over long length scales.
We predict the propagation speed of a disturbance to the surfactant flux and investigate how excess surfactant can be advected out of a channel to maximise the space- and time-averaged drag reduction. 

The paper is arranged as follows. 
In \S\ref{sec:Formulation}, the problem is formulated and homogenised to derive an unsteady advection--diffusion equation for surfactant flux transport through the channel. 
At leading order, we derive a purely advective transport equation for the surfactant flux, valid at the channel scale.
In \S\ref{sec:results}, results are presented for the surfactant flux, drag reduction, propagation speed, slip velocity and concentration field. 
We describe the parameter space and identify regions of high and low drag reduction. 
We detail results for two (bell-shaped) canonical distributions of surfactant flux. 
In particular, one profile induces a transition between the high and low drag reduction regions of the parameter space, giving rise to shock formation. 
We study these cases using both theoretical and numerical methods, providing closed-form asymptotic predictions of drag reduction.
In \S\ref{sec:discussion}, we summarise and discuss our main results. 
We provide a table with closed-form asymptotic predictions for the flux propagation speed in the different parts of the parameter space. 
These predictions are expressed both using relevant non-dimensional and dimensional parameters, and are intended as a useful guide for applications. 
%
 
\section{Formulation} \label{sec:Formulation}

\subsection{Governing equations}

\begin{figure}
    \centering
\includegraphics[width=.95\textwidth]{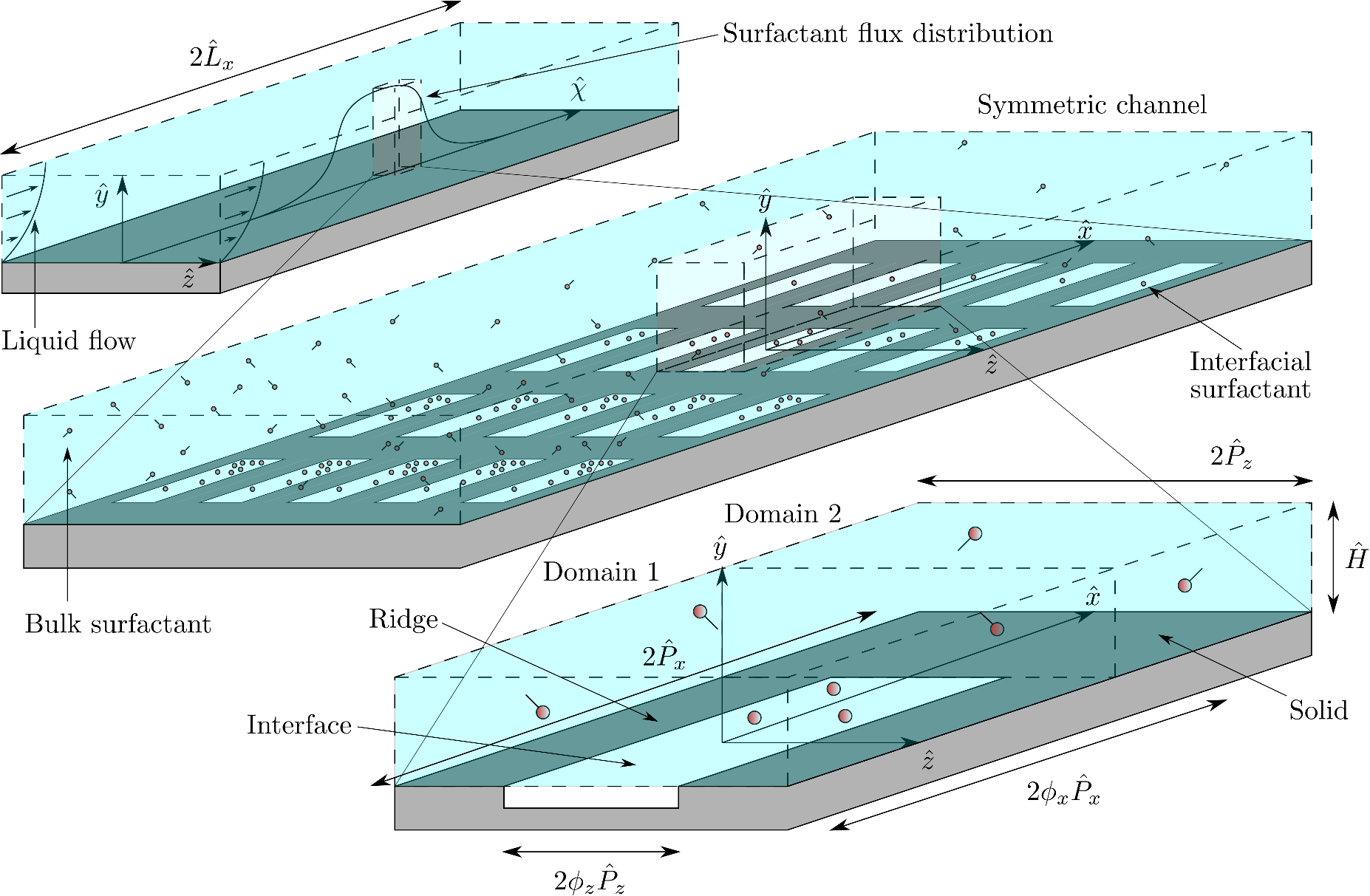}
    \caption{A schematic illustrating the multiple length scales present in a channel flow bounded by streamwise and spanwise-periodic SHSs.
    Top figure: the surfactant flux varies in the streamwise direction, $\hat{\chi}$, and time, $\hat{\tau}$, over the length of the channel, $2\hat{L}_x$.
    Middle figure: over each period, the concentration of surfactant varies quasi-steadily in the streamwise direction, $\hat{x}$.
    Multiple periodic cells are shown, each with a varying concentration of surfactant.
    The origin of the Cartesian coordinate system, $\hat{\boldsymbol{x}}=\boldsymbol{0}$, is placed in the middle of the interface of the central period, at a distance $\hat{L}_x$ along the channel.
    Bottom figure: a periodic cell, identifying domains 1 and 2 and, at the SHS, the ridge, solid wall and liquid--gas interface.
    For each periodic cell, the half channel height is $\hat{H}$, the streamwise (spanwise) gas fraction is $\phi_z$ ($\phi_z$) and the streamwise (spanwise) period length is $2\hat{P}_x$ ($2\hat{P}_z$).
    }
    \label{fig:my_label}
\end{figure}

We consider a laminar pressure-driven fluid flow, contaminated with soluble surfactant, in a channel bounded between two SHSs that are periodic in the streamwise and spanwise directions, as illustrated in figure \ref{fig:my_label}.
We use hats to indicate dimensional quantities.
The streamwise, wall-normal and spanwise directions are denoted by \text{$\hat{x}$-}, $\hat{y}$- and $\hat{z}$-coordinates, where $\hat{\boldsymbol{x}} = (\hat{x},\,\hat{y},\,\hat{z})$ is the space vector and $\hat{t}$ is time.
Assuming that the fluid is incompressible and Newtonian, we introduce the velocity vector $\hat{\boldsymbol{u}}=(\hat{u}(\hat{\boldsymbol{x}},\, \hat{t}),\,\hat{v}(\hat{\boldsymbol{x}},\, \hat{t}),\,\hat{w}(\hat{\boldsymbol{x}},\, \hat{t}))$, pressure field $\hat{p}(\hat{\boldsymbol{x}},\, \hat{t})$, bulk surfactant concentration field $\hat{c}(\hat{\boldsymbol{x}},\, \hat{t})$ and interfacial surfactant concentration field $\hat{\Gamma}(\hat{x},\,\hat{z},\, \hat{t})$.
The streamwise length of the channel is $2 \hat{L}_x$ and the periodic cell has streamwise (spanwise) period length $2 \hat{P}_x$ ($2 \hat{P}_z$), liquid--gas interface length (width) $2 \phi_x \hat{P}_x $ ($2 \phi_z \hat{P}_z$) and gas fraction $\phi_x$ ($\phi_z$).
The channel height is $2\hat{H}$. 
\st{The liquid–gas interfaces (plastrons) can protrude into or out of the gas cavities by an amount that varies along the channel with the liquid and gas pressures (see Lee \textit{et al.} 2016); however, for the sake of simplicity, we assume here that all the plastrons are flat.}
The SHSs are made up of $2N+1$ periodic cells in the streamwise direction. 
For $n \in \{-N, \, ..., \, N\}$, the $n$th periodic cell is split into two subdomains along the streamwise direction, similarly to \cite{temprano2023single},
\begin{subequations} \label{eq:dimensional_domains}
    \begin{align}
    \hat{\mathcal{D}}^n_1 &= \{\hat{x}- 2n\hat{P}_x\in [-\phi_x\hat{P}_x,\, \phi_x\hat{P}_x]\} \times \{\hat{y}\in [0, \,2 \hat{H}]\} \times \{\hat{z} \in [- \hat{P}_z, \,\hat{P}_z]\}, \\
    \hat{\mathcal{D}}^n_2 &= \{\hat{x}- 2n\hat{P}_x\in [\phi_x \hat{P}_x, \,(2 - \phi_x)\hat{P}_x]\} \times \{\hat{y}\in [0,\, 2 \hat{H}]\} \times \{\hat{z} \in [- \hat{P}_z, \,\hat{P}_z]\}.
    \end{align}
\end{subequations}
At the SHS, $\hat{y} = 0$ and $\hat{y}= 2\hat{H}$, we define the $n$th interface, ridge and solid region, as
\begin{subequations} \label{eq:dimensional_interface}
    \begin{align}
    \hat{\mathcal{I}}^n &= \{\hat{x} - 2n\hat{P}_x \in [-\phi_x\hat{P}_x, \,\phi_x\hat{P}_x]\} \times \{\hat{z} \in [- \phi_z \hat{P}_z, \,\phi_z \hat{P}_z]\}, \\
    \hat{\mathcal{R}}^n &= \{\hat{x}- 2n\hat{P}_x \in [-\phi_x\hat{P}_x,\,\phi_x\hat{P}_x]\} \times \{\hat{z} \in [- \hat{P}_z,\,-\phi_z \hat{P}_z]\cup [\phi_z \hat{P}_z,\,\hat{P}_z]\}, \\
    \hat{\mathcal{S}}^n &= \{\hat{x}- 2n\hat{P}_x \in [\phi_x\hat{P}_x,\, (2 - \phi_x)\hat{P}_x]\} \times \{\hat{z} \in [- \hat{P}_z, \,\hat{P}_z]\}.
    \end{align}
\end{subequations}

The steady equations that govern the fluid and surfactant in each periodic cell are described in detail by \citet{tomlinson2023laminar}. 
Here, we highlight differences due to the unsteady transport of surfactant, whilst allowing the total flux of surfactant to vary over the long length scale and slow time scale associated with the channel. 
The bulk surfactant is coupled to the steady incompressible flow through an unsteady advection--diffusion equation. 
In $\hat{\mathcal{D}}_1^n$ and $\hat{\mathcal{D}}_2^n$, 
\refstepcounter{equation} \label{eq:dimensional_equations} 
\begin{equation}
	\hat{\boldsymbol{\nabla}} \cdot \hat{\boldsymbol{u}} = 0, \quad
    \hat{\mu} \hat{\nabla}^2 \hat{\boldsymbol{u}}- \hat{\boldsymbol{\nabla}} \hat{p} = \boldsymbol{0}, \quad
    \hat{D} \hat{\nabla}^2 \hat{c} - \hat{\boldsymbol{u}}\cdot \hat{\boldsymbol{\nabla}} \hat{c}  - \hat{c}_{\hat{t}} = 0, \tag{\theequation\textit{a--c}}
\end{equation}
where $\hat{\mu}$ is dynamic viscosity and $\hat{D}$ is the surfactant bulk diffusivity. 
The interfacial surfactant is coupled to the flow through an unsteady advection--diffusion equation and a linear equation of state and adsorption--desorption kinetics, such that $(\hat{\sigma}_{\hat{x}}, \, \hat{\sigma}_{\hat{z}}) = (-\hat{A}\hat{\Gamma}_{\hat{x}}, \, -\hat{A}\hat{\Gamma}_{\hat{z}})$, where $\hat{\sigma}$ is the surface tension and $\hat{A}$ is the surface activity \citep{manikantan2020surfactant}.
\st{We linearise the equation of state and adsorption--desorption kinetics, as surfactant concentrations are generally small when surfactant is not artificially added \citep{manikantan2020surfactant}.}
At $\hat{y} = 0$ ($\hat{y}= 2\hat{H}$) and along $\hat{\mathcal{I}}^n$, the boundary and bulk--interface coupling conditions for surfactant and flow and the interfacial surfactant transport equation are
\refstepcounter{equation} \label{eq:dimensional_interface_bcs} 
\begin{multline}  
	\hat{\mu} \boldsymbol{n}\cdot \hat{\boldsymbol{\nabla}} \hat{u} - \hat{A} \hat{\Gamma}_{\hat{x}} = 0, \quad
    \hat{v}=0, \quad
    \hat{\mu} \boldsymbol{n}\cdot \hat{\boldsymbol{\nabla}} \hat{w} - \hat{A} \hat{\Gamma}_{\hat{z}} = 0, \quad \hat{D} \boldsymbol{n}\cdot \hat{\boldsymbol{\nabla}} \hat{c} - \hat{K}_a \hat{c} + \hat{K}_d \hat{\Gamma} =0, \\ 
    \hat{D}_I ( \hat{\Gamma}_{\hat{x}\hat{x}} + \hat{\Gamma}_{\hat{z}\hat{z}} ) + \hat{K}_a \hat{c} - \hat{K}_d \hat{\Gamma} -(\hat{u} \hat{\Gamma})_{\hat{x}} - (\hat{w} \hat{\Gamma})_{\hat{z}}  - \hat{\Gamma}_{\hat{t}} = 0, \tag{\theequation\textit{a--e}}
\end{multline} 
where $\boldsymbol{n}$ is the unit normal to the interface (pointing into the channel), $\hat{D}_I$ is the surfactant interfacial diffusivity, $\hat{K}_a$ is the adsorption rate coefficient and $\hat{K}_d$ is the desorption rate coefficient. 
At $\hat{y} = 0$ ($\hat{y} = 2\hat{H}$) on $\partial \hat{\mathcal{I}}^n$, there is no flux of surfactant
\refstepcounter{equation} \label{eq:dimensional_interface_bnd_bcs}
\begin{equation}
    \hat{u}\hat{\Gamma} - \hat{D}_I \hat{\Gamma}_{\hat{x}} = 0 \quad \text{at} \quad \hat{x} = \pm \phi_x \hat{P}_x, \quad \hat{w}\hat{\Gamma} - \hat{D}_I \hat{\Gamma}_{\hat{z}} = 0 \quad \text{at} \quad \hat{z} = \pm \phi_z \hat{P}_z. \tag{\theequation\textit{a,\,b}}
\end{equation}
At $\hat{y} = 0$ ($\hat{y} = 2\hat{H}$) along $\hat{\mathcal{R}}^n\cup\hat{\mathcal{S}}^n$, the flow and surfactant boundary conditions are
\refstepcounter{equation} \label{eq:dimensional_solid_bcs} 
\begin{equation} 
	\hat{u} = 0, \quad \hat{v} =0, \quad \hat{w} = 0, \quad \hat{c}_{\hat{y}} =0.
	\tag{\theequation\textit{a--d}}
\end{equation}
\st{Next, we define $\hat{\boldsymbol{q}}(\hat{\boldsymbol{x}},\,\hat{t}) = (\hat{u}, \, \hat{v}, \, \hat{w}, \, \hat{p}_{\hat{x}}, \, \hat{c})$.}
Throughout $\hat{\mathcal{D}}^n_1$ and $\hat{\mathcal{D}}^n_2$, we assume that the flow and concentration fields are periodic in the spanwise directions,
\begin{equation} \label{eq:dimensional_periodicity} 
    \hat{\boldsymbol{q}}(\hat{x},\,\hat{y},\,-\hat{P}_z,\,\hat{t}) = \hat{\boldsymbol{q}}(\hat{x},\,\hat{y},\,\hat{P}_z,\,\hat{t}).
\end{equation}
Across interfaces between $\hat{\mathcal{D}}_1^n$, $\hat{\mathcal{D}}_2^n$ and $\hat{\mathcal{D}}_1^{n+1}$ the flow and concentration fields are assumed to be continuous between subdomains,
\begin{subequations} \label{eq:dimensional_continuity}
    \begin{align}
    \hat{\boldsymbol{q}}(((2n + \phi_x)\hat{P}_x)^-,\,\hat{y},\,\hat{z},\,\hat{t}) &= \hat{\boldsymbol{q}}(((2n + \phi_x)\hat{P}_x)^+,\,\hat{y},\,\hat{z} ,\,\hat{t}), \\
    \hat{\boldsymbol{q}}(((2(n+1) -\phi_x)\hat{P}_x)^-,\,\hat{y},\,\hat{z},\,\hat{t}) &= \hat{\boldsymbol{q}}(((2(n+1) - \phi_x )\hat{P}_x)^+,\,\hat{y},\,\hat{z},\,\hat{t}).
    \end{align}
\end{subequations}
\st{In addition to \eqref{eq:dimensional_periodicity}--\eqref{eq:dimensional_continuity}, derivatives of $\hat{\boldsymbol{q}}$ should be continuous across subdomains and spanwise periods.}

We can integrate \eqref{eq:dimensional_equations}--\eqref{eq:dimensional_periodicity} across the channel to derive equations relating the bulk and interfacial flux of fluid and surfactant
\begin{subequations} \label{eq:dim_cons_1}
\begin{align}
     \int_{\hat{\mathscr{A}}_n} \hat{c}_{\hat{t}} \, \text{d}\hat{A} + \frac{\text{d}}{\text{d}\hat{x}} \int_{\hat{\mathscr{A}}_n} (\hat{u}\hat{c} - \hat{D} \hat{c}_{\hat{x}}) \, \text{d}\hat{A} - 2 \int_{\hat{\mathscr{I}}_n} (\hat{K}_d \hat{\Gamma} - \hat{K}_a \hat{c}) \, \text{d}\hat{z} = 0 \quad \text{in} \quad \mathcal{D}_1^n, \\ 
    \int_{\hat{\mathscr{I}}_n} \hat{\Gamma}_{\hat{t}}\, \text{d}\hat{z} + \frac{\text{d}}{\text{d}\hat{x}}\int_{\hat{\mathscr{I}}_n} (\hat{u}\hat{\Gamma} - \hat{D}_I \hat{\Gamma}_{\hat{x}}) \, \text{d}\hat{z} +    \int_{\hat{\mathscr{I}}_n} (\hat{K}_d \hat{\Gamma} - \hat{K}_a \hat{c}) \, \text{d}\hat{z} = 0 \quad \text{in} \quad \mathcal{D}_1^n, \\
    \int_{\hat{\mathscr{A}}_n} \hat{c}_{\hat{t}} \, \text{d}\hat{A} + \frac{\text{d}}{\text{d}\hat{x}}  \int_{\hat{\mathscr{A}}_n} (\hat{u}\hat{c} - \hat{D} \hat{c}_{\hat{x}}) \, \text{d}\hat{A}  = 0 \quad \text{in} \quad \mathcal{D}_2^n,
\end{align}
\end{subequations}
where $\int_{\hat{\mathscr{A}}_n}\cdot\, \text{d}\hat{A} \equiv \int_{\hat{z}=-\hat{P}_z}^{\hat{P}_z}\int_{\hat{y}=0}^{2\hat{H}} \cdot \, \text{d} \hat{y} \, \text{d} \hat{z}$ and $\int_{\hat{\mathscr{I}}_n} \cdot \, \text{d}\hat{z} \equiv \int_{\hat{z}=-\hat{P}_z}^{\hat{P}_z} \cdot \, \text{d} \hat{z}$ for $\hat{x} - 2n\hat{P}_x \in [-\phi_x\hat{P}_x, \, (2 - \phi_x)\hat{P}_x]$.
\st{The unsteady surfactant transport equations, \eqref{eq:dim_cons_1}, model how the bulk and interfacial surfactant fluxes (second term in \ref{eq:dim_cons_1}\text{a--c}) change as surfactants adsorb and desorb at the liquid--gas interface (third term in \ref{eq:dim_cons_1}\text{a,\,b}) and the concentration field evolves in time (first term in \ref{eq:dim_cons_1}\text{a--c}).} 
For a flow driven in the streamwise direction, the cross-channel integrated streamwise velocity field, referred to hereafter as the flux of fluid, $\hat{Q}$, is uniform along the length of the channel,
\begin{equation} \label{eq:dimensional_velocity_flux}
    \hat{Q} = \int_{\hat{\mathscr{A}}_n} \hat{u} \, \text{d}\hat{A}.
\end{equation}
In contrast, the cross-channel integrated total flux of surfactant, referred to hereafter as the flux of surfactant, $\hat{K}=\hat{K}(\hat{x}, \, \hat{t})$, can vary along the length of the channel due to unsteady effects, according to
\refstepcounter{equation} \label{eq:dimensional_surfactant_unsteady}
\begin{equation}
    \int_{\hat{\mathscr{A}}_n} \hat{c}_{\hat{t}} \, \text{d}\hat{A} + 2 \int_{\hat{\mathscr{I}}_n} \hat{\Gamma}_{\hat{t}} \, \text{d}\hat{z} + \hat{K}_{\hat{x}} = 0 \quad \text{in} \quad \hat{\mathcal{D}}_1^n, \quad \int_{\hat{\mathscr{A}}_n} \hat{c}_{\hat{t}} \, \text{d}\hat{A} + \hat{K}_{\hat{x}} = 0 \quad \text{in} \quad \hat{\mathcal{D}}_2^n, \tag{\theequation\textit{a,\,b}}
\end{equation}
where we have reformulated \eqref{eq:dim_cons_1} and defined
\begin{subequations} \label{eq:dimensional_surfactant_flux}
\begin{align}
    \hat{K} &= \int_{\hat{\mathscr{A}}_n} (\hat{u}\hat{c} - \hat{D} \hat{c}_{\hat{x}}) \, \text{d}\hat{A} + 2 \int_{\hat{\mathscr{I}}_n} (\hat{u}\hat{\Gamma} - \hat{D}_I \hat{\Gamma}_{\hat{x}}) \, \text{d}\hat{z} \quad \text{in} \quad \hat{\mathcal{D}}_1^n, \\ \hat{K} &= \int_{\hat{\mathscr{A}}_n} (\hat{u}\hat{c} - \hat{D} \hat{c}_{\hat{x}}) \, \text{d}\hat{A} \quad \text{in} \quad \hat{\mathcal{D}}_2^n.
\end{align}
\end{subequations}
We also define $\hat{K}_m = \max(\hat{K}(\hat{x}, \, 0))$ to be the maximum initial surfactant flux along the length of the channel.

Defining the cross-channel-averaged pressure drop per period $\Delta_n \hat{p}(\hat{t}) \equiv \langle \hat{p}\rangle((2n-\phi_x) \hat{P}_x) - \langle \hat{p}\rangle((2(n+1)-\phi_x)\hat{P}_x) > 0$ where $\langle \cdot \rangle \equiv \int_{\hat{z}=-\hat{P}_z}^{\hat{P}_z} \int_{\hat{y}=0}^{2 \hat{H}} \cdot \, \text{d}\hat{y} \, \text{d}\hat{z} / (4 \hat{P}_z \hat{H})$ is the cross-channel average, we can define the normalised drag reduction over the $n$th cell as
\begin{equation}  \label{eq:dimensional_drag}
{DR}_n(\hat{t}) = \frac{\Delta_n \hat{p}_I - \Delta_n \hat{p}}{\Delta_n \hat{p}_I - \Delta_n \hat{p}_U},
\end{equation}
where $\Delta_n \hat{p} = \Delta_n \hat{p}_I$ when the liquid--gas interface is immobilised by surfactant and is no-slip (${DR}_n=0$) and $\Delta_n \hat{p} = \Delta_n \hat{p}_U$ when the liquid--gas interface is unaffected by surfactant and is shear-free (${DR}_n=1$).

\subsection{Non-dimensionalisation and scalings} 

In table \ref{tab:0}, we summarise the different length, time and velocity scales of interest in the transport problem described in \eqref{eq:dimensional_domains}--\eqref{eq:dimensional_surfactant_flux} and figure \ref{fig:my_label}, assuming that the channel has an order-one cross-channel aspect ratio $\hat{H} \sim \hat{P}_z$, but small channel-height-to-streamwise-period ratio $\epsilon = \hat{H}/\hat{P}_x \ll 1$ and small streamwise-period-to-channel-length ratio $\mathcal{E} = \hat{P}_x/\hat{L}_x \ll 1$.
\st{Our aim is to explore a limit that exhibits a balance of dominant physical effects, is relevant to applications, and which leads to a tractable mathematical and numerical problem, avoiding intensive computations of the whole channel and every cell of the SHS.}
Defining $\epsilon \hat{U} = \hat{Q}/(\hat{H}^2)$ as a velocity scale, 
$\hat{C} = \hat{K}_m/\hat{Q}$ as a bulk concentration scale and $\hat{G} = \hat{K}_a \hat{C}/\hat{K}_d$ as an interfacial concentration scale, we non-dimensionalise \eqref{eq:dimensional_domains}--\eqref{eq:dimensional_drag} using multiple time  and spatial scales:
\refstepcounter{equation} \label{eq:nondimensionalisation}
\begin{multline}
    t = \frac{\hat{t}}{\epsilon \hat{P}_x / \hat{U}}, \quad T = \frac{\hat{t}}{\hat{P}_x / \epsilon \hat{U}}, \quad \tau = \frac{\hat{t}}{\hat{P}_x / (\epsilon \mathcal{E} \hat{U})}, \quad \boldsymbol{x}_\perp = \frac{(\hat{y},\,\hat{z})}{\epsilon \hat{P}_x}, \quad x = \frac{\hat{x}}{\hat{P}_x}, \quad \chi = \frac{\hat{x}}{\hat{P}_x / \mathcal{E}}, \\ \boldsymbol{u}_\perp = \frac{(\hat{v},\,\hat{w})}{\hat{U}}, \quad u = \frac{\hat{u}}{\epsilon \hat{U}}, \quad \st{p = \frac{\hat{p}}{\hat{\mu} \hat{U}/\hat{H}},} \quad c = \frac{\hat{c}}{\hat{C}}, \quad \Gamma = \frac{\hat{\Gamma}}{\hat{G}}, \quad K = \frac{\hat{K}}{\hat{K}_m},
    \tag{\theequation\textit{a--l}}
\end{multline}
where $\boldsymbol{x}_\perp \equiv (y_\perp, \, z_\perp)$ and $\boldsymbol{u}_\perp \equiv (v_\perp,\, w_\perp)$.
In this paper, we focus on the distinguished limit in which $\mathcal{E} = \lambda \epsilon^2$ as $\epsilon \rightarrow 0$, where $\lambda$ is an $O(1)$ constant (this scaling clarifies the asymptotics and is reasonable from an applications point of view).
\st{This non-dimensionalisation yields a long-wave theory with rapid cross-channel transport of surfactant over each periodic cell, which will be homogenised to describe the slow transport of surfactant over multiple periods. 
The cross-channel flow decays exponentially fast \citep{mcnair2022surfactant}, but it must  be formally retained to develop a consistent asymptotic model.}
For $n \in \{-N, \, ..., \, N\}$, the $n$th periodic cell becomes, in dimensionless form (using quantities without hats),
\begin{subequations}
    \begin{align}
    \mathcal{D}_1^n &= \{x- 2n\in [-\phi_x, \,\phi_x]\} \times \{y_\perp\in [0, \,2]\} \times \{z_\perp \in [- P_z, \,P_z]\}, \\ 
    \mathcal{D}_2^n &= \{x- 2n\in [\phi_x, \,2 - \phi_x]\} \times \{y_\perp\in [0,\, 2]\} \times \{z_\perp \in [- P_z,\, P_z]\},
    \end{align}
\end{subequations}
where $P_z = \hat{P}_z/\hat{H}$. 
At $y = 0$ ($y= 2$), the regions of the SHS are given by
\begin{subequations}
    \begin{align}
    \mathcal{I}^n &= \{x - 2n \in [-\phi_x, \,\phi_x]\} \times \{z_\perp \in [- \phi_z P_z, \,\phi_z P_z]\}, \\
    \mathcal{R}^n &= \{x- 2n\in [-\phi_x, \,\phi_x]\} \times \{z_\perp \in [- P_z, \,-\phi_z P_z]\cup [\phi_z P_z, \,P_z]\}, \\
    \mathcal{S}^n &= \{x- 2n\in [\phi_x,\, 2 - \phi_x]\} \times \{z_\perp \in [- P_z, \,P_z]\}.
    \end{align}
\end{subequations}
The length of the channel becomes $2 \hat{L}_x / \hat{P}_x = 2/\mathcal{E} = 2 / (\lambda \epsilon^2)$.

\setlength{\tabcolsep}{0.75em}
\begin{table}
\resizebox{\columnwidth}{!}{%
\centering
    \begin{tabular}{c c c c}
    \hline
    Length scale & Time scale & Velocity scale & Interpretation \\[6pt]
    $\epsilon \hat{P}_x$ & $\displaystyle\frac{\epsilon \hat{P}_x}{\hat{U}}$ & $\hat{U}$ & Spanwise flow across a unit cell \\[10pt]
    $\hat{P}_x$ & $\displaystyle\frac{\hat{P}_x}{\epsilon \hat{U}}$ & $\epsilon \hat{U}$  & Streamwise flow along a unit cell \\[10pt]
    $\displaystyle\frac{\hat{P}_x}{\mathcal{E}}$ & $\displaystyle\frac{\hat{P}_x}{\epsilon \mathcal{E} \hat{U}}$ & $\epsilon \hat{U} $ & Streamwise flow over multiple cells \\
    \hline
    \end{tabular}%
    }
    \caption{Summary of the length, time and velocity scales of interest involved in the problem described in \eqref{eq:dimensional_domains}--\eqref{eq:dimensional_surfactant_flux} and figure \ref{fig:my_label},
    where $\epsilon \hat{U}$ is the moderate streamwise velocity scale at the scale of individual cells (owing to continuity of flux), $\epsilon = \hat{H}/\hat{P}_x$ is the ratio of channel height to streamwise period length and $\st{\mathcal{E} = \hat{P}_x/\hat{L}_x = \lambda \epsilon^2}$ is the ratio of streamwise period length to channel length.
    }
    \label{tab:0}
\end{table}

We then assume that the flow and surfactant variables are functions of the short length scale and rapid time scale ($\boldsymbol{x}_{\perp}=(y_{\perp},z_{\perp})$ and $t$, respectively), moderate length scale and intermediate time scale ($x$ and $T$, respectively) and long length scale and slow time scale ($\chi$ and $\tau$, respectively), where these six variables are treated as independent of each other.
In $\mathcal{D}_1^n$ and $\mathcal{D}_2^n$, the incompressible Stokes and surfactant transport equations in \eqref{eq:dimensional_equations} become
\begin{subequations} \label{eq:nondimensional_equations} 
\begin{align}
	\epsilon^2 (u_x + \lambda \epsilon^2 u_\chi) + \boldsymbol{\nabla}_\perp\cdot \boldsymbol{u}_\perp = 0, \\
    \epsilon^2 (u_{xx} + 2  \lambda \epsilon^2 u_{x\chi} +  \lambda^2 \epsilon^4 u_{\chi\chi}) + \nabla^2_\perp u -  p_x - \lambda \epsilon^2 p_\chi = 0, \\ 
    \epsilon^2 (\boldsymbol{u}_{\perp xx} + 2  \lambda \epsilon^2 \boldsymbol{u}_{\perp x\chi} +  \lambda^2 \epsilon^4 \boldsymbol{u}_{\perp \chi\chi}) + \nabla^2_\perp \boldsymbol{u}_{\perp} -  \boldsymbol{\nabla}_{\perp} p = \boldsymbol{0}, \\ 
	(\epsilon^2 (c_{xx} + 2 \lambda \epsilon^2 c_{x\chi} +  \lambda^2 \epsilon^4 c_{\chi\chi})+ \nabla^2_\perp c)/\Pen - \epsilon^2 u (c_x + \lambda \epsilon^2 c_\chi) - \boldsymbol{u}_\perp \cdot \boldsymbol{\nabla}_{\perp} c  \hspace{1.0cm} \nonumber \\ - c_t - \epsilon^2(c_T +  \lambda \epsilon^2 c_\tau) = 0,
\end{align}
\end{subequations}
with $\Pen = \hat{U} \hat{H}/\hat{D}$ the bulk P\'{e}clet number, $\boldsymbol{\nabla}_\perp \equiv (\partial_{y_\perp},\, \partial_{z_\perp})$ and $\nabla^2_\perp \equiv \partial_{y_\perp y_\perp} + \partial_{z_\perp z_\perp}$. 
At $y_\perp = 0$ ($y_\perp=2$) and along $\mathcal{I}^n$, the boundary conditions for flow and surfactant, the coupling conditions, and the interfacial surfactant transport equations in \eqref{eq:dimensional_interface_bcs} give
\begin{subequations}  \label{eq:nondimensional_interface_bcs}
\begin{align}
	\boldsymbol{n}\cdot \boldsymbol{\nabla} u - \Ma (\Gamma_{x} + \lambda \epsilon^2 \Gamma_\chi) = 0, \\
     v_\perp =0, \\
    \boldsymbol{n}\cdot \boldsymbol{\nabla} w_\perp - \Ma \Gamma_{z_\perp} = 0, \\ 
    \boldsymbol{n}\cdot \boldsymbol{\nabla} c - Da (c -\Gamma) = 0, \\ 
    (\epsilon^2 (\Gamma_{xx} + 2  \lambda \epsilon^2 \Gamma_{x\chi} +  \lambda^2 \epsilon^4 \Gamma_{\chi\chi}) + \Gamma_{z_\perp z_\perp} )/\Pen_I  -	\epsilon^2 (u \Gamma)_{x} - \lambda \epsilon^4 (u \Gamma)_{\chi} \hspace{1.5cm} \nonumber \\ - (w_\perp \Gamma)_{z_\perp} - \Gamma_t - \epsilon^2(\Gamma_T +  \lambda \epsilon^2 \Gamma_\tau) - \Bi( c - \Gamma) = 0,
\end{align}
\end{subequations}
with $\Ma = \hat{A}\hat{G}/\hat{\mu}\hat{U}$ the Marangoni number, $\Da= \hat{K}_a \hat{H}/\hat{D}$ the Damk\"{o}hler number, $\Pen_I= \hat{H} \hat{U}/\hat{D}_I$ the interfacial P\'{e}clet number and $\Bi = \hat{K}_d \hat{H}/\hat{U}$ the Biot number. 
At $y_\perp = 0$ ($y_\perp=2$) on $\partial \mathcal{I}^n$, the no-flux interfacial surfactant boundary conditions in \eqref{eq:dimensional_interface_bnd_bcs} become
\begin{subequations}\label{eq:nondimensional_noflux}
\begin{align} 
    u \Gamma - (\Gamma_{x} + \lambda \epsilon^2 \Gamma_\chi)/\Pen_I &= 0 \ \ \text{at} \ \  x = \pm \phi_x, \\ 
    w_\perp \Gamma - \Gamma_{z_\perp}/\Pen_I &= 0 \ \  \text{at} \  \ z_\perp = \pm \phi_z P_z.
\end{align}
\end{subequations}
At $y_\perp = 0$ ($y_\perp=2$)  along $\mathcal{R}^n\cup\mathcal{S}^n$, the no-flux bulk flow and bulk surfactant boundary conditions in \eqref{eq:dimensional_solid_bcs} give
\refstepcounter{equation} \label{eq:nondimensional_solid_bcs}
\begin{equation}
	u = 0, \quad v_\perp = 0, \quad w_\perp =0, \quad \ c_{y_\perp} =0.
    \tag{\theequation\textit{a--d}}
\end{equation}
Defining $\boldsymbol{q} = (u, \, v_\perp, \, w_\perp, \, p_x, \, c)$, across $\mathcal{D}_1^n$ and $\mathcal{D}_2^n$, the spanwise \eqref{eq:dimensional_periodicity} and streamwise (\ref{eq:dimensional_continuity}\textit{a}) continuity conditions for the flow and surfactant become
\begin{subequations} \label{eq:nondimensional_periodicity}
\begin{align}
    \boldsymbol{q}(x,\,y_\perp,\,-P_z,\,t,\,T,\,\chi,\,\tau) &= \boldsymbol{q}(x,\,y_\perp,\,P_z,\,t,\,T,\,\chi,\,\tau), \\
    \boldsymbol{q}((2n + \phi_x)^-,\,y_\perp,\,z_\perp,\,t,\,T,\,\chi,\,\tau) &= \boldsymbol{q}((2n + \phi_x)^+,\,y_\perp,\,z_\perp,\,t,\,T,\,\chi,\,\tau).
\end{align}
\end{subequations}
The streamwise flow and surfactant continuity condition for $\boldsymbol{q}$ between one cell and the next, i.e. between $\mathcal{D}_2^n$ and $\mathcal{D}_1^{n+1}$, in (\ref{eq:dimensional_continuity}\textit{b}) is replaced by a stronger assumption to allow the use of homogenisation theory \citep{bottaro2019flow}, namely that $\boldsymbol{q}$ is a periodic function of the moderate length scale $x$, such that
\begin{equation} \label{eq:nondimensional_continuity}
    \boldsymbol{q}(2n -\phi_x,\,y_\perp,\,z_\perp,\,t,\,T,\,\chi,\,\tau) = \boldsymbol{q}(2(n + 1) - \phi_x,\,y_\perp,\,z_\perp,\,t,\,T,\,\chi,\,\tau).
\end{equation}
Slow variations of flow properties from cell to cell will be accommodated via dependence of the flow and surfactant variables on the long length scale $\chi$ and slow time scale $\tau$.

The bulk and interfacial surfactant fluxes (\ref{eq:dim_cons_1}) satisfy 
\begin{subequations} \label{eq:nondim_cons_1}
\begin{align}
     \int_{\mathscr{A}_n} (c_t + \epsilon^2(c_T + \lambda \epsilon^2 c_\tau))  \, \text{d}A + \epsilon^2 \frac{\text{d}}{\text{d}x} \int_{\mathscr{A}_n} \left( u c -  \frac{c_{ x}}{\Pen} -  \lambda \epsilon^2 \frac{c_{\chi}}{\Pen} \right) \, \text{d} A \hspace{1cm} \nonumber \\ + \lambda \epsilon^4 \frac{\text{d}}{\text{d}\chi} \int_{\mathscr{A}_n} \left( u c -  \frac{c_{ x}}{\Pen} - \lambda \epsilon^2 \frac{ c_{\chi}}{\Pen}\right) \, \text{d} A - \frac{2 \Da}{\Pen} \int_{ \mathscr{I}_n} ( \Gamma -  c) \, \text{d} z_\perp = 0 \ \ \text{in} \ \ \mathcal{D}_1, \\ 
    \int_{ \mathscr{I}_n }  (\Gamma_t + \epsilon^2(\Gamma_T + \lambda \epsilon^2 \Gamma_\tau)) \, \text{d} z_\perp  + \epsilon^2 \frac{\text{d}}{\text{d} x }\int_{ \mathscr{I}_n } \left(u \Gamma - \frac{\Gamma_x}{\Pen_I}- \lambda \epsilon^2 \frac{\Gamma_\chi}{\Pen_I}\right) \, \text{d} z_\perp \hspace{1cm} \nonumber \\ 
    + \lambda \epsilon^4 \frac{\text{d}}{\text{d} \chi}\int_{ \mathscr{I}_n } \left(u \Gamma - \frac{\Gamma_x}{\Pen_I}- \lambda \epsilon^2 \frac{\Gamma_\chi}{\Pen_I}\right) \, \text{d} z_\perp +   \Bi \int_{ \mathscr{I}_n } ( \Gamma -  c)\, \text{d} z_\perp = 0 \ \ \text{in} \ \ \mathcal{D}_1, \\
    \int_{\mathscr{A}_n} (c_t + \epsilon^2(c_T + \lambda \epsilon^2 c_\tau))  \, \text{d}A + \epsilon^2 \frac{\text{d}}{\text{d}x} \int_{\mathscr{A}_n} \left( u c -  \frac{c_{ x}}{\Pen} - \lambda \epsilon^2 \frac{ c_{\chi}}{\Pen} \right) \, \text{d} A \hspace{1cm} \nonumber \\ + \lambda \epsilon^4 \frac{\text{d}}{\text{d}\chi} \int_{\mathscr{A}_n} \left( u c -  \frac{c_{ x}}{\Pen} - \lambda \epsilon^2 \frac{c_{\chi}}{\Pen} \right) \, \text{d} A = 0  \ \ \text{in} \ \ \mathcal{D}_2,
\end{align}
\end{subequations}
where $\int_{\mathscr{A}_n}\cdot\, \text{d}A \equiv \int_{z_\perp=-P_z}^{P_z}\int_{y_\perp=0}^{2H} \cdot \, \text{d} y_\perp \, \text{d} z_\perp$ and $\int_{\mathscr{I}_n} \cdot \, \text{d}z_\perp \equiv \int_{z_\perp=-P_z}^{P_z} \cdot \,  \text{d} z_\perp$ for $x -2n \in [\phi_x, \, 2 - \phi_x]$.
In $\mathcal{D}_1^n$ and $\mathcal{D}_2^n$, the flux of fluid \eqref{eq:dimensional_velocity_flux} is given by
\begin{equation} \label{eq:nondimensional_velocity_flux}
    \int_{\mathscr{A}_n} u \, \text{d} A = 1.
\end{equation}
The flux of surfactant, $K = K(x, \, t, \, T, \, \chi, \, \tau)$, is related to changes in the bulk and surface concentration via \eqref{eq:dimensional_surfactant_unsteady}, which becomes
\begin{subequations} \label{eq:nondimensional_surfactant_unsteady}
\begin{align}
    \int_{\mathscr{A}_n} (c_t + \epsilon^2(c_T + \lambda \epsilon^2 c_\tau)) \, \text{d}A + \frac{2\Da}{\Bi\Pen} \int_{\mathscr{I}_n} (\Gamma_t + \epsilon^2(\Gamma_T + \lambda \epsilon^2 \Gamma_\tau)) \, \text{d}z_\perp \hspace{1.6cm} \nonumber \\  + \epsilon^2(K_{x} + \lambda \epsilon^2 K_\chi) = 0 \quad \text{in} \quad \mathcal{D}_1^n, \\ \int_{\mathscr{A}_n} (c_t + \epsilon^2(c_T + \lambda \epsilon^2 c_\tau)) \, \text{d} A + \epsilon^2(K_{x} + \lambda \epsilon^2 K_\chi) = 0 \quad \text{in} \quad  \mathcal{D}_2^n,
\end{align}
\end{subequations}
where the flux of surfactant \eqref{eq:dimensional_surfactant_flux} is given by
\begin{subequations} \label{eq:nondimensional_surfactant_flux}
\begin{align}
    K &= \int_{ \mathscr{A}_n } \left( u c -  \frac{c_{ x}}{\Pen} - \lambda \epsilon^2 \frac{c_{\chi}}{\Pen} \right) \, \text{d} A\nonumber \\ & \hspace{3.6cm} + \frac{2\Da}{\Bi\Pen} \int_{ \mathscr{I}_n } \left(u \Gamma - \frac{\Gamma_x}{\Pen_I} - \lambda \epsilon^2 \frac{\Gamma_\chi}{\Pen_I}\right) \, \text{d} z_\perp  \quad \text{in} \quad \mathcal{D}_1^n, 
    \\ K &= \int_{ \mathscr{A}_n } \left( u c -  \frac{c_{ x}}{\Pen} - \lambda \epsilon^2 \frac{ c_{\chi}}{\Pen} \right) \, \text{d} A \quad \text{in} \quad \mathcal{D}_2^n,
\end{align}
\end{subequations}
and $\max(K(x, \, t, \, T, \, \chi, \, \tau)) = 1$ at $t = T = \tau = 0$.

The normalised drag reduction \eqref{eq:dimensional_drag} over the $n$th periodic cell becomes
\begin{equation}  \label{eq:nondimensional_drag}
{DR}_n(t, \, T, \, \chi, \, \tau) = \frac{\Delta^n p_I - \Delta^n p}{\Delta^n p_I - \Delta^n p_U},
\end{equation}
where $\Delta^n {p} \equiv \langle p\rangle(2n-\phi_x) - \langle p\rangle(2(n+1)-\phi_x)$ and $\langle \cdot \rangle \equiv \int_{z_\perp=- P_z}^{ P_z} \int_{y_\perp=0}^{2} \cdot \, \text{d} y_\perp \text{d} z_\perp/ (4 P_z)$. 

\subsection{Asymptotic homogenisation} \label{subsec:Asymptotic homogenisation}

We assume that $\Pen \sim \Pen_I \sim \Ma \sim O(1)$ and $\Bi \sim \Da \sim O(\epsilon^2)$ in the limit $\epsilon \ll 1$, so that bulk--surface exchange is comparable to advection, diffusion and Marangoni effects in $\mathcal{D}_1^n$ and $\mathcal{D}_2^n$ for $n \in \{-N, \, ..., \, N\}$. 
As discussed in \citet{tomlinson2023laminar}, this scaling means that we arrive at the most general form of the surfactant transport equations with moderate exchange, whereas, if we had assumed that $\Bi \sim \Da \sim O(1)$, then we would arrive at a sublimit with strong exchange.
In the limit $\epsilon \rightarrow 0$, we rescale $\Bi = \epsilon^2 \mathscr{B}$ and $\Da = \epsilon^2 \mathscr{D}$, where $\mathscr{B}$ and $\mathscr{D}$ are positive $O(1)$ constants.
We then substitute the asymptotic expansion
\begin{equation} \label{eq:asymptotic_expansion_1}
    \begin{pmatrix} u \\ v_\perp \\ w_\perp \\ p \\ c \\ \Gamma \\ K \end{pmatrix}	 = \begin{pmatrix} u_0 \\ v_{0\perp} \\ w_{0\perp} \\ p_0 \\ c_0 \\ \Gamma_0 \\ K_0 \end{pmatrix} + \epsilon^2 \begin{pmatrix} u_1 \\ v_{1\perp} \\ w_{1\perp} \\ p_1 \\ c_1 \\ \Gamma_1 \\ K_1 \end{pmatrix} + \epsilon^4 \begin{pmatrix} u_2 \\ v_{2\perp} \\ w_{2\perp} \\ p_2 \\ c_2 \\ \Gamma_2 \\ K_2 \end{pmatrix} + ...,
\end{equation}
into \eqref{eq:nondimensional_equations}--\eqref{eq:nondimensional_surfactant_flux}. 
The leading-order, first-order and second-order problems are addressed in \S\ref{subsubsec:Leading-order problem}--\ref{subsubsec:Second-order problem} respectively.
\st{At the start of each subsection, we direct the reader who is not interested in the details towards the main equations and results derived in each subsection.}

\subsubsection{Leading-order problem}
\label{subsubsec:Leading-order problem}

\st{First, we simplify the dependence of the velocity, pressure and concentration fields on the space and time variables. 
The streamwise velocity and volume flux are written (in \eqref{eq:u_def}--\eqref{eq:bulk_flux_constraints} below) in terms of the interfacial concentration and pressure gradient, which are independent cross-plane variables ($y_\perp$ and $z_\perp$).}

In $\mathcal{D}_1^n$ and $\mathcal{D}_2^n$, streamwise gradients of the velocity and bulk concentration are small compared to cross-channel gradients. 
Hence, cross-channel diffusion balances advection and unsteady effects in the bulk equation, through the two-dimensional problem
\refstepcounter{equation} \label{eq:nondimensional_equations_1} 
\begin{equation} 
	\boldsymbol{\nabla}_\perp\cdot \boldsymbol{u}_{0\perp} = 0, \quad
    \nabla^2_\perp \boldsymbol{u}_0 -  \boldsymbol{\nabla}p_0 = \boldsymbol{0}, \quad 
	\nabla^2_\perp c_0/\Pen - \boldsymbol{u}_{0\perp} \cdot \boldsymbol{\nabla}_{\perp} c_0  - c_{0t} = 0. \tag{\theequation\textit{a--c}}
\end{equation}
At $y_\perp = 0$ ($y_\perp = 2$) and along $\mathcal{I}^n$, streamwise gradients of the streamwise velocity and surface concentration are small compared to spanwise gradients. 
Hence, spanwise diffusion balances advection and unsteady effects in the interfacial equation, via
\refstepcounter{equation}  \label{eq:nondimensional_interface_bcs_1}
\begin{multline}
	\boldsymbol{n}\cdot \boldsymbol{\nabla} u_0 - \Ma \Gamma_{0x} = 0, \quad
     v_{0\perp} =0, \quad
    \boldsymbol{n}\cdot \boldsymbol{\nabla} w_{0\perp} - \Ma \Gamma_{0z_\perp} = 0, \\
    \boldsymbol{n}\cdot \boldsymbol{\nabla} c_0 = 0, \quad 
    \Gamma_{0z_\perp z_\perp}/\Pen_I  - (w_{0\perp} \Gamma_0)_{z_\perp} - \Gamma_{0t} = 0.   \tag{\theequation\textit{a--e}}
\end{multline}
At $y_\perp = 0$ ($y_\perp=2$) and on $\partial \mathcal{I}^n$, 
\begin{subequations} \label{eq:nondimensional_noflux_1}
\begin{align}
    u_0 \Gamma_0 - \Gamma_{0x}/\Pen_I &= 0 \quad \text{at} \quad x = \pm \phi_x, \\
    w_{0\perp} \Gamma_0 - \Gamma_{0z_\perp}/\Pen_I &= 0 \quad \text{at} \quad z_\perp = \pm \phi_z P_z.
\end{align}
\end{subequations}
At $y_\perp = 0$ ($y_\perp=2$) and along $\mathcal{R}^n \cup \mathcal{S}^n$, 
\refstepcounter{equation} \label{eq:nondimensional_solid_bcs_1}
\begin{equation}
	u_0 = 0, \quad v_{0\perp} = 0,  \quad w_{0\perp} =0, \quad \ c_{0y_\perp} =0.
    \tag{\theequation\textit{a--d}}
\end{equation}
\st{As there are no streamwise gradients of $\boldsymbol{u}_0$ in \eqref{eq:nondimensional_equations_1}--\eqref{eq:nondimensional_interface_bcs_1}, the two-dimensional problem does not capture inner regions near the stagnation points $x=2n \pm \phi_x$. These inner regions are governed by the three-dimensional Stokes equations and guarantee continuity of $\boldsymbol{u}_0$ across domains $\mathcal{D}_1^n$ and $\mathcal{D}_2^n$. }

In $\mathcal{D}_1^n$ and $\mathcal{D}_2^n$, the surfactant field evolves faster in time $t$ than any changes to the flux of surfactant and bulk--surface exchange at leading-order, so that \eqref{eq:nondim_cons_1}--\eqref{eq:nondimensional_velocity_flux} give 
\refstepcounter{equation} \label{eq:nondim_cons_1_0}
\begin{equation}
     \int_{\mathscr{A}_n} c_{0t} \, \text{d}A = 0, \quad 
    \int_{ \mathscr{I}_n }  \Gamma_{0t} \, \text{d} z_\perp  = 0, \quad \int_{\mathscr{A}_n} u_0 \, \text{d} A = 1. \tag{\theequation\textit{a--c}}
\end{equation}
According to \eqref{eq:nondimensional_surfactant_flux}, the flux of surfactant, $K_0 = K_0(x, \, t, \, T,\, \chi, \, \tau)$, is given by,
\begin{subequations} \label{eq:nondimensional_surfactant_flux_1}
\begin{align}
    K_0 &= \int_{ \mathscr{A}_n} \left( u_0 c_0 -  \frac{c_{0 x}}{\Pen} \right) \, \text{d} A  + \frac{2\mathscr{D}}{\mathscr{B} \Pen}\int_{ \mathscr{I}_n } \left(u_0 \Gamma_0 - \frac{\Gamma_{0x}}{\Pen_I}\right) \, \text{d} z_\perp  \quad \text{in} \quad \mathcal{D}_1^n, \\ K_0 &= \int_{ \mathscr{A}_n } \left( u_0 c_0 -  \frac{c_{0 x}}{\Pen} \right) \, \text{d} A \quad \text{in} \quad \mathcal{D}_2^n.
\end{align}
\end{subequations}

\st{The leading-order solution can be expected to decay exponentially fast in $t$ to $\Gamma_{0}=\Gamma_{0}(x, \, T,\, \chi, \, \tau)$, $c_{0} = c_{0}(x, \, T,\, \chi, \, \tau)$, $p_{0} = p_{0}(x, \, T,\, \chi, \, \tau)$, $v_{0\perp}=w_{0\perp}=0$ and $K_0 = K_0(x, \, T, \, \chi, \, \tau)$ \citep{mcnair2022surfactant}, satisfying \eqref{eq:nondim_cons_1_0}.}
\st{That is, at intermediate ($T$) and slow ($\tau$) time scales, the concentration field does not vary in the spanwise direction and there are no concentration gradients associated with velocities in the cross-plane.}
Using linear superposition, we can decompose $u_0$ into a contribution from the streamwise pressure gradient $p_{0x}$ which drives the flow and the streamwise interfacial concentration gradient $\Gamma_{0x}$ which inhibits it owing to adverse Marangoni forces, via
\refstepcounter{equation} \label{eq:u_def}
\begin{equation}
	u_{0} = \tilde{U} p_{0x} + \Ma \bar{U}\Gamma_{0x} \quad \text{in} \quad \mathcal{D}_1^n \quad \text{and} \quad u_{0} = \breve{U} p_{0x} \quad \text{in} \quad \mathcal{D}^n_2,
	\tag{\theequation\textit{a,\,b}}
\end{equation}
where the steady velocity profiles $\tilde{U}(y_\perp, \, z_\perp)$, $\bar{U}(y_\perp, \, z_\perp)$ and $\breve{U}(y_\perp, \, z_\perp)$ are given in Appendix \ref{app:v}.
Substituting \eqref{eq:u_def} into (\ref{eq:nondim_cons_1_0}\textit{c}), we obtain relations between the volume flux, pressure gradient and interfacial surfactant gradient in $\mathcal{D}_1$ and $\mathcal{D}_2$,
\refstepcounter{equation} \label{eq:bulk_flux_constraints} 
\begin{equation} 
    \tilde{Q} p_{0x} + \Ma \bar{Q} \Gamma_{0x} =1, \quad q = \tilde{q} p_{0x} + \Ma \bar{q} \Gamma_{0x} \quad \text{in} \quad \mathcal{D}_1^n, \quad \breve{Q} p_{0x} = 1 \quad \text{in} \quad \mathcal{D}_2^n,
	\tag{\theequation\textit{a--c}}
\end{equation}
where the fluxes $\tilde{Q}$, $\bar{Q}$, $\breve{Q}$, $\tilde{q}$, $\bar{q}$ and $q$ are given in Appendix \ref{app:v}. 
\st{Briefly, $\tilde{Q}$ and $\tilde{q}$ are bulk volume and surface fluxes, respectively, of the flow $\tilde{U}$ driven by the pressure gradient in $\mathcal{D}_1$; $\bar{Q}$ and $\bar{q}$ are bulk volume and surface fluxes, respectively, of the flow $\bar{U}$ driven by the surfactant-induced Marangoni shear stress gradient in $\mathcal{D}_1$; and $\breve{Q}$ is the bulk volume flux of the flow $\breve{U}$ driven by the pressure gradient in $\mathcal{D}_2$.}  

\subsubsection{First-order problem} \label{subsubsec:First-order problem}

\st{Next, we relate the $x$-distributions of the leading-order surfactant concentrations ($c_0$ and $\Gamma_0$) at the scale of the periodic cell to the surfactant-flux distribution $K_0(\chi, \, \tau)$ that varies over long length and slow time scales. 
Assuming the periodic-cell problem is quasi-steady, 
we derive advection--diffusion equations for $c_0(x)$ and $\Gamma_0(x)$ (see \eqref{eq:strong_diff_2} below), parameterised by $K_0$ (via \eqref{eq:strong_diff_flux_2} below) and physical parameters given in \eqref{eq:coefficients}.
}

Solvability conditions are imposed on the first-order problem to constrain $u_{0}$, $c_{0}$ and $\Gamma_{0}$. 
These conditions are provided by the conservation arguments that result in the surfactant transport equations at $O(\epsilon^2)$.
Hence, \eqref{eq:nondim_cons_1} gives
\begin{subequations} \label{eq:nondim_cons_1_1i}
\begin{align}
     \int_{\mathscr{A}_n} c_{1t} \, \text{d}A &= - \int_{\mathscr{A}_n} c_{0T}  \, \text{d}A -\frac{\text{d}}{\text{d}x} \int_{\mathscr{A}_n} \left( u_{0} c_{0} -  \frac{c_{0 x}}{\Pen} \right) \, \text{d} A \nonumber \\ & \hspace{5cm} + \frac{2 \mathscr{D}}{\Pen} \int_{ \mathscr{I}_n } ( \Gamma_{0} -  c_{0}) \, \text{d} z_\perp  \quad \text{in} \quad \mathcal{D}_1, \\ 
    \int_{ \mathscr{I}_n }  \Gamma_{1t} \, \text{d} z_\perp &= - \int_{ \mathscr{I}_n }  \Gamma_{0T} \, \text{d} z_\perp - \frac{\text{d}}{\text{d} x }\int_{ \mathscr{I}_n } \left(u_{0} \Gamma_{0} - \frac{\Gamma_{0x}}{\Pen_I}\right) \, \text{d} z_\perp \hspace{3cm} \nonumber \\ & \hspace{5.25cm} -  \mathscr{B}  \int_{ \mathscr{I}_n } ( \Gamma_{0} -  c_{0})\, \text{d} z_\perp  \quad \text{in} \quad \mathcal{D}_1, \\
    \int_{\mathscr{A}_n} c_{1t}  \, \text{d}A &= - \int_{\mathscr{A}_n} c_{0T}  \, \text{d}A -\frac{\text{d}}{\text{d}x} \int_{\mathscr{A}_n} \left( u_{0} c_{0} -  \frac{c_{0 x}}{\Pen} \right) \, \text{d} A \quad \text{in} \quad \mathcal{D}_2,
\end{align}
\end{subequations}
and \eqref{eq:nondimensional_surfactant_flux} becomes
\begin{subequations} \label{eq:nondimensional_surfactant_flux_2}
\begin{align}
    K_1 &= \int_{ \mathscr{A}_n } \left( u_0 c_1 + u_1 c_0 -  \frac{c_{1 x}}{\Pen} -  \lambda \frac{c_{0 \chi}}{\Pen} \right) \, \text{d} A \hspace{5cm} \nonumber \\ & \hspace{2.4cm} + \frac{2\mathscr{D}}{\mathscr{B} \Pen} \int_{ \mathscr{I}_n } \left(u_0 \Gamma_1 + u_1\Gamma_0 - \frac{\Gamma_{1x}}{\Pen_I}- \lambda \frac{\Gamma_{0\chi}}{\Pen_I} \right) \, \text{d} z_\perp  \quad \text{in} \quad  \mathcal{D}^n_1, \\ K_1 &= \int_{ \mathscr{A}_n } \left( u_0 c_1 + u_1 c_0 -  \frac{c_{1 x}}{\Pen} -  \lambda \frac{c_{0 \chi}}{\Pen} \right) \, \text{d} A \quad \text{in} \quad \mathcal{D}^n_2.
\end{align}
\end{subequations}

To avoid secular growth of the net mass of surfactant in \eqref{eq:nondim_cons_1_1i}, we require that the right-hand sides of \eqref{eq:nondim_cons_1_1i} are zero \jl{\citep{bender2013advanced}}.
Hence, the bulk and interfacial concentrations evolve over intermediate time scales according to
\begin{subequations} \label{eq:nondim_cons_1_1_f}
\begin{align}
     \int_{\mathscr{A}_n} c_{0T}  \, \text{d}A + \frac{\text{d}}{\text{d}x} \int_{\mathscr{A}_n} \left( u_{0} c_{0} -  \frac{c_{0 x}}{\Pen} \right) \, \text{d} A - \frac{2 \mathscr{D}}{\Pen} \int_{ \mathscr{I}_n } ( \Gamma_{0} -  c_{0}) \, \text{d} z_\perp = 0  \ \  \text{in}  \ \  \mathcal{D}_1^n, \\
      \int_{\mathscr{I}_n} \Gamma_{0T}  \, \text{d}z_\perp + \frac{\text{d}}{\text{d} x }\int_{ \mathscr{I}_n } \left(u_{0} \Gamma_{0} - \frac{\Gamma_{0x}}{\Pen_I}\right) \, \text{d} z_\perp  +  \mathscr{B}  \int_{ \mathscr{I} } ( \Gamma_{0} -  c_{0})\, \text{d} z_\perp = 0 \ \ \text{in} \ \ \mathcal{D}_1^n, \\
      \int_{\mathscr{A}_n} c_{0T}  \, \text{d}A + \frac{\text{d}}{\text{d}x} \int_{\mathscr{A}_n} \left( u_{0} c_{0} -  \frac{c_{0 x}}{\Pen} \right) \, \text{d} A =0  \ \  \text{in} \ \ \mathcal{D}_2^n.
\end{align}
\end{subequations}
Substituting the velocity and flux conditions \eqref{eq:u_def}--\eqref{eq:bulk_flux_constraints} into the surfactant transport equations \eqref{eq:nondim_cons_1_1_f} gives us ODEs that govern the unsteady advection, diffusion and exchange of surfactant over one period, 
\begin{subequations} \label{eq:strong_diff}
\begin{align}
      \theta c_{0T} + c_{0x} - \alpha c_{0xx} - \nu(\Gamma_0 - c_0) &= 0 \quad \text{in} \quad \mathcal{D}_1^n, \\
      \zeta \Gamma_{0T} + \beta \Gamma_{0x} - \gamma (\Gamma_{0} \Gamma_{0x})_x - \delta \Gamma_{0xx} - \nu(c_0 - \Gamma_0) &= 0 \quad \text{in} \quad \mathcal{D}_1^n, \\
      \theta c_{0T} + c_{0x} - \alpha c_{0xx}  &= 0 \quad \text{in} \quad \mathcal{D}_2^n.
\end{align}
\end{subequations}
We describe \eqref{eq:strong_diff} as the unsteady moderate-exchange equations. 
The steady-state problem was solved in \citet{tomlinson2023laminar}, where the transport coefficients $\alpha$, $\beta$, $\gamma$, $\delta$ and $\nu$ (specified below) were defined in terms of physical and geometrical parameters and the fluxes $\tilde{Q}$, $\bar{Q}$, $\tilde{q}$ and $\bar{q}$. 
\jl{The new transport coefficients $\theta$ and $\zeta$ are associated with unsteady effects for the bulk and interfacial surfactant concentration, respectively (see more details below).}
Combining \eqref{eq:strong_diff} with \eqref{eq:nondimensional_surfactant_flux_1} gives a set of constraints on the total surfactant flux over one period, 
\begin{subequations} \label{eq:strong_diff_flux}
\begin{align}
      \theta c_{0T} + \zeta \Gamma_{0T} + K_{0x} &= 0, \ \text{where} \ K_0 = c_{0} - \alpha c_{0x} + \beta \Gamma_{0} - \gamma \Gamma_{0} \Gamma_{0x} - \delta \Gamma_{0x} \ \text{in} \ \mathcal{D}_1^n, \\
       \theta c_{0T} + K_{0x} &= 0, \ \text{where} \  K_0 = c_{0} - \alpha c_{0x} \ \text{in} \ \mathcal{D}_2^n.
\end{align}
\end{subequations}
We solve \eqref{eq:strong_diff}--\eqref{eq:strong_diff_flux} subject to boundary conditions which enforce continuity and periodicity, of both the surfactant concentration and flux, between subdomains
\begin{subequations} \label{eq:model3}
\begin{align}
        c_{0}(2n+\phi_x^{-}, \, T,\, \chi,\,\tau) &= c_{0}(2n+\phi_x^{+}, \, T,\, \chi,\,\tau), \\ 
        c_{0}(2n-\phi_x, \, T,\, \chi,\,\tau) &= c_{0}(2(n+1) - \phi_x, \, T,\, \chi,\,\tau), \\  
        K_0(2n+\phi_x^{-}, \, T,\, \chi,\,\tau) &= K_0(2n +\phi_x^{+}, \, T,\, \chi,\,\tau), \\
        K_0(2n -\phi_x, \, T,\, \chi,\,\tau) &= K_0(2(n+1) - \phi_x, \, T,\, \chi,\,\tau), \\ 
        [\beta\Gamma_{0} - \gamma \Gamma_0 \Gamma_{0x} - \delta \Gamma_{0x}](2n \pm \phi_x, \, T,\, \chi,\,\tau) &= 0.
\end{align}
\end{subequations}
In \eqref{eq:strong_diff}--\eqref{eq:model3}, we have introduced the following transport coefficients: 
\begin{subequations} \label{eq:coefficients}
\begin{alignat}{2}
    \alpha &= \frac{4 P_z}{\Pen} \quad \quad \quad \quad &&\text{(bulk diffusion)}; \label{eq:coefficients1} \\ 
   \beta &= \frac{2 \mathscr{D}  \tilde{q}}{\mathscr{B} \Pen \tilde{Q}} \quad \quad \quad \quad &&\text{(partition coefficient)}; \label{eq:coefficients2} \\  
   \gamma &=  \frac{2 \Ma \mathscr{D}(\tilde{q}\bar{Q}/\tilde{Q} - \bar{q})}{\mathscr{B}\Pen} \quad \quad \quad \quad &&\text{(surfactant strength)}; \label{eq:coefficients3} \\
    \delta &= \frac{4 \phi_z P_z \mathscr{D}}{\mathscr{B} \Pen \Pen_I} \quad \quad \quad \quad &&\text{(surface diffusion)}; \label{eq:coefficients4} \\
   \nu &= \frac{4\phi_z P_z \mathscr{D}}{\Pen} \quad \quad \quad \quad &&\text{(exchange strength)}; \label{eq:coefficients5} \\
   \theta &= 4 P_z \quad \quad \quad \quad &&\text{(bulk capacitance)}; \label{eq:coefficients6} \\
   \zeta &= \frac{4 \phi_z P_z \mathscr{D}}{\mathscr{B} \Pen} \quad \quad \quad \quad &&\text{(surface capacitance)}. \label{eq:coefficients7}
\end{alignat}
\end{subequations}
The bulk (surface) diffusion coefficient $\alpha > 0$ ($\delta>0$) compares the strength of bulk (interfacial) streamwise diffusion to advection.
The partition coefficient $\beta > 0$ characterises the distribution of the surfactant flux, where for $\beta \gg 1$ ($\beta \ll 1$) the interfacial (bulk) surfactant flux dominates.
The surfactant strength $\gamma > 0$ characterises the impact of Marangoni stresses on the interfacial surfactant flux.
The exchange strength $\nu > 0$ compares the rate of adsorption to advection.
The remaining parameters, $\theta$ and $\zeta$, are associated with time-dependent variations and were not reported in \citet{tomlinson2023laminar}.
The bulk capacitance coefficient $\theta>0$ characterises the transverse aspect ratio of the channel and specifies the bulk response to time-dependent changes in the surfactant flux. 
The surface capacitance $\zeta>0$ is the rescaled (by $4\phi_z P_z$) surfactant depletion depth $L_d = \mathscr{D}/(\mathscr{B} \Pen)$; $\zeta$ captures the manner in which solubility regulates the surface response to gradients in the surfactant flux. 
The dependence of the transport coefficients on dimensional parameters will be discussed later in \S\ref{sec:discussion}.

\begin{figure}
    \centering
    \raisebox{4cm}{(\textit{a})}\includegraphics[width=.45\textwidth]{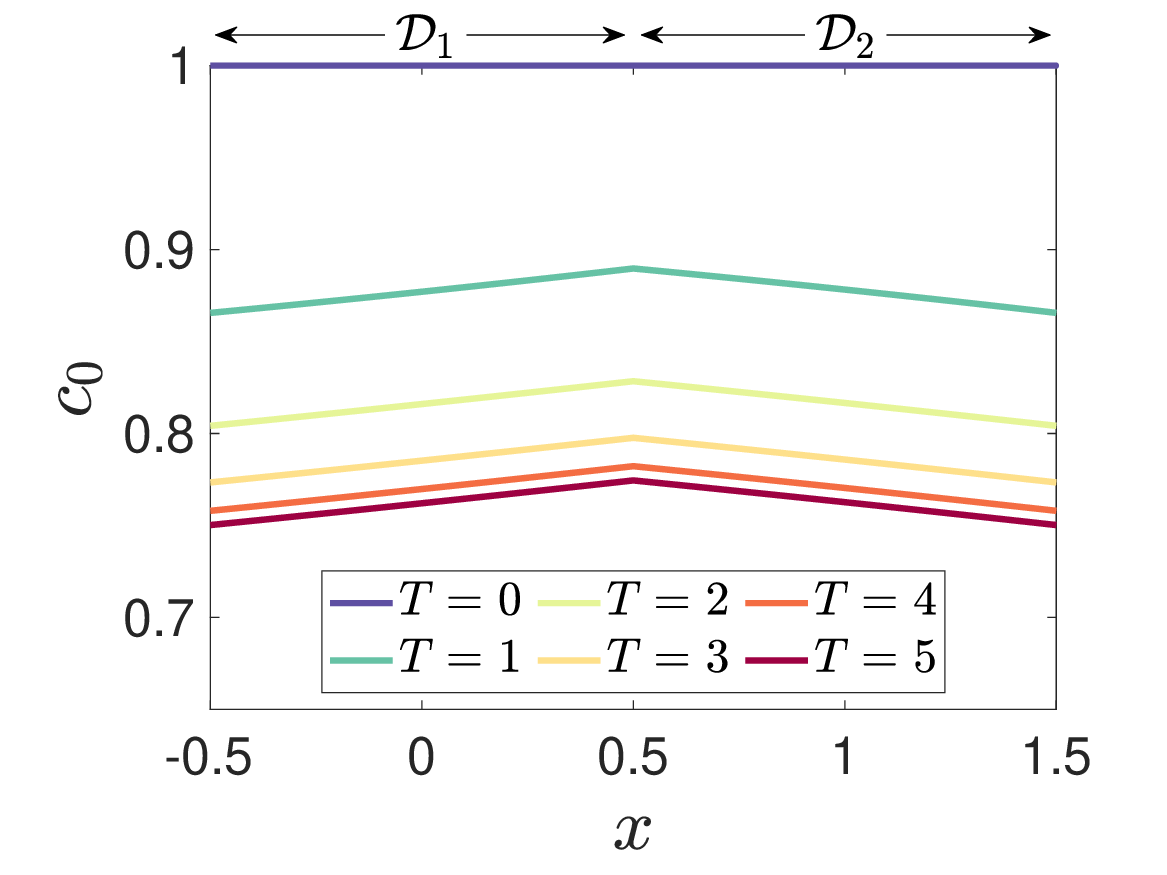}\raisebox{4cm}{(\textit{b})}\includegraphics[width=.45\textwidth]{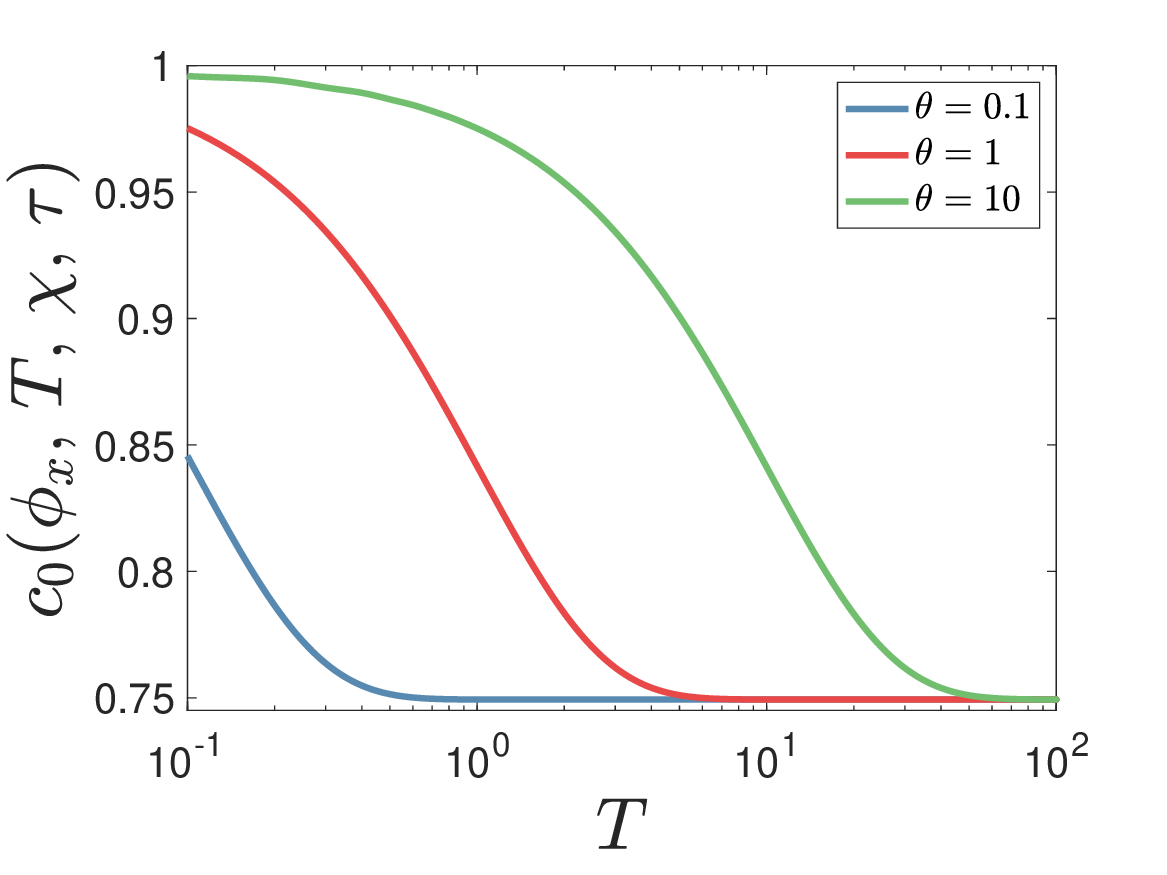}\hfill \hfill \hfill \\
    \caption{
    \st{The unsteady bulk concentration field ($c_0$) in a unit cell ($\mathcal{D}_1\cup\mathcal{D}_2$), computed using \eqref{eq:strong_diff}--\eqref{eq:model3} with $c_0(x, \, 0, \, \chi, \, 0) = 1$, where bulk--surface exchange is strong ($c_0\approx\Gamma_0$).} 
    The drag reduction depends on $\Delta c_0 = c_0(\phi_x, \, T, \, \chi, \, \tau) -  c_0(-\phi_x, \, T, \, \chi, \, \tau)$ and a large gradient of $c_0$ in $\mathcal{D}_1$ can partially immobilise the liquid--gas interface. 
    Plot of (\textit{a}) $c_0(x, \, T, \, \chi, \, \tau)$ at different times $T$ for $\theta = \zeta = 1$; and (\textit{b}) leading-order bulk concentration at the end of the plastron $c_0(\phi_x, \, T, \, \chi, \, \tau)$ for varying bulk capacitance $\theta = \zeta$, computed using \eqref{eq:strong_diff_2} where $\alpha = \delta = 10$ and $\beta = \gamma = 1$.   
    }
    \label{fig:int}
\end{figure}

We solve \eqref{eq:strong_diff}--\eqref{eq:model3} subject to the initial condition $c_0(x, \, 0, \, \chi, \, 0) = 1$ to illustrate the dependence of the bulk concentration field on the intermediate time scale $T$ in figure \ref{fig:int}(\textit{a}); convergence to a steady state for different values of $\theta = \zeta$ is illustrated using $c_0(\phi_x, \, T, \, \chi, \, \tau)$ in figure \ref{fig:int}(\textit{b}).
The initially uniform concentration falls to an equilibrium state, periodic over the unit cell, in which the positive gradient ($c_{0x}> 0$) in $\mathcal{D}_1$ generates an interfacial stress opposing the mean flow.
\st{The discontinuities in $c_{0x}$ are related to the use of a long-wave theory, which does not resolve the inner problems near the contact lines.}
The time taken to reach a steady state increases with $\theta$ and $\zeta$. 
However, as our primary objective is to investigate surfactant transport over the full length of the channel, we assume that the leading-order solution is close to equilibrium and the concentration field no longer depends on the intermediate time $T$, i.e. $\Gamma_{0} = \Gamma_{0}(x, \, \chi, \, \tau)$ and $c_{0} = c_{0}(x, \, \chi, \, \tau)$. 
\st{As the concentration field does not depend on $T$, from  \eqref{eq:strong_diff_flux}, $K_{0x} = 0$ in $\mathcal{D}_1$ and $\mathcal{D}_2$, and therefore the problem in each period simplifies to finding $c_{0} = c_{0}(x; \, K_0)$ and $\Gamma_{0} = \Gamma_{0}(x; \, K_0)$ for a given surfactant flux $K_0 = K_0(\chi, \, \tau)$ that is uniform along each periodic cell.}  
Hence, we solve the steady moderate-exchange equations from \citet{tomlinson2023laminar}, given by
\begin{subequations} \label{eq:strong_diff_2}
\begin{align} 
      c_{0x} - \alpha c_{0xx} - \nu(\Gamma_0 - c_0) &= 0 \quad \text{in} \quad \mathcal{D}_1^n, \\ 
      \beta \Gamma_{0x} - \gamma (\Gamma_{0} \Gamma_{0x})_x - \delta \Gamma_{0xx} - \nu(c_0 - \Gamma_0) &= 0 \quad \text{in} \quad \mathcal{D}_1^n, \\
      c_{0x} - \alpha c_{0xx} &= 0 \quad \text{in} \quad \mathcal{D}_2^n,
\end{align} 
\end{subequations}
subject to the steady surfactant flux conditions 
\refstepcounter{equation} \label{eq:strong_diff_flux_2}
\begin{equation} 
      K_0 = c_{0} - \alpha c_{0x} + \beta \Gamma_{0} - \gamma \Gamma_{0} \Gamma_{0x} - \delta \Gamma_{0x} \quad \text{in} \quad \mathcal{D}_1^n, \quad
      K_0 = c_{0} - \alpha c_{0x} \quad \text{in} \quad \mathcal{D}_2^n, \tag{\theequation\textit{a,\,b}}
\end{equation}
and boundary conditions given in \eqref{eq:model3}. 

The solution to the surfactant concentration transport equations (\ref{eq:model3}, \ref{eq:strong_diff_2}, \ref{eq:strong_diff_flux_2}) exhibits multiple asymptotic regimes, which are discussed in detail in \citet{tomlinson2023laminar}.
Briefly, we distinguish a strong-exchange problem ($\nu \gg \max(1, \, \alpha, \, \delta)$), where the $c_0$ and $\Gamma_0$ fields are in equilibrium ($c_0 \approx \Gamma_0$), from a moderate-exchange problem ($\nu = O(1, \, \alpha, \, \delta)$), where $c_0$ and $\Gamma_0$ are distinct.
In the strong-exchange problem, we identify three primary areas of parameter space and two significant boundaries between them; these are summarised in figure \ref{fig:dr}(\textit{a}).
In the Marangoni-dominated (M) region (analysed in Appendix \ref{subsec:a_strong_surfactant_strength}), the interfacial surfactant gradient immobilises the liquid--gas interface (leading to low drag reduction); in the advection-dominated (A) region (Appendix \ref{subsec:a_strong_advection}), the interfacial surfactant is swept to the downstream stagnation point of each plastron and the liquid--gas interface is mostly shear-free (high drag reduction); and in the diffusion-dominated (D) region (Appendix \ref{subsec:a_strong_diffusion}), the surfactant gradient is attenuated by diffusion and the liquid--gas interface is mostly shear-free (high drag reduction).
Across the advection--Marangoni (AM) (Appendix \ref{subsec:a_strong_diff_strong_surfactant_strength}) and the diffusion--Marangoni (DM) boundaries (Appendix \ref{subsec:a_strong_diff_and_surf}), these effects compete to partially immobilise the liquid--gas interface (moderate drag reduction).
Each of these regions has an analogue when exchange is weak ($\nu \ll \min(1, \, \alpha, \, \delta)$; see figure \ref{fig:dr}\textit{b}). 
These sub-regions have the same leading-order physics as regions M, D and A and are referred to as Marangoni--exchange (M$_\text{E}$), diffusion--exchange (D$_\text{E}$) and advection--exchange  (A$_\text{E}$) sub-regions. 

\begin{figure}
    \centering
    \raisebox{4cm}{(\textit{a})}\includegraphics[width=.45\textwidth]{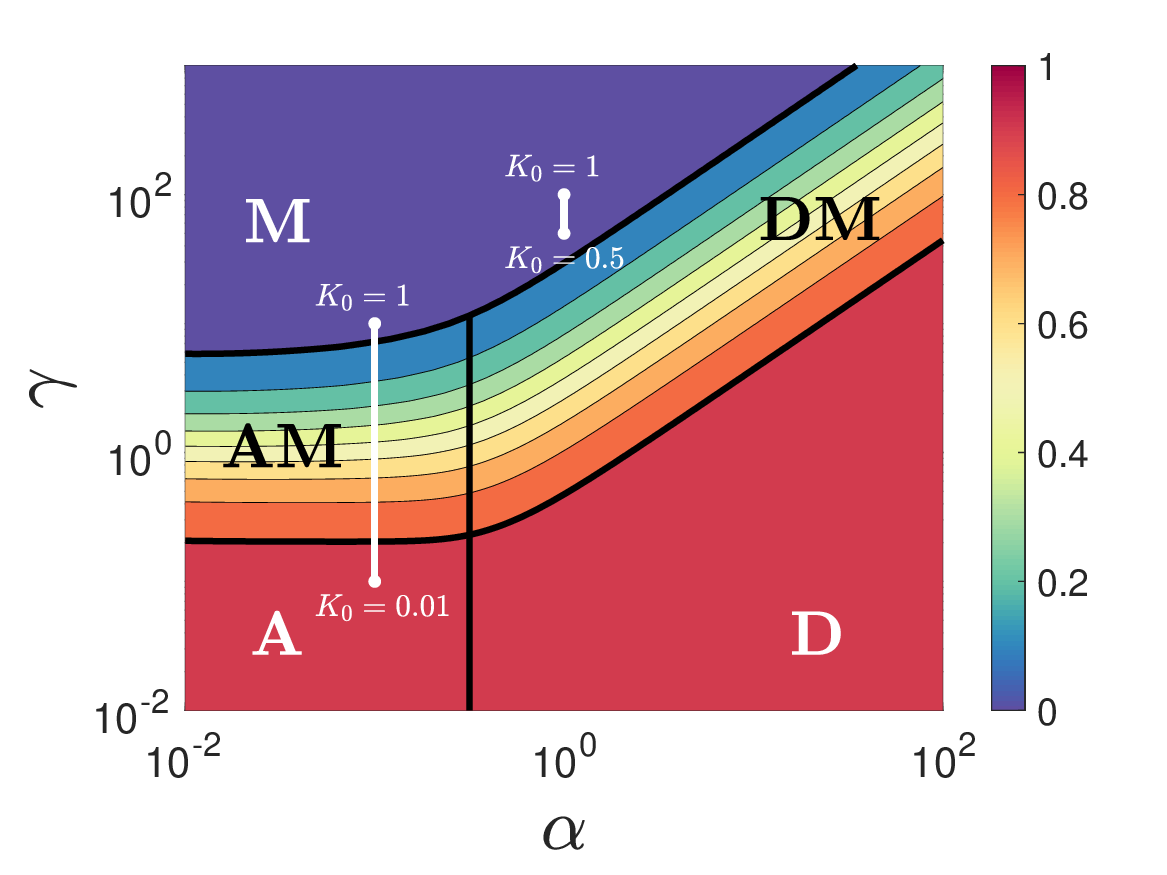}\raisebox{4cm}{(\textit{b})}\includegraphics[width=.45\textwidth]{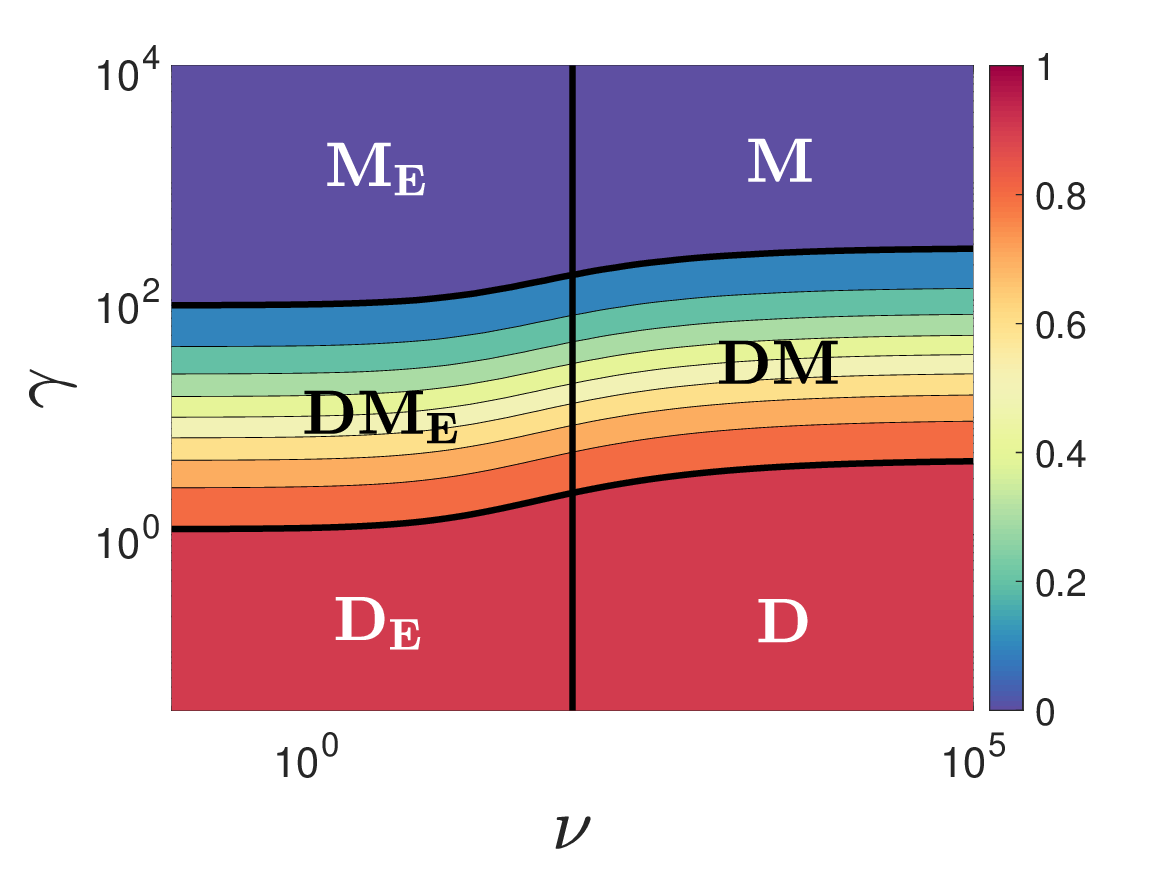} \hfill \hfill \hfill \\
    \caption{
    Contours of the leading-order drag reduction (${DR}_0$) in the (\textit{a}) $(\alpha, \, \gamma)$-plane for $\nu = 100$ (strong-exchange problem) and  (\textit{b}) in the $(\nu, \, \gamma)$-plane for $\alpha = 10$ (illustrating the transition from weak to strong exchange as $\nu$ increases), computed using \eqref{eq:strong_diff_2} where $\alpha = \delta$, $\beta = 1$, $K_0 = 1$, $\phi_x = 0.5$ and $P_z=1$ in both panels.
    When ${DR}_0 = 0$ the liquid-gas interface is no-slip and when ${DR}_0 = 1$ the liquid-gas interface is shear-free.
    The Marangoni (M), advection (A) and diffusion (D) regions, the advection--Marangoni (AM) and diffusion--Marangoni (DM) boundaries (taken to lie in between the ${DR}=0.1$ and ${DR}_0=0.9$ contours), the Marangoni--exchange (M$_\text{E}$) and diffusion--exchange (D$_\text{E}$) sub-regions, and the diffusion--Marangoni--exchange boundary (DM$_\text{E}$), are separated by black lines. 
    The advection--diffusion and moderate-exchange boundaries are illustrated using black lines at $\alpha = 0.3$ and $\nu = 100$ in (\textit{a}) and (\textit{b}), respectively.
    The white lines illustrate the relationship between $K_0$ and $\gamma$ that is discussed at the end of \S\ref{subsubsec:First-order problem}.
    }
    \label{fig:dr}
\end{figure}

The link between surfactant flux $K_0$ and surfactant concentration is evident from \eqref{eq:strong_diff_flux_2}, by noting that $K_0$ can be scaled to unity under the mapping $c_0\rightarrow K_0 c^*_0$, $\Gamma_0 \rightarrow K_0 \Gamma^*_0$ and $\gamma \rightarrow \gamma^*/K_0$. 
Equivalently, by solving the surfactant transport equations (\ref{eq:strong_diff_2}, \ref{eq:strong_diff_flux_2}) with $K_0=1$, we can capture variations in the surfactant flux parametrically through variations in ${\gamma}$. For instance, increasing $K_0$ from $1/2$ to $1$ for fixed $\alpha = 1$ and $\gamma=100$ is equivalent (for the rescaled concentrations $c^*_0$ and $\Gamma^*_0$) to setting $K_0=1$ and increasing $\gamma$ from 50 to 100 (illustrated by the right white line in figure \ref{fig:dr}\textit{a}), thus moving away from the DM boundary and further into the M region. Similarly, increasing $K_0$ from $0.01$ to $1$ for fixed $\alpha = 0.1$ and $\gamma=10$ has the more dramatic effect of moving from the A region (high drag reduction) to the M region (low drag reduction), by varying $\gamma$ from $0.1$ to $10$ with $K_0=1$ (illustrated by the left white line in figure \ref{fig:dr}\textit{a}).  
While we could reduce the number of parameters in the plastron-scale problem by using the rescaled concentrations $c^*_0$ and $\Gamma^*_0$ and by subsuming the parameter $K_0$ into the rescaled surfactant strength parameter $\gamma^*=\gamma K_0$, we choose to retain $K_0$ explicitly and use the concentrations $c_0$ and $\Gamma_0$, because it provides a crucial link between the plastron-scale and large channel-scale problems. 
We will return to these examples in \S\ref{sec:results}.

\subsubsection{Second-order problem}
\label{subsubsec:Second-order problem}

\st{Finally, we relate the surfactant-flux distribution $K_0(\chi,\,\tau)$ to the surfactant concentrations $c_0(x;\, K_0)$ and $\Gamma_0(x;\, K_0)$.
We derive an unsteady advection--diffusion equation for $K_0$ at the channel scale (see \eqref{eq:advection} below), which is coupled to $c_0$ and $\Gamma_0$ through nonlinear coefficients (see \eqref{eq:Rs} below).
}

Solvability conditions are imposed on the second-order problem to constrain $u_{1}$, $c_{1}$ and $\Gamma_{1}$ appearing in \eqref{eq:nondim_cons_1_1i}--\eqref{eq:nondimensional_surfactant_flux_2}. 
These conditions are provided by the conservation arguments that result in the surfactant transport equations at $O(\epsilon^4)$.
In $\mathcal{D}_1^n$, (\ref{eq:nondim_cons_1}\textit{a,\,b}) give bulk and interfacial equations, which can be combined into
\begin{multline} \label{eq:nondim_cons_2_1}
     \int_{\mathscr{A}_n} (c_{2t} + c_{1T})  \, \text{d}A + \frac{2\mathscr{D}}{\mathscr{B} \Pen} \int_{ \mathscr{I}_n } (\Gamma_{2t} + \Gamma_{1T}) \, \text{d} z_\perp + K_{1x} = - \lambda \bigg( \int_{\mathscr{A}_n} c_{0\tau}  \, \text{d}A \\ +  \frac{2\mathscr{D}}{\mathscr{B} \Pen} \int_{ \mathscr{I}_n }  \Gamma_{0\tau} \, \text{d} z_\perp + K_{0\chi} 
     - \lambda \epsilon^2\bigg(\int_{\mathscr{A}_n} \frac{c_{0\chi\chi}}{\Pen}  \, \text{d}A +  \frac{2\mathscr{D}}{\mathscr{B} \Pen} \int_{ \mathscr{I}_n }  \frac{\Gamma_{0\chi\chi}}{\Pen_I} \, \text{d} z_\perp \bigg)\bigg),
\end{multline}
and in $\mathcal{D}_2$, (\ref{eq:nondim_cons_1}\textit{c}) gives
\begin{equation} \label{eq:nondim_cons_2_2}
       \int_{\mathscr{A}_n} (c_{2t}  + c_{1T}) \, \text{d}A + K_{1x} = - \lambda \bigg(\int_{\mathscr{A}_n} c_{0\tau} \, \text{d}A +  K_{0\chi} - \lambda \epsilon^2 \int_{\mathscr{A}_n} \frac{c_{0\chi\chi}}{\Pen} \, \text{d}A\bigg),
\end{equation}
using the definition of the leading- and first-order surfactant fluxes $K_0$ and $K_1$ given in \eqref{eq:nondimensional_surfactant_flux_1} and \eqref{eq:nondimensional_surfactant_flux_2}, respectively.
\st{In \eqref{eq:nondim_cons_2_1}--\eqref{eq:nondim_cons_2_2}, we have retained only the 
 $O(\epsilon^2)$ diffusion terms in order to regularise any shocks that may arise in the numerical solution of these equations, because of the nonlinear dependence of $c_0$ and $\Gamma_0$ on $K_0$ (discussed further in \S\ref{sec:results} below).}

\begin{figure}
    \centering
    \includegraphics[trim={0cm 0cm 0cm 0cm},clip,width=\linewidth]{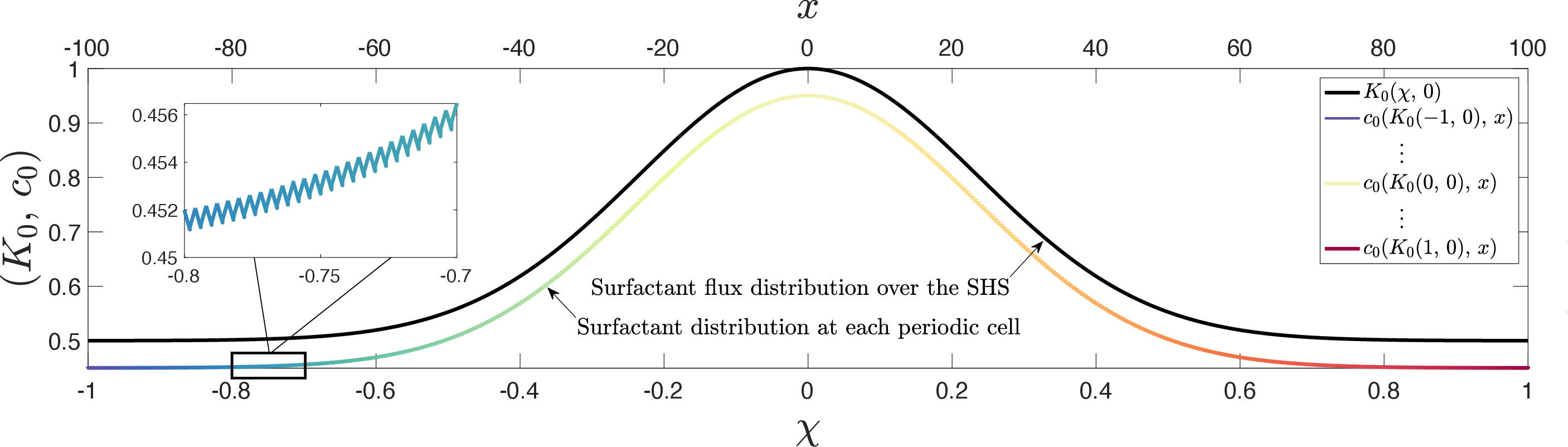}
    \caption{A plot illustrating the relationship between the total flux of surfactant ($K_0 = (1+\exp(-9\chi^2))/2$) over the length of the SHS $(\chi\in[-1,\,1])$ and the corresponding concentration field ($c_0$ and $\Gamma_0$) over $2N = 500$ periods $(x\in[-100, \, 100])$, computed using \eqref{eq:marangoni_limited_results} where $\alpha = \delta = 0.1$, $\beta = 1$, $\gamma = 10$, $\phi_x = 0.5$ and bulk--surface exchange is strong ($c_0 = \Gamma_0$).}
    \label{fig:new_fig}
\end{figure}

To avoid secular growth of the net mass of surfactant, we require that the combined right-hand sides (i.e. source/sink terms) of the surfactant transport equations, \eqref{eq:nondim_cons_2_1}--\eqref{eq:nondim_cons_2_2}, integrate to zero along one period.
We know from \S\ref{subsubsec:First-order problem} that $c_0 = c_{0}(x; \, K_0)$ and $\Gamma_0 = \Gamma_{0}(x; \, K_0)$  where $K_0 = K_0(\chi, \, \tau)$.
As $c_0$ is assumed to be periodic in $x$, we can use the cell with $n=0$ as representative of all others.
Hence, integrating the surfactant transport equations \eqref{eq:nondim_cons_2_1}--\eqref{eq:nondim_cons_2_2} over one period, using the velocity fields and fluxes from \eqref{eq:u_def}--\eqref{eq:bulk_flux_constraints} and using the definition of the transport coefficients in \eqref{eq:coefficients}, we obtain
\begin{equation} \label{eq:advection_1}
    \frac{\text{d} C_0}{\text{d} \tau} + \frac{\text{d} A_0}{\text{d} \chi} - \frac{\text{d} M_0}{\text{d} \chi} - \frac{\text{d} D_0}{\text{d} \chi}
    -\lambda \epsilon^2 \frac{\text{d}^2 D_1}{\text{d} \chi^2} = 0,
\end{equation}
where 
\begin{subequations} \label{eq:Rs}
\begin{alignat}{2}
    C_0(K_0) &=  \theta \int_{x=-\phi_x}^{2-\phi_x} c_{0}\, \text{d}x + \zeta\int_{x=-\phi_x}^{\phi_x} \Gamma_{0} \, \text{d}x, \quad &&\text{(total weighted concentration)}, \\
    A_0(K_0) &= \int_{x=-\phi_x}^{2-\phi_x} c_{0}\, \text{d}x + \beta \int_{x=\phi_x}^{\phi_x} \Gamma_{0}\, \text{d}x \quad &&\text{(advective flux)}, \\
    M_0(K_0) &= \gamma\bigg[\frac{\Gamma_{0}^2}{2}\bigg]_{x = -\phi_x}^{\phi_x} \quad &&\text{(Marangoni flux)},\\
    D_0(K_0) &= \alpha \bigg[c_{0}\bigg]_{x = -\phi_x}^{2-\phi_x}  + \delta \bigg[\Gamma_{0}\bigg]_{x = -\phi_x}^{\phi_x} \quad &&\text{(primary diffusive flux)}, \\
    D_1(K_0) &= \alpha \int_{x=-\phi_x}^{2-\phi_x} c_{0}\, \text{d}x + \delta\int_{x=-\phi_x}^{\phi_x} \Gamma_{0} \, \text{d}x, \quad &&\text{(secondary diffusive flux)}.
\end{alignat}
\end{subequations}
We can express \eqref{eq:advection_1} as a nonlinear advection--diffusion equation for the leading-order surfactant flux:
\begin{equation} \label{eq:advection}
    \frac{\partial C_0}{\partial K_0} \frac{\partial K_0}{\partial \tau}  + \left(\frac{\partial A_0}{\partial K_0} - \frac{\partial M_0}{\partial K_0} - \frac{\partial D_0}{\partial K_0}\right)\frac{\partial K_0}{\partial \chi} - \lambda \epsilon^2 \frac{\partial}{\partial \chi} \left(\frac{\partial D_1}{\partial K_0} \frac{\partial K_0}{\partial \chi}\right) = 0.
\end{equation}
Equation \eqref{eq:advection} describes the spatio--temporal evolution of a disturbance to the flux of surfactant. 
It predicts how such disturbances are advected and spread by the flow  over the long length scale ($\chi$) and slow time scale ($\tau$) that are characteristic of the channel flow. 
This equation is motivated by environmental surfactant concentrations that can vary significantly in space and time across the large length scales involved in applications \citep{frossard2019properties}. 
\st{As the surfactant flux $K_0$ evolves with respect to $\chi$ and $\tau$, its transport \eqref{eq:advection} is coupled to smaller-scale surfactant concentration transport (\ref{eq:strong_diff_2}, \ref{eq:strong_diff_flux_2}) over a given periodic cell. 
The local steady surfactant flux condition (\ref{eq:strong_diff_flux_2}) (where $K_0$ remains uniform over a given periodic cell) shows how $K_0$ decomposes into local advective and diffusive fluxes of $c_0$ and $\Gamma_0$, as well as the nonlinear flux associated with the Marangoni stress.}
We illustrate the relationship between the surfactant flux \jl{$K_0$} and the resulting surfactant bulk concentration \jl{$c_0$} in figure \ref{fig:new_fig}, which shows how they both vary over $500$ plastrons, in a case which is representative of the numerical simulations we perform in this study.
Where $K_0$ is elevated, stronger adsorption can be expected to lead to interfacial rigidification, reducing the proportion of net flux $K_0$ carried by the interface and increasing the local drag. 
The impact of these changes on the evolution of the flux field is described by \eqref{eq:advection}.
While $K_0$ varies smoothly with $\chi$, the underlying concentration field has a wavy multiscale structure (inset).
We now aim to solve this coupled transport problem to compute the time- and space-varying leading-order drag reduction and slip associated for specific initial and boundary conditions of relevance to applications.

We can solve \eqref{eq:advection} using analytical methods and numerical techniques. 
Analytically, we will neglect $O(\epsilon^2)$ terms. 
This yields a hyperbolic problem for which the solution is found using the method of characteristics, generating simple formulae that can be used by experimentalists and practitioners.
Numerically, we will retain secondary diffusion terms to regularise any shocks that may arise in hyperbolic problem and to validate the analytical results.

We solve \eqref{eq:advection} subject to an initial condition, $K_0(\chi, \,0)$, satisfying the constraint $\max(K_0(\chi, \, 0)) = 1$.
However, bar this constraint, we can choose any initial condition for the distribution of surfactant flux in the channel.
Here, we choose a Gaussian distribution, rather than a step or a ramp function, as it constitutes a classical example for which behaviours in purely advective transport systems are observed, such as: wave steepening,  wave expansion or shock formation \citep{strauss2007partial}. 
We solve \eqref{eq:advection} subject to the initial and boundary conditions
\refstepcounter{equation} \label{eq:ic1}
\begin{equation} 
    K_0(\chi, \, 0) = K_{b} + (1 - K_b)\exp(- (10 \chi + 15/2)^2), \quad K_0(\st{- 1}, \, \tau) = K_b, \tag{\theequation\textit{a,\,b}}
\end{equation}
where 
$K_b$ is the background surfactant flux.
Taking $K_b = 1/2$, the distribution defined in \eqref{eq:ic1} is characteristic of a channel that is contaminated with a bolus of surfactant that locally doubles the background surfactant flux.
In this case, the drag-reduction values remain in region M (see figure \ref{fig:dr}), as we will show in \S\ref{sec:results}.
We also take $K_b = 0.01$,
representing an almost clean channel that is contaminated with a bolus of surfactant.
We investigate $K_b=0.01$, as an alternative to $K_b=1/2$, as in this case values of the drag reduction can transition between region A (or D) and region M (figure \ref{fig:dr}). 
As we show in \S\ref{sec:results}, the surfactant-flux distribution can exhibit shocks with such initial conditions.

\begin{figure}
    \centering
    (\textit{a}) \hfill (\textit{b}) \hfill \hfill \hfill \\ \includegraphics[width=.45\textwidth]{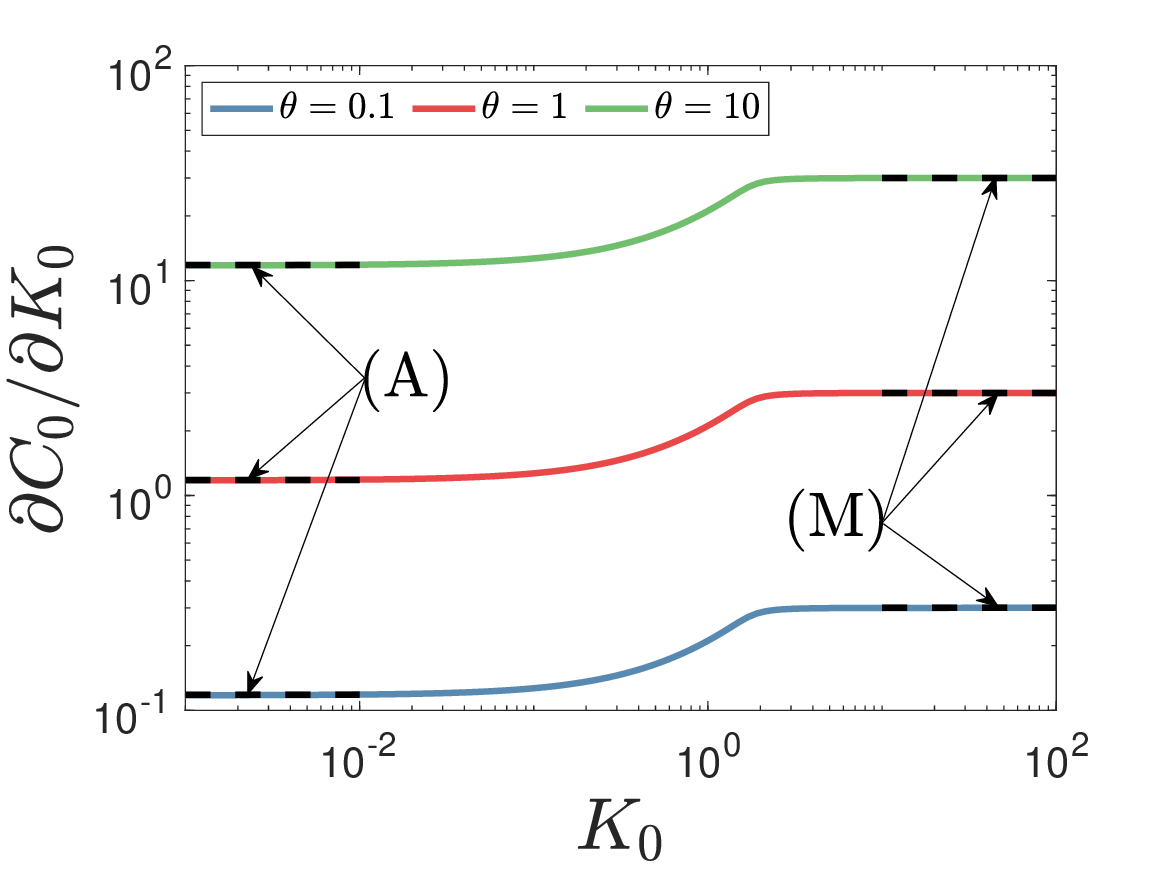} \includegraphics[width=.45\textwidth]{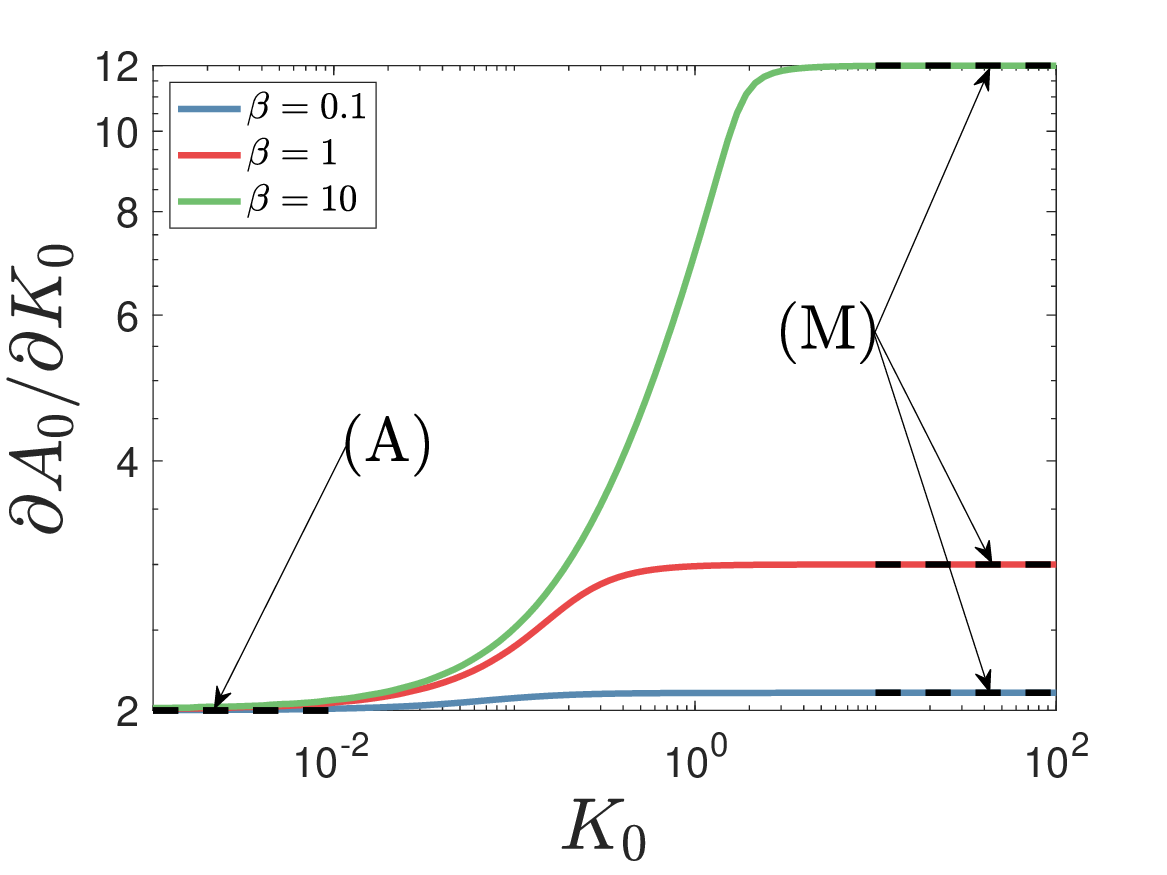} \\     (\textit{c}) \hfill (\textit{d}) \hfill \hfill \hfill \\ \includegraphics[width=.45\textwidth]{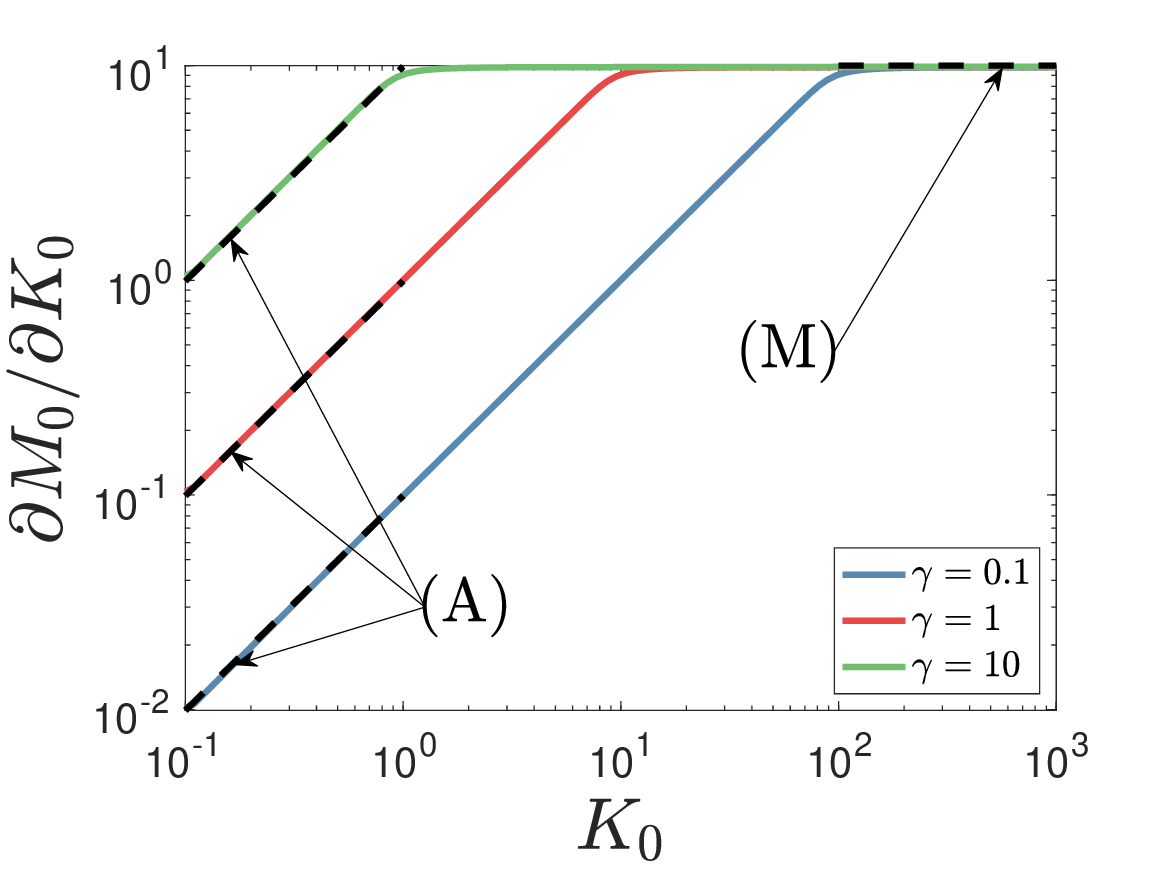} \includegraphics[width=.45\textwidth]{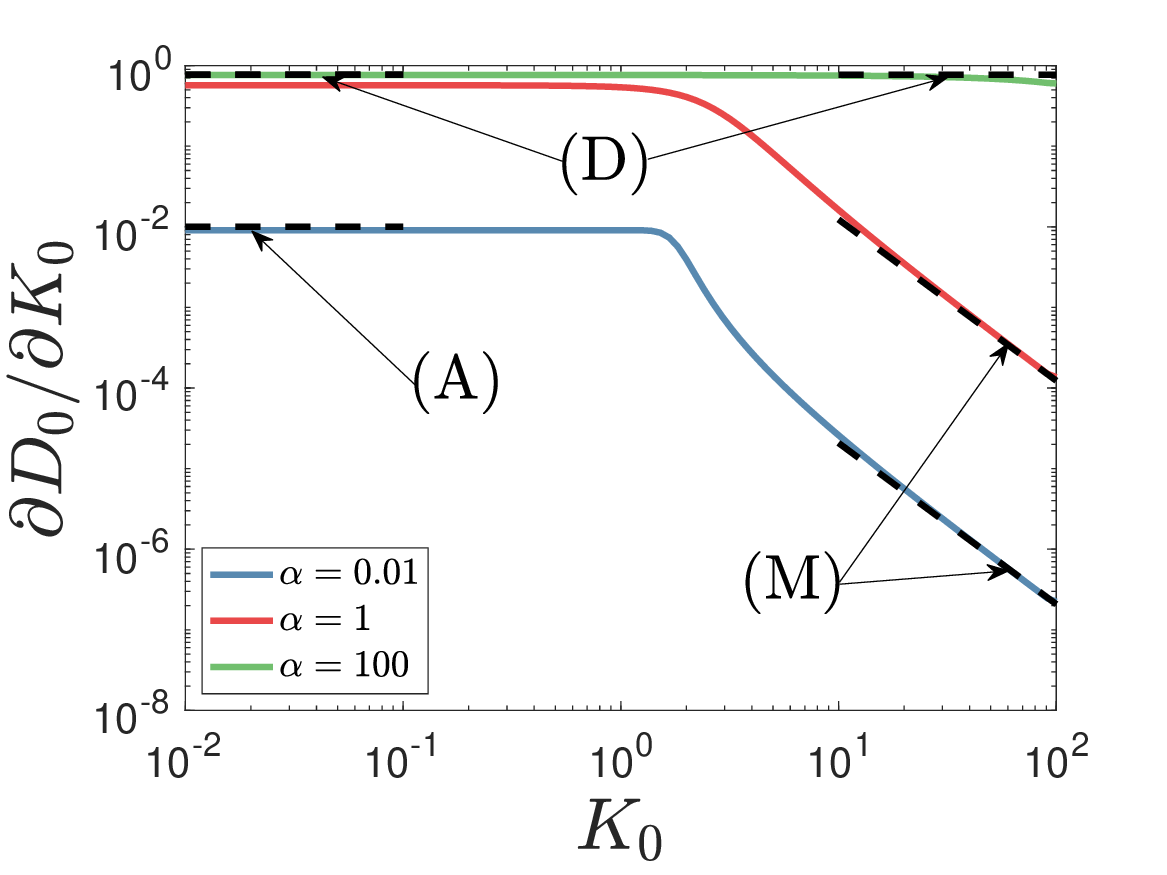}
    \caption{Coefficients in the nonlinear advection--diffusion equation \eqref{eq:advection} for the leading order surfactant flux $K_0$: (\textit{a}) $\partial C_0 / \partial K_0$ defined in (\ref{eq:Rs}\textit{a}) for different $\theta = \zeta$, $\alpha = \delta = 0.01$, $\beta = 10$ and $\gamma = 5$; (\textit{b}) $\partial A_0 / \partial K_0$ defined in (\ref{eq:Rs}\textit{b}) for different $\beta$, $\alpha = \delta = 0.01$, $\theta = \zeta = 1$ and $\gamma = 5$; (\textit{c}) $\partial M_0 / \partial K_0$ defined in (\ref{eq:Rs}\textit{c}) for different $\gamma$, $\alpha = \delta = 0.01$, $\theta = \zeta = 1$ and $\beta = 10$; (\textit{d}) $\partial D_0 / \partial K_0$ defined in (\ref{eq:Rs}\textit{d}) for different $\alpha = \delta$, $\beta = 10$, $\gamma = 5$ and $\theta = \zeta = 1$, where we vary the surfactant flux ($K_0$) and $\phi_x = 0.5$.
    Numerical solutions are plotted using solid coloured lines and asymptotic solutions are plotted using dashed black lines using \eqref{eq:eq_marangoni}, \eqref{eq:Rs_advection} and \eqref{eq:Rs_diffusion} in regions M, A and D respectively.
    }
    \label{fig:rs}
\end{figure}

The dependence of $C_0$, $A_0$, $M_0$ and $D_0$ in \eqref{eq:Rs} on $K_0$, for different values of $\alpha$, $\beta$, $\gamma$, $\delta$, $\theta$ and $\zeta$, is illustrated in figure \ref{fig:rs}. 
We choose parameter values for $\alpha$, $\beta$, $\gamma$ and $\delta$ such that drag reduction values are generally at the AM boundary in the parameter space (see figure \ref{fig:dr}). 
By increasing $K_0$, we transition from regions A to M when $K_0 = O(2\phi_x \beta/\gamma)$ (see Appendix \ref{subsec:a_strong_diff_strong_surfactant_strength}).
In figure \ref{fig:rs}(\textit{a,\,b}), the relationship between $C_0$ and $A_0$ with $K_0$ is linear in regions M (see (\ref{eq:Rs_marangoni}\textit{a,\,b})) and A (see (\ref{eq:Rs_advection}\textit{a,\,b})), and nonlinear in between.
In figure \ref{fig:rs}(\textit{c,\,d}), the relationship between $M_0$ and  $D_0$ with $K_0$ can be nonlinear in regions M and A, however, this nonlinearity does not affect the leading-order surfactant-flux distribution and the drag reduction in regions M and A, because $\gamma \gg \max(1,\,\alpha,\,\delta)$ and $\max(\alpha, \, \delta, \, \gamma) \ll 1$, respectively (see Appendices \ref{subsec:a_strong_surfactant_strength} and \ref{subsec:a_strong_advection}).
In figure \ref{fig:rs}(\textit{d}), when $\alpha = 100$, we transition to region D because $\min(\alpha,\,\delta)\gg\max(1,\,\gamma)$ (see Appendix \ref{subsec:a_strong_diffusion}).
In figure \ref{fig:rs}, we see that all the coefficients $C_0$, $A_0$, $M_0$ and $D_0$ have a nonlinear dependence on $K_0$ at the AM boundary, which will affect the leading-order surfactant-flux distribution and drag reduction. 
We will discuss this further in \S\ref{sec:results}. 

\subsection{Solving the surfactant-flux evolution equation} \label{methods}

\st{To solve \eqref{eq:advection} using the method of characteristics, we neglect the $O(\epsilon^2)$ terms and seek closed-form expressions for the surfactant-flux distribution. 
$\text{d}K_0/\text{d}\tau = 0$ on the characteristic curves of \eqref{eq:advection}, which are solutions of \citep{strauss2007partial}
\begin{equation}
    \frac{\text{d} \chi}{\text{d} \tau} = a(K_0(\chi, \, \tau)) = \frac{A'_{0}-M'_{0} - D'_{0}}{C'_{0}},
\end{equation}
where primes denote derivatives of the functions defined in \eqref{eq:Rs} with respect to $K_0$.
The propagation speed, $a$, characterises how fast changes in the surfactant-flux distribution will be transported in space and time along the length of the channel.
The characteristic curves $\chi = \chi(\tau)$ are straight lines and $K_0$ is uniform along each characteristic.
We can solve \eqref{eq:advection} subject to \eqref{eq:ic1} provided that the characteristics do not intersect. 
The characteristics are given by
\begin{equation} \label{eq:ccurves}
    \chi = \xi + a (K_0(\xi, \, 0)) \tau \quad \text{for} \quad \xi \in \mathbb{R},
\end{equation}
which gives $\xi$ implicitly as a function of $\chi$ and $\tau$, i.e. $\xi =\xi(\chi, \, \tau)$.
Hence, the solution to \eqref{eq:advection} subject to \eqref{eq:ic1} is given by $K_0(\chi, \, \tau) = K_0(\xi, \, 0)$.
The time $\tau_b$ when a shock first forms is given by \citep{strauss2007partial}
\begin{equation} \label{eq:shock}
    \tau_b = \min_{\xi\in\mathbb{R}}\left\{\frac{-1}{a_\xi(K_0(\xi, \, 0))}\right\},
\end{equation}
and \eqref{eq:ccurves} gives the streamwise location $\chi_b$ where a shock first forms. 
The integral form of \eqref{eq:advection_1} can be used to derive the Rankine--Hugoniot condition, $u_s = [[A_0 - M_0 - D_0]]/[[C_0]]$, where $u_s$ is the shock speed, with the jump bracket defined as $[[q]] = q(\chi_s^+, \, \tau) - q(\chi_s^-, \, \tau)$ and $\chi_s$ is the location of the shock \citep{strauss2007partial}.
The jump condition can then be integrated to determine the location of the shock for times $\tau > \tau_b$, for admissible shocks that satisfy the entropy condition,
\begin{equation} \label{eq:shock_location}
    \chi_s(\tau) = u_s \tau + B,
\end{equation}
where the integration constant $B$ is determined using $\chi_s(\tau_b) = \chi_b$ from \eqref{eq:ccurves}--\eqref{eq:shock}.}

Alternatively, we retain $O(\epsilon^2)$ diffusive terms in \eqref{eq:advection} and solve it numerically subject to the initial and boundary conditions \eqref{eq:ic1}. 
We use the method of lines and a backwards-in-time and centered-in-space scheme. 
As discussed earlier, retaining small diffusive terms avoids numerical difficulties associated with shock formation and regularises the shocks through a small amount of diffusion. 
This procedure is outlined in detail in Appendix \ref{app:0}.

\subsection{Quantities of interest for applications}

As discussed in \S\ref{sec:introduction}, the main quantities of interest in SHS applications are the effective slip length and drag reduction.
\st{The pressure drop across a plastron can be expressed in terms of $\Delta p_U = -2\phi_x/\tilde{Q} - 2(1-\phi_x)/\breve{Q}$ (its value when the interface is shear free), $\Delta p_I = - 2 / \breve{Q}$ (its value when the interface is immobilised) and $\Delta p_0 = - 2\phi_x/\tilde{Q} - 2(1-\phi_x)/\breve{Q} + \Ma \bar{Q}\Delta \Gamma_0/\tilde{Q}$ (its value in general, where $\Delta \Gamma_0 \equiv \Gamma_0(\phi_x; \, K_0) - \Gamma_0(-\phi_x; \, K_0)$).
Following \citet{tomlinson2023laminar}, integrating $-p_{0x}$ across the period, substituting $\Delta p_U$, $\Delta p_I$ and $\Delta p_0$ into \eqref{eq:nondimensional_drag} and using the definition of $\beta$ and $\gamma$ in (\ref{eq:coefficients}\textit{b,\,c}), the leading-order drag reduction (over a plastron) depends on the total flux of surfactant via}
\begin{equation} \label{eq:dD_0ef}
        {DR}_0(\chi, \, \tau) = 1 - \frac{\gamma \Delta \Gamma_0}{2\phi_x \beta}.
\end{equation} 
Given a surfactant-flux distribution $K_0$, we can calculate the corresponding drag-reduction distribution ${DR}_0$ by solving the surfactant transport equations (\ref{eq:model3}, \ref{eq:strong_diff_2}, \ref{eq:strong_diff_flux_2}) for each $K_0$ to get $\Gamma_0(x; \, K_0)$, and then use \eqref{eq:dD_0ef} to calculate ${DR}_0$.
The drag reduction inherits a dependence on $\tau$ and $\chi$ from $K_0$; we therefore define the space- and time-averaged drag reduction as
\refstepcounter{equation} \label{eq:average_DR}
\begin{equation}
\langle {DR}_0 \rangle_\chi (\tau) = \frac{1}{\st{2}}\int_{\chi = \st{-1}}^{\st{1}} {DR}_0 \, \text{d}\chi, \quad \overline{DR}_0(\chi) = \frac{1}{\mathcal{T}} \int_{\tau = 0}^{\mathcal{T}} {DR}_0 \, \text{d}\tau, \tag{\theequation\textit{a,\,b}}
\end{equation}
where $\st{[-1, \, 1]}$ covers the length of the channel and $[0, \, \mathcal{T}]$ is the time interval over which the drag reduction is measured.
To evaluate the effective slip length $\lambda_e$ over a plastron \citep{tomlinson2023laminar}, we integrate the leading-order streamwise momentum equation (\ref{eq:nondimensional_equations_1}\textit{b}) for an equivalent channel with the mixed boundary conditions (\ref{eq:nondimensional_interface_bcs_1}\textit{a}, \ref{eq:nondimensional_solid_bcs_1}\textit{a}) replaced by $\lambda_e u_{0y_\perp} - u_0 = 0$. 
We obtain $u_0 = \check{U} p_{0x}$, where $\check{U} = y_\perp(y_\perp-2)/2 - \lambda_e$ and $p_{0x}$ is the same pressure gradient as in the SHS channel. 
The corresponding volume flux is $\check{Q} = (\breve{Q} - 2 P_z \lambda_e)p_{0x}$, or by integrating over one period $\check{Q} = (\breve{Q} - 2 P_z \lambda_e) \Delta p_{0} / 2$.
Equating the volume flux of the equivalent channel with the volume flux \eqref{eq:bulk_flux_constraints}, we find 
\begin{equation} \label{eq:lambda_e}
    \lambda_e = \frac{{DR}_0(\Delta p_I - \Delta p_U)}{P_z \Delta p_I(\Delta p_U {DR}_0  + \Delta p_I (1-{DR}_0))},
\end{equation}
which can be used to convert results from ${DR}_0$ to $\lambda_e$.
Therefore, with $\lambda_e$ and ${DR}_0$ being directly related to $K_0$, the governing equations, \eqref{eq:advection}--\eqref{eq:ic1}, can also be rewritten as initial boundary value problems for either the effective slip length or drag reduction, that vary over the long length scale and slow time scale of the channel. 

\section{Results} \label{sec:results}

In \S\ref{subsec:Marangoni--dominated region}--\ref{subsec:Diffusion--Marangoni boundary}, we investigate how the surfactant flux ($K_0$), drag reduction (${DR}_0$), propagation speed ($a$), streamwise velocity ($u_0$) and surfactant concentration ($c_0$ and $\Gamma_0$) vary with the bulk diffusion ($\alpha$), partition coefficient ($\beta$), surfactant strength ($\gamma$), surface diffusion ($\delta$), exchange strength ($\nu$), bulk capacitance ($\theta$), surface capacitance ($\zeta$) and streamwise gas fraction ($\phi_x$).
We discuss the quantities that vary over the length of the channel ($K_0$, ${DR}_0$ and $a$) and their effect on the flow and surfactant transport ($u_0$, $c_0$ and $\Gamma_0$) over each period. 
Using asymptotic (detailed in Appendix \ref{app:A}) and numerical (Appendix \ref{app:0}) techniques, we evaluate the solution to (\ref{eq:advection}, \ref{eq:ic1}) in the main regions and boundaries illustrated in figure \ref{fig:dr} and discussed in \S\ref{subsubsec:Second-order problem}: the Marangoni-dominated (M) region (\S\ref{subsec:Marangoni--dominated region}); the advection-dominated (A) region (\S\ref{subsec:Advection--dominated region}); the diffusion-dominated (D) region (\S\ref{subsec:Advection--dominated region}); the advection--Marangoni (AM) boundary (\S\ref{subsec:Advection--Marangoni boundary}); and the diffusion--Marangoni (DM) boundary (\S\ref{subsec:Diffusion--Marangoni boundary}). 
\st{Shocks in the surfactant-flux and drag-reduction distribution can arise in regions AM and DM.}
Throughout \S\ref{sec:results}, we fix the
channel-height-to-streamwise-period ratio $\epsilon = 0.1$, spanwise gas fraction $\phi_z = 0.5$ and spanwise period width $P_z=1$.
We construct asymptotic solutions for any
$\phi_z$ and $P_z$ in Appendix \ref{app:A} where $\epsilon \ll 1$. 

\subsection{Marangoni--dominated region} \label{subsec:Marangoni--dominated region}

\subsubsection{Flow and surfactant flux transport at the channel scale}

\begin{figure}
    \centering
    \raisebox{4cm}
    {(\textit{a})}\includegraphics[width=.45\textwidth]{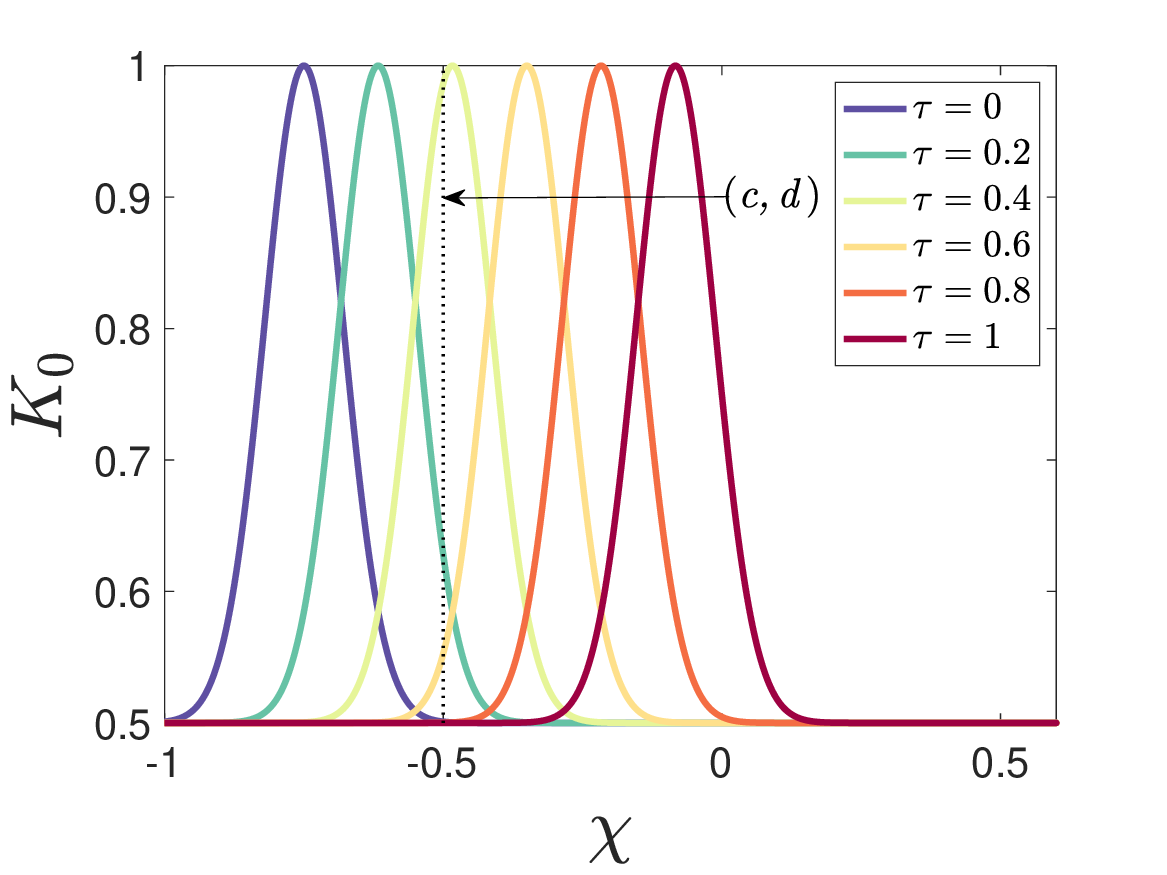}\raisebox{4cm}{(\textit{b})}\includegraphics[width=.45\textwidth]{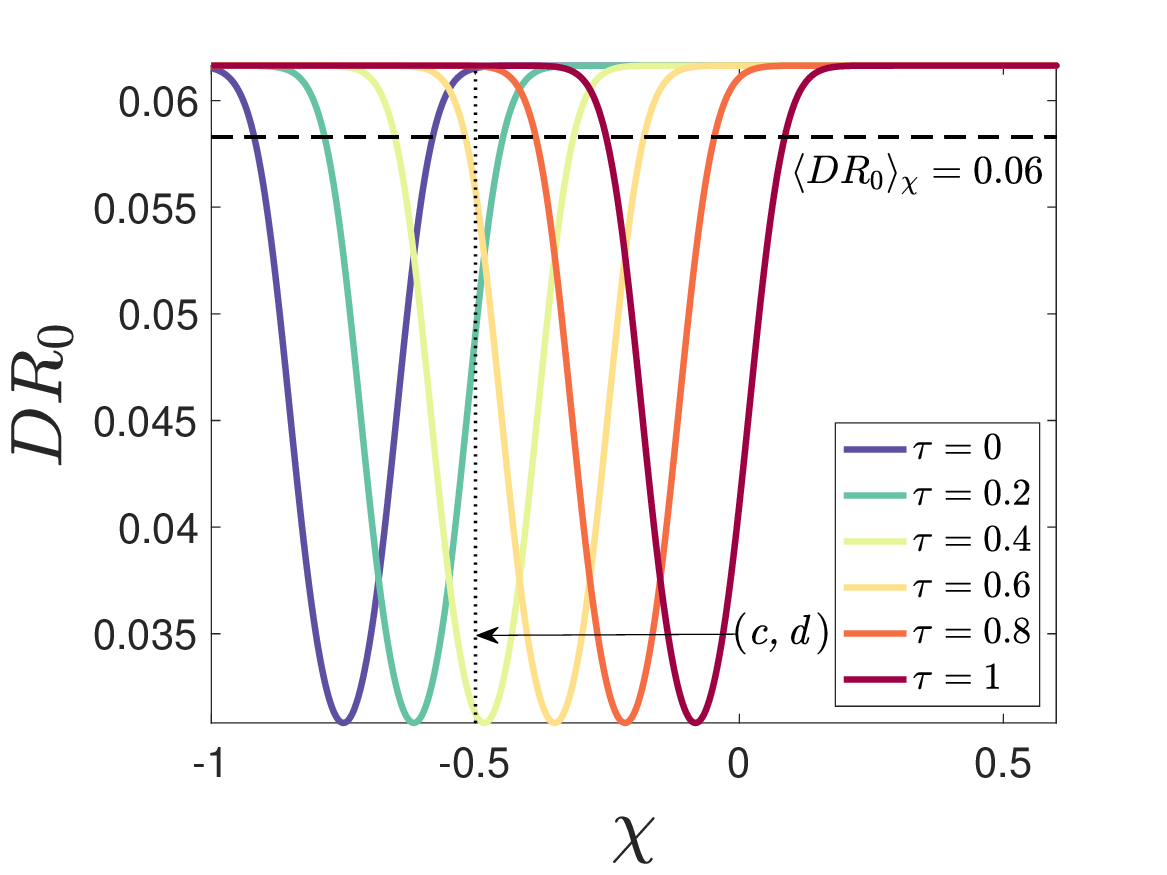} \hfill \hfill \hfill \\
    \raisebox{4cm}{(\textit{c})}\includegraphics[width=.45\textwidth]{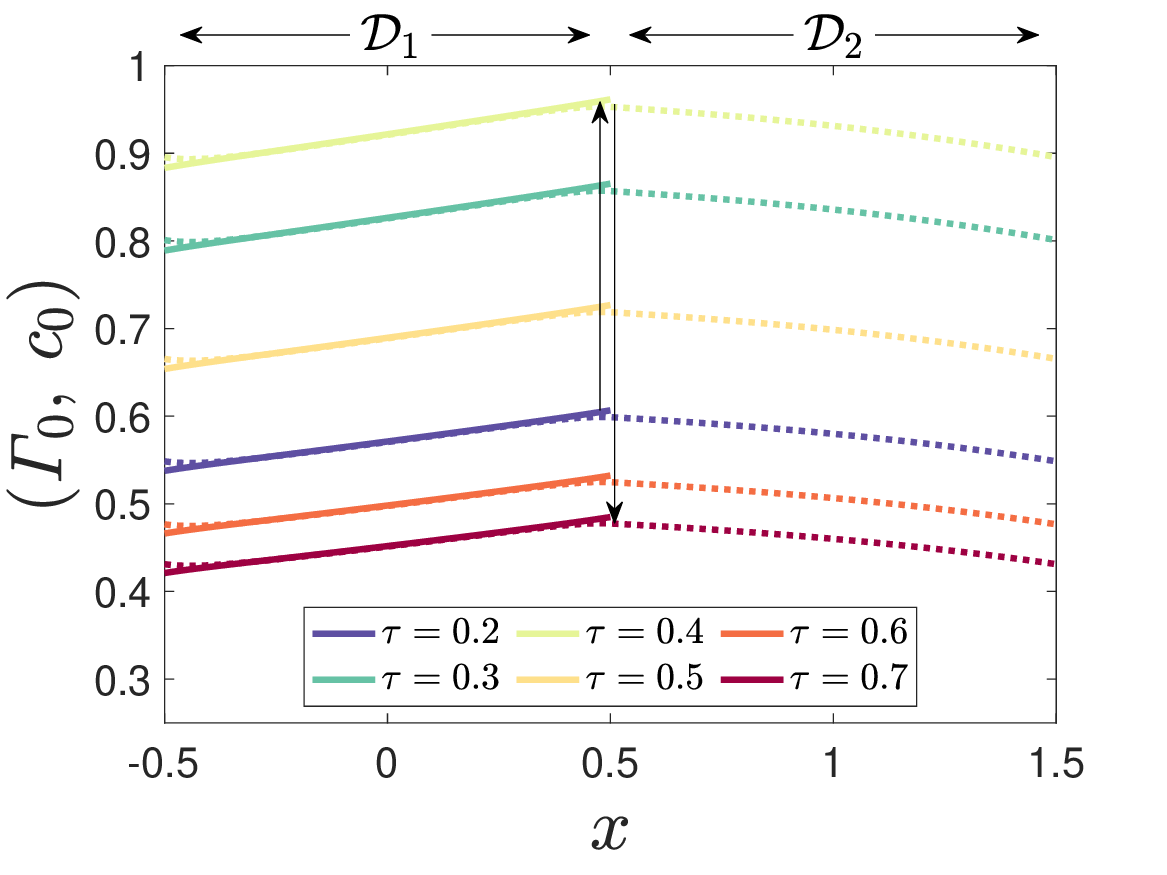}\raisebox{4cm}{(\textit{d})}\includegraphics[width=.45\textwidth]{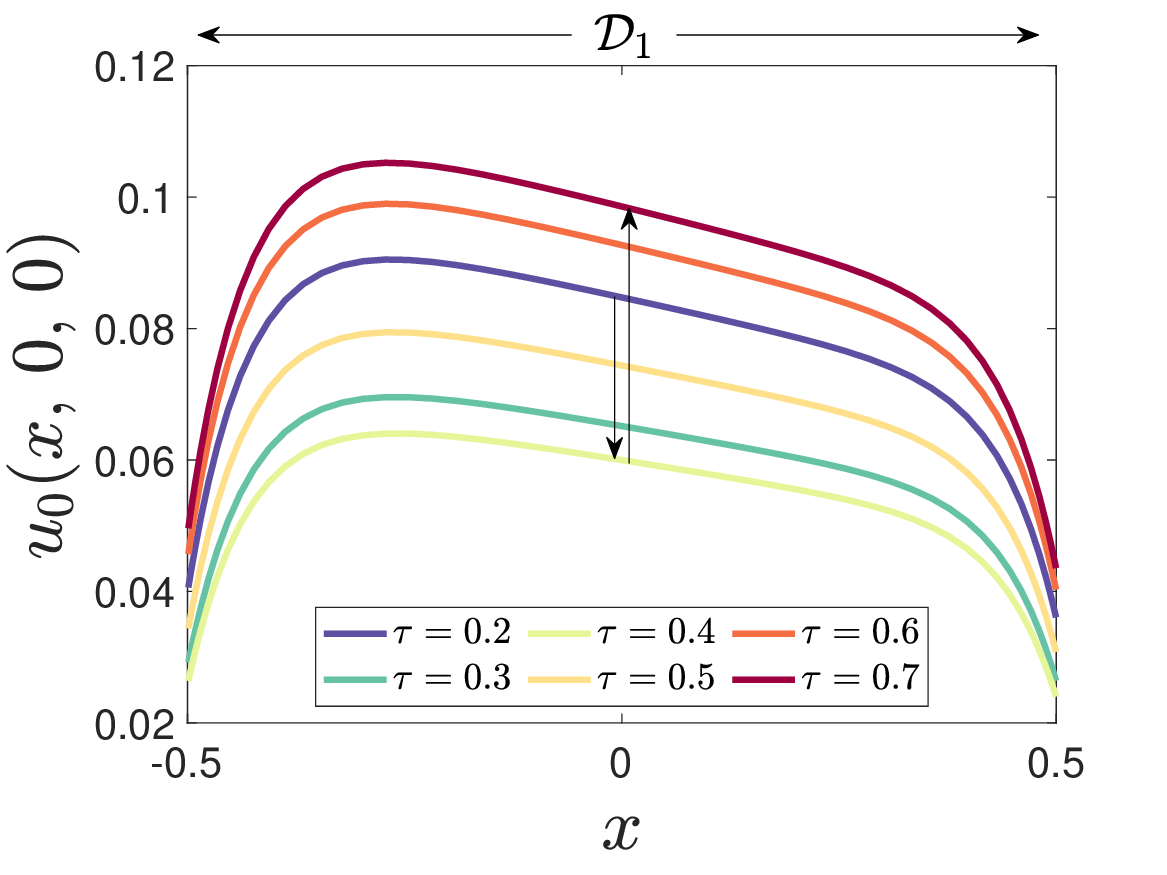} \hfill \hfill \hfill \\
    \caption{
    Quantities of interest as a bolus of surfactant flux, \eqref{eq:ic1}, passes through the channel in the Marangoni-dominated region (M), where $\alpha = \beta = \delta = 1$, $\gamma = 10$, $\nu = 100$, $\phi_x = 0.5$
    and $K_b = 1/2$. 
    (\textit{a}) Surfactant flux $K_0$ and (\textit{b}) drag reduction ${DR}_0$ versus distance along the channel $\chi$ at different times $\tau = 0$, $0.2$, $0.4$, $0.6$, $0.8$ and $1$, computed using (\ref{eq:dD_0ef}, \ref{eq:ic1}, \ref{eq:eq_marangoni}) for $\theta = \zeta = 1$. 
    The dashed horizontal line in panel (\textit{b}) is the space-averaged drag reduction $\langle {DR}_0 \rangle_\chi$, calculated using \eqref{eq:average_DR}. 
    (\textit{c}) Bulk $c_0$ (dotted lines) and interfacial $\Gamma_0$ (solid lines) concentration fields across a unit cell $\mathcal{D}_1\cup\mathcal{D}_2$ and (\textit{d}) streamwise slip velocity $u_0(x, \, 0, \, 0)$ across a plastron (domain $\mathcal{D}_1$), located at $\chi = -0.5$ (dotted vertical line in panels \textit{a} and \textit{b}) when $\tau = 0.2$, $0.3$, $0.4$, $0.5$, $0.6$ and $0.7$, and calculated using (\ref{eq:u_def}, \ref{eq:strong_diff_2}).
    }
    \label{fig:m}
\end{figure}

Figure \ref{fig:m}(\textit{a}) shows how a bolus of surfactant flux, using initial and boundary conditions \eqref{eq:ic1}, is advected along the length of the channel at increasing times.
Taking $K_b= 1/2$ in \eqref{eq:ic1}, $K_0$ remains sufficiently large for the flow to be everywhere in the Marangoni-dominated (M) region. 
Where the surfactant flux and concentration increase, the leading-order drag reduction in figure \ref{fig:m}(\textit{b}) decreases, implying that the liquid--gas interface is more immobilised.
We plot in figure \ref{fig:m}(\textit{a,\,b}) asymptotic solutions for the leading-order surfactant flux and drag reduction (derived in Appendix \ref{subsec:a_strong_surfactant_strength}),
\begin{subequations} 
\label{eq:eq_marangoni_results}
\begin{align}
    \st{K_0(\chi, \, \tau)} &\approx\st{K_b + (1-K_b)\exp(-(10(\chi - a_{\text{M}} \tau)+15/2)^2),} \\ \st{{DR}_0(\chi, \, \tau)} &\approx\st{(\alpha + \delta + \phi_x(E+1)/(E-1)) / (\gamma K_0),}  \\
    \st{a_{\text{M}}}&\approx\st{1/(\theta + \zeta \phi_x)}
\end{align}
\end{subequations} 
\st{derived in the limit where both Marangoni effects and bulk--surface exchange are strong compared to advection and diffusion (see region M in figure \ref{fig:dr}\textit{a}); $a_{\text{M}}$ is the propagation speed in region M (see further discussion about its physical meaning in the next paragraph) and $E \equiv \exp(2(1-\phi_x)/\alpha)$.}
As the flux of surfactant is conserved in \eqref{eq:advection}, the space-averaged drag reduction ($\langle{{DR}_0}\rangle_\chi$) defined in (\ref{eq:average_DR}\textit{a}) is constant provided the bolus of surfactant flux remains in the channel for the given time interval. 
However, once the bolus of surfactant flux is advected out of the channel, the space- and time-averaged drag reduction $\langle \overline{{DR}_0}\rangle_\chi$ defined in (\ref{eq:average_DR}) is maximised by \jl{increasing} the propagation speed ($a_{\text{M}}$) in region M.
In figure \ref{fig:m}(\textit{b}), $\langle \overline{{DR}_0}\rangle_\chi=\langle {DR}_0 \rangle_\chi = 0.06$ for $\chi \in [-1, \, 1]$ and $\tau \in [0, \, 1]$; the drag reduction is small as the liquid--gas interface is mostly immobilised.

In region M, the bolus of surfactant flux, \eqref{eq:ic1}, is advected downstream by the flow with a constant propagation speed, (\ref{eq:eq_marangoni_results}\textit{c}), and therefore the distribution of surfactant flux and drag reduction in figure \ref{fig:m}(\textit{a,\,b}) do not change shape as the bolus moves through the channel.
For $\zeta \ll \theta$, $a_{\text{M}} \approx 1/\theta$, which corresponds to a dimensional speed $\hat{U}_m = \hat{Q}/ (4\hat{P}_z\hat{H})$.
This is the cross-channel-averaged bulk propagation speed and indicates bulk-dominated surfactant transport. 
For fixed $\theta$, adsorption at the liquid--gas interface (via an increase in $\zeta$) causes the propagation speed of the bolus of surfactant flux to fall significantly compared to the cross-channel-averaged bulk propagation speed, reducing to $a_{\text{M}} \approx 1 / (\zeta \phi_x)$ for $\zeta \gg \theta$.
Dimensionally, this corresponds to a reduction in the bulk propagation speed by a factor $\phi_x \phi_z L_d$ and indicates surface-dominated transport, where $L_d = \hat{K}_a / (\hat{H} \hat{K}_d)$ is the normalised surfactant depletion length and $\phi_x \phi_z$ is the area gas fraction of the SHS. 
Hence, the propagation of disturbances to the surfactant concentration field is significantly slower for more insoluble surfactants (large $L_d$) and when the area of adsorption, i.e. the liquid--gas interface $0<\phi_x \phi_z<1$, is maximised. 
%

\subsubsection{Flow and surfactant transport at the scale of the periodic cell}

The magnitude of the background surfactant flux in figure \ref{fig:m}, $K_b= 1/2$, means that the liquid--gas interface is almost immobile along the entire SHS and the leading-order bulk and interfacial concentrations ($\Gamma_0$ and $c_0$) in each period are approximately linear with a shallow gradient (Appendix \ref{subsec:a_strong_surfactant_strength}), 
\begin{equation} 
\label{eq:marangoni_limited_results}
    \st{c_{0}(x; \, K_0) \approx \Gamma_0(x; \, K_0) \approx K_0 + \beta (x - \phi_x (E + 1)/(E - 1))/\gamma,}
\end{equation}
as shown in figure \ref{fig:m}(\textit{c}). 
In \eqref{eq:marangoni_limited_results}, we see that $c_0$ depends linearly on $K_0$ at leading-order, however, $\Delta c_0$ does not, as the liquid--gas interface is already immobilised when $K_0=O(1)$.
As the bolus of surfactant flux passes over an individual plastron and $K_0$ varies from $1/2$ up to $1$ and back down to $1/2$, the concentration rises (from times $\tau=0.2$ to $\tau=0.4$) and then falls (from times $\tau=0.4$ to $\tau=0.7$).
We observe adsorption and desorption inside boundary layers around $x = \pm \phi_x=\pm 0.5$ where the bulk and interfacial concentrations deviate from each other, generating local surfactant gradients that reduce the streamwise slip velocity ($u_0$) close to the stagnation points ($x=\pm \phi_x=\pm 0.5$) in figure \ref{fig:m}(\textit{d}). 
The streamwise slip velocity inherits a dependence on $K_0$ through $c_0$, as more surfactant increases the amount of immobilisation at the liquid--gas interface. 
Thus, $u_0(x, \, 0, \, 0)$ falls and then rises as the bolus passes over an individual plastron (see curves from $\tau=0.2$ to $\tau=0.7$ in the graph in figure \ref{fig:m}\textit{d}). 
As mentioned in \S\ref{subsubsec:Leading-order problem}, the present long-wave theory does not capture inner regions close to the stagnation points where the streamwise velocity satisfies $u_0 = 0$, explaining the non-zero value exhibited by $u_0(x,0,0)$ in figure \ref{fig:m}(\textit{d}) near $x=\pm \phi_x=\pm 0.5$.
As described in (\ref{eq:nondimensional_noflux_1}\textit{a}), we have instead imposed no flux of surfactant in the streamwise direction at these stagnation points.

\subsection{Advection and diffusion--dominated regions}
\label{subsec:Advection--dominated region}

We also briefly discuss asymptotic results (derived in Appendix \ref{subsec:a_strong_advection} and \ref{subsec:a_strong_diffusion}) for the advection-dominated (A) and diffusion--dominated (D) regions (see regions A and D in figure \ref{fig:dr}\textit{a}). 
When bulk--surface exchange is strong, the bolus of surfactant flux in \eqref{eq:ic1} propagates in a similar manner to figure \ref{fig:m}(\textit{a}), but is advected with speeds $a_{\text{A}} \approx (\beta+1)/(\phi_{x}  \left(\zeta +\theta \right)+ \theta (1 - \phi_{x})(\beta+1))$ and $a_{\text{D}} \approx (\alpha (1+\beta) + \delta(1-\phi_x))/((\theta+\zeta\phi_x)(\alpha + \delta(1-\phi_x)))$ in the A and D regions, respectively.
In both regions, the propagation speed increases with the partition coefficient $\beta = 2 L_d \tilde{q} / \tilde{Q}$. 
For $\beta \gg 1$ ($\beta \ll 1$), the flux of surfactant along the liquid--gas interface is greater (smaller) than the flux of surfactant in the bulk, which is non-dimensionalised to unity in \eqref{eq:strong_diff_2}.
Hence, as $L_d$ grows, the localised concentration of surfactant will be advected faster along the liquid--gas interface, and therefore, the surfactant will be advected faster throughout $\mathcal{D}_1$.
When bulk--surface exchange is weak, e.g. regions M$_\text{E}$ and D$_\text{E}$ depicted in figure \ref{fig:dr}(\textit{b}), the propagation speed is the same in all regions M, A and D (see Appendices \ref{subsec:a_strong_surfactant_strength}, \ref{subsec:a_strong_advection} and \ref{subsec:a_strong_diffusion}), with $a_{\text{M}} = a_{\text{D}} = a_{\text{A}} \approx 1 / (\theta + \zeta \phi_x)$. 

\subsection{Advection--Marangoni boundary}
\label{subsec:Advection--Marangoni boundary}

\subsubsection{Flow and surfactant flux transport  at the channel scale}

Figure \ref{fig:am}(\textit{a}) shows asymptotic predictions (derived in Appendix \ref{subsec:a_strong_diff_strong_surfactant_strength}),
\begin{subequations}
\label{eq:eq_AM_results}
\begin{align}
    \st{K_0(\chi, \, \tau)} &\approx\st{K_b + (1 - K_b)\exp(-(10(\chi - a_{\text{AM}} \tau)+15/2)^2),} \\ \st{{DR}_0(\chi, \, \tau)}&\approx\st{1 - \gamma K_0 / (2\phi_x\beta),}  \\
    \st{a_{\text{AM}}(K_0)}&\approx\st{2 \beta / (\gamma (\zeta + \theta) K_0 + 2 \beta \theta (1 - \phi_x))}
\end{align}
\end{subequations} 
and numerical solutions (outlined in Appendix \ref{app:0}) of the spatio--temporal evolution of ${DR}_0$ along the length of the channel at the advection--Marangoni (AM) boundary (where Marangoni effects, interfacial advection and bulk--surface exchange are strong compared to diffusion, see the AM boundary in figure \ref{fig:dr}\textit{a}). 
Taking low background flux $K_b = 0.01$, we focus on the case where $\gamma K_0 \leq 2\phi_x \beta$, which places the surfactant profile at  the plastron scale in the stagnant-cap regime \citep{tomlinson2023laminar}. 
The flow advects the bolus of surfactant flux \eqref{eq:ic1} through the channel with a $K_0$-dependent propagation speed (\ref{eq:eq_AM_results}\textit{c}). 
As the bolus propagates in $(\chi, \, \tau)$-space, the wave steepens at its rear side and ultimately a shock forms at some location and time along the channel, which we discuss further in \S\ref{ref:subsubsec:shock}.
For the chosen parameters, Marangoni effects are sufficiently weak for the space-averaged drag reduction ($\langle {DR}_0 \rangle_\chi$) to remain close to the shear-free value.

\begin{figure}
    \centering
    \raisebox{4cm}{(\textit{a})}\includegraphics[width=.45\textwidth]{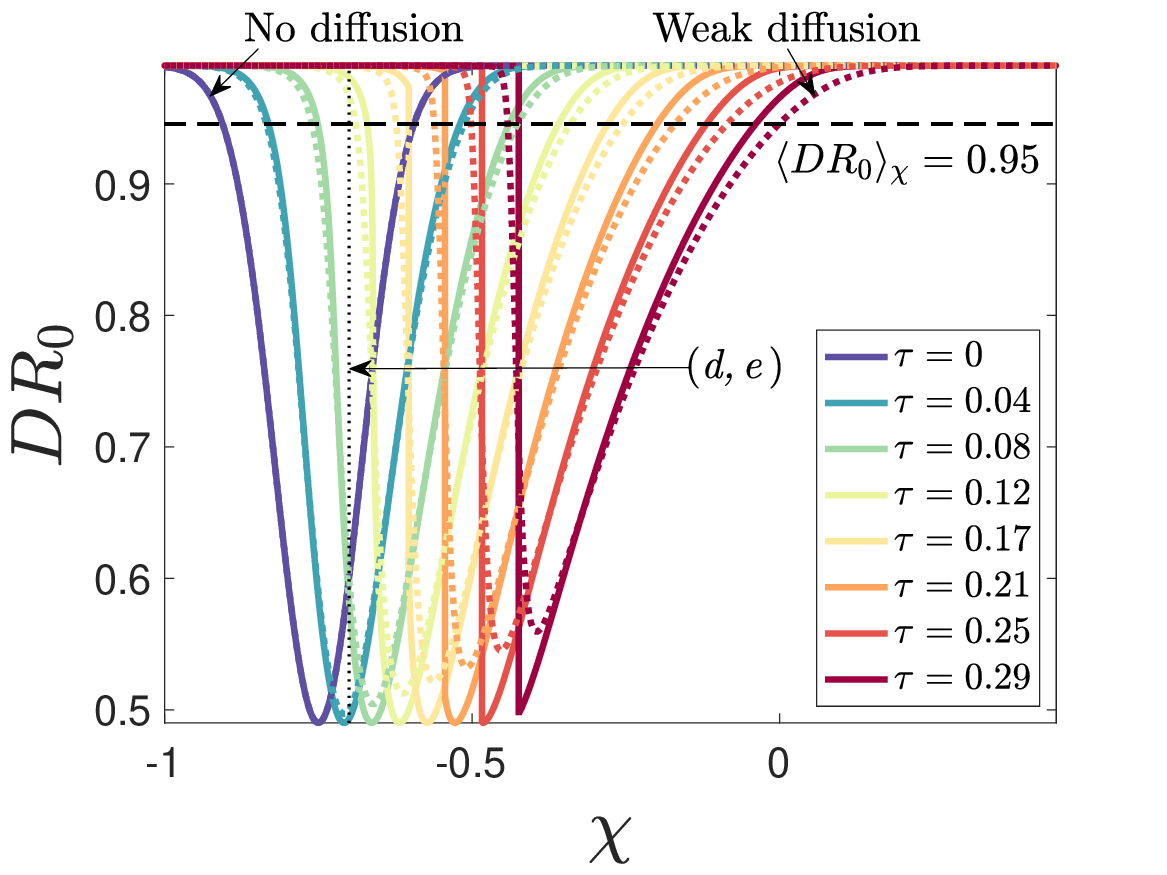}\raisebox{4cm}{(\textit{b})}\includegraphics[width=.45\textwidth]{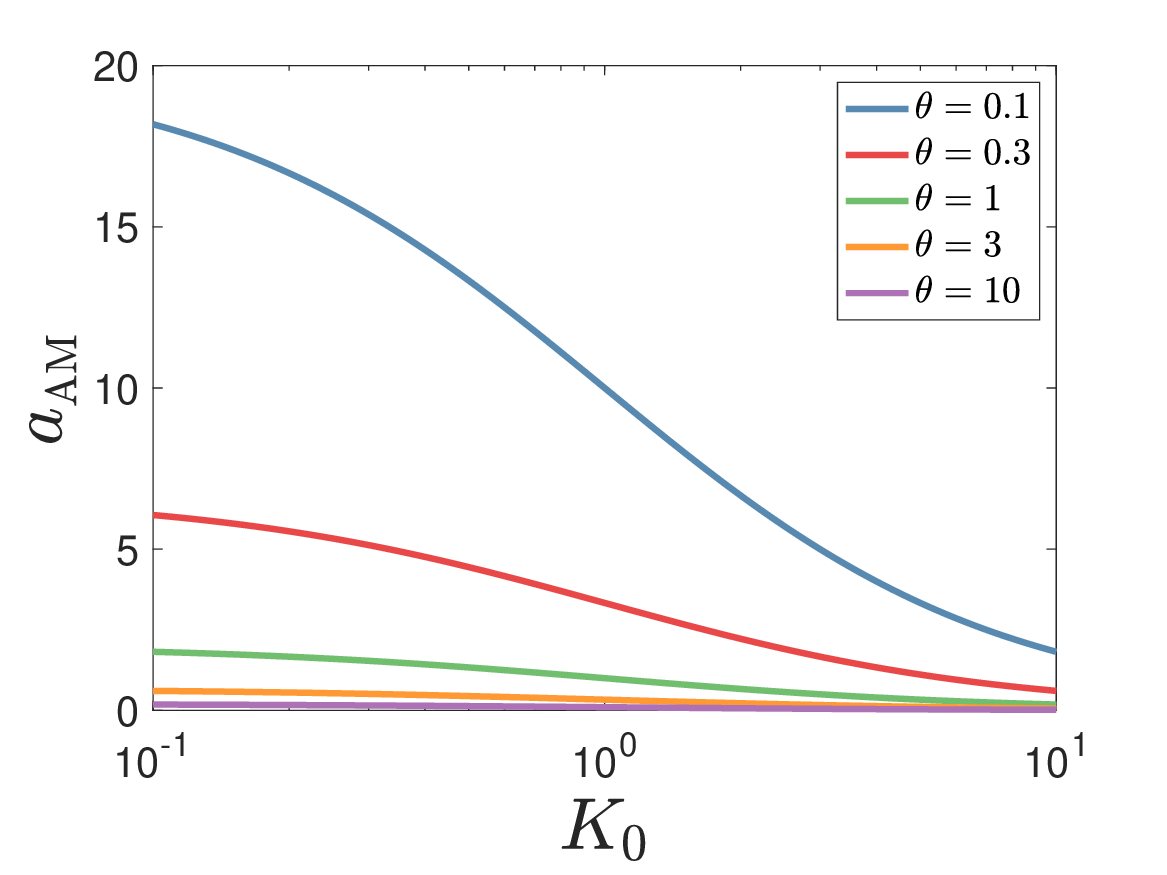} \hfill \hfill \hfill \\
    \raisebox{3.6cm}{(\textit{c})}\includegraphics[trim={3.2cm 0cm 2.4cm 0cm},clip,width=.945\textwidth]{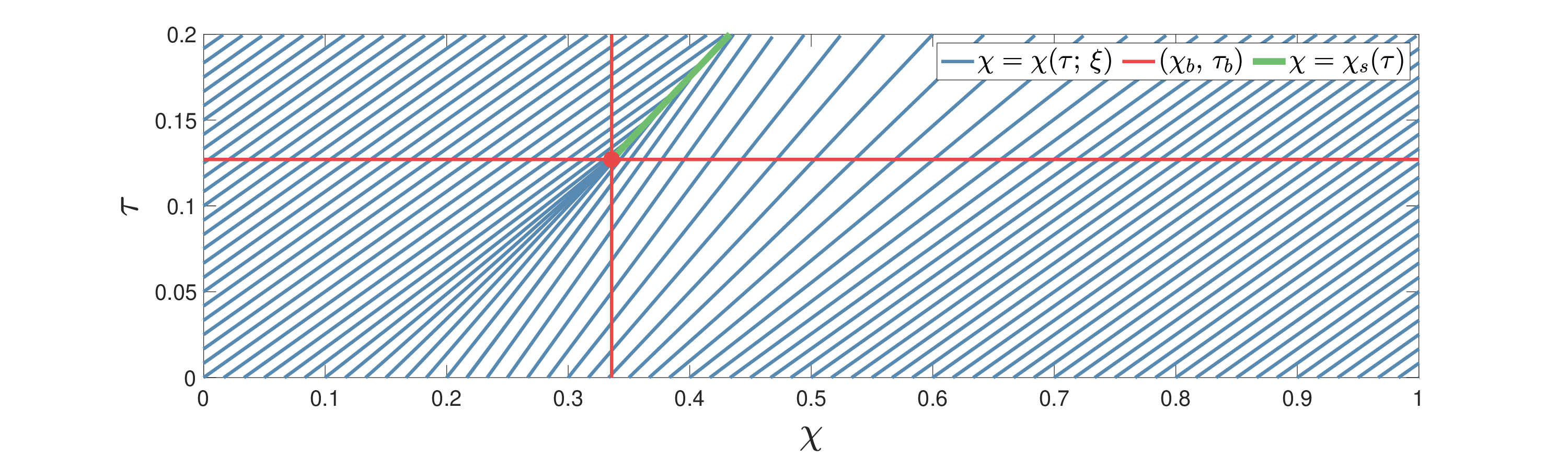} \hfill \hfill \hfill \\
    \raisebox{4cm}{(\textit{d})}\includegraphics[width=.45\textwidth]{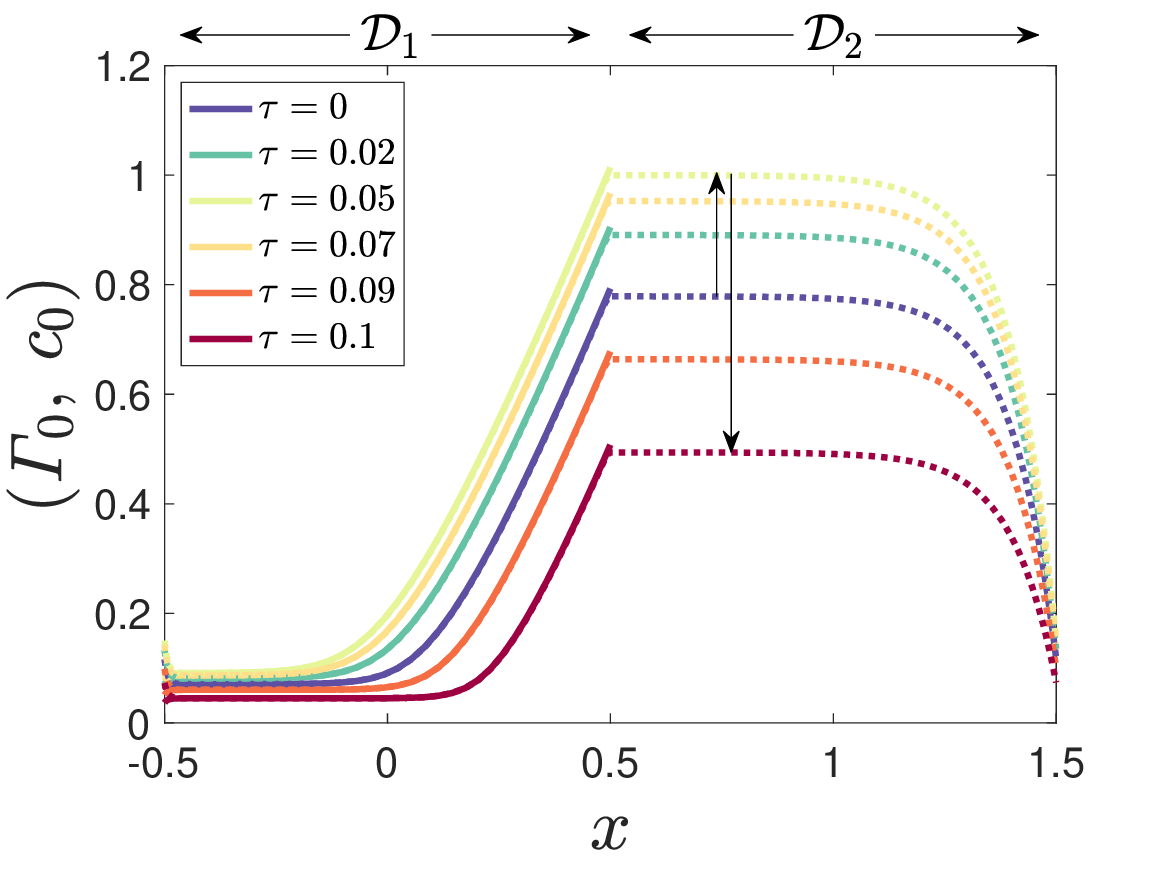}\raisebox{4cm}{(\textit{e})}\includegraphics[width=.45\textwidth]{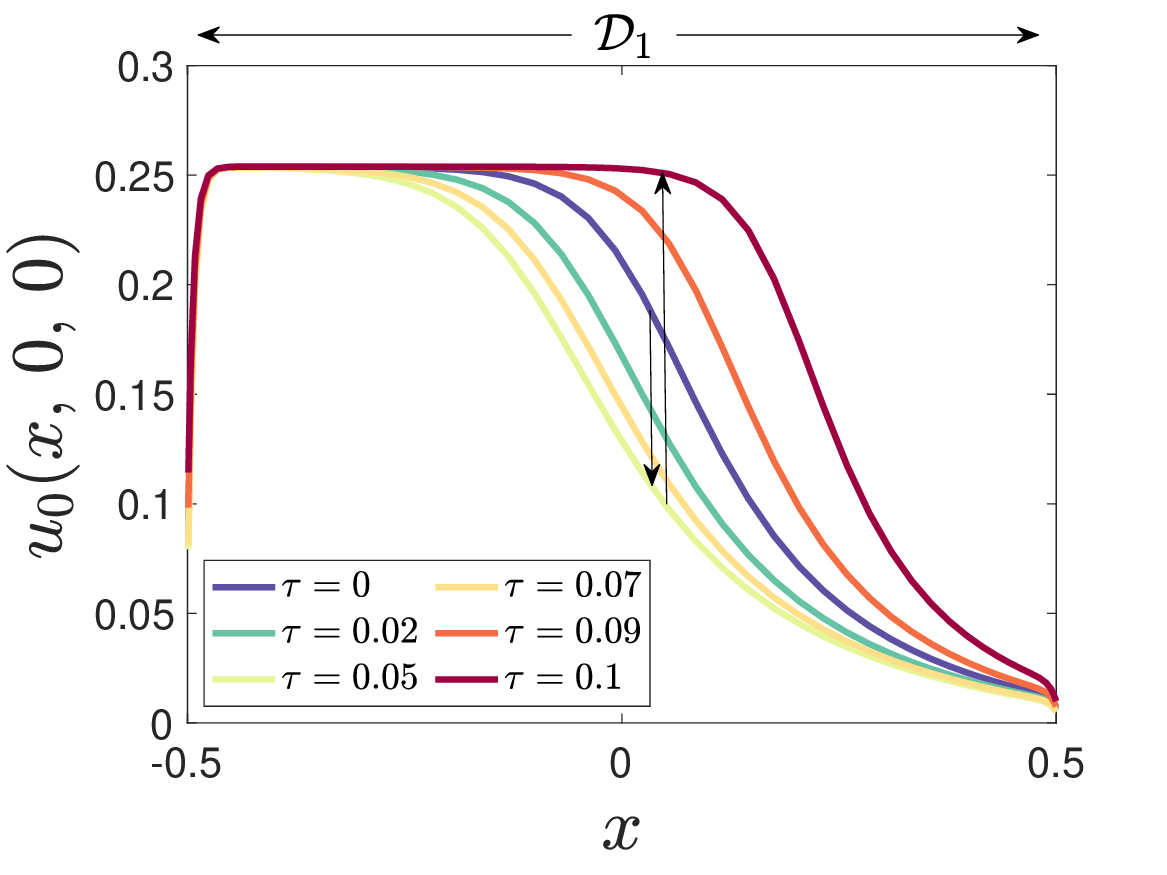} \hfill \hfill \hfill \\
    \caption{
    Quantities of interest as a bolus of surfactant flux, \eqref{eq:ic1}, forms a shock while passing through the channel at the advection--Marangoni boundary (AM), where $\alpha = \delta = 0.1$, $\beta = 10$, $\epsilon=0.1$, $\gamma = 5$, $\nu = 1000$, $\phi_x = 0.5$
    and $K_b = 0.01$. 
    (\textit{a}) Drag reduction ${DR}_0$ versus distance along the channel $\chi$ at different times $\tau = 0$, $0.04$, $0.08$, $0.12$, $0.17$, $0.21$, $0.25$ and $0.29$, computed using (\ref{eq:dD_0ef}, \ref{eq:ic1}), \eqref{eq:eq_am} (solid lines, no streamwise diffusion) and \eqref{eq:strong_diff_2} (dashed lines, weak streamwise diffusion), for $\theta = \zeta = 1$. 
    The dashed horizontal line in panel (\textit{a}) is the space-averaged drag reduction $\langle {DR}_0 \rangle_\chi$, calculated using \eqref{eq:average_DR}.
    (\textit{b}) Propagation speed $a_{\text{AM}}$ for varying surfactant flux $K_0$ and capacitance $\theta=\zeta$, computed using \eqref{eq:eq_am}.
    (\textit{c}) Characteristic curves $\chi = \chi(\tau; \, \xi)$ intersecting and forming a shock at $(\chi_b,\,\tau_b)$ that propagates along $\chi = \chi_s(\tau)$, computed using (\ref{eq:ic1}, \ref{eq:ccurves}, \ref{eq:shock}, \ref{eq:shock_location}).
    (\textit{d}) Bulk $c_0$ (dotted lines) and interfacial $\Gamma_0$ (solid lines) concentration fields across a unit cell $\mathcal{D}_1\cup\mathcal{D}_2$ and (\textit{e}) streamwise slip velocity $u_0(x, \, 0, \, 0)$ across a plastron (domain $\mathcal{D}_1$), located at $\chi = - 0.7$ (dotted line in panel \textit{a}) when $\tau = 0$, $0.02$, $0.05$, $0.07$, $0.08$ and $0.1$, and calculated using (\ref{eq:u_def}, \ref{eq:strong_diff_2}).
    }
    \label{fig:am}
\end{figure}

Figure \ref{fig:am}(\textit{b}) shows how the propagation speed depends on the spatio--temporal evolution of the bolus of surfactant at the AM boundary.
The wave steepening in the ${DR}_0$-distribution observed in figure \ref{fig:am}(\textit{a}) occurs because there is wave steepening in the surfactant-flux distribution: a small surfactant flux is advected faster than a large surfactant flux.
Physically, greater surfactant concentrations at a given location in the channel imply that the liquid--gas interface is more immobilised than it is for small surfactant concentrations. 
This implies that the streamwise slip velocity and thus the propagation speed decreases with increasing $K_0$. 
The dependence of $a_{\text{AM}}$ on $\beta$, $\theta$ and $\zeta$ is similar as for $a_{\text{M}}$, which is discussed in \S\ref{subsec:Marangoni--dominated region}.
When $K_0 \ll 1$, $a_{\text{AM}} \approx 1 / (\theta(1-\phi_x))$, the cross-channel-averaged bulk propagation speed is enhanced by a factor proportional to the streamwise groove length $1-\phi_x$.

\subsubsection{Shock formation and regularisation via streamwise diffusion}
\label{ref:subsubsec:shock}

The time taken for the shock to form in the surfactant-flux distribution, $\tau_b$, increases as the partition coefficient ($\beta$) decreases and the bulk and surface capacitance parameters ($\theta$ and $\zeta$) increase. 
The location in space where the shock in the surfactant-flux distribution forms, $\chi_b$, does not vary with respect to $\theta$ and $\zeta$; however, $\chi_b$ increases with decreasing $\beta$, for reasons that we now explain.
Recall that, as $\beta$ increases, the propagation speed increases because the interfacial surfactant flux increases and the bulk surfactant flux is fixed; as $\theta$ and $\zeta$ decrease, the propagation speed increases because the cross-sectional area reduces and the volume flux is fixed.
A larger propagation speed means that the difference between the total flux of surfactant when the concentration is small and large is greater (see figure \ref{fig:am}\textit{b}), and therefore, the larger difference causes the wave to steepen faster and the shock forms earlier.  
Taking $K_b= 0.01$ in the initial distribution of surfactant flux \eqref{eq:ic1}, we use \eqref{eq:ccurves}--\eqref{eq:shock} to calculate that a shock will form when $(\chi_b, \, \tau_b) \approx (0.34,\,0.13)$; as illustrated by the red dot in figure \ref{fig:am}(\textit{c}).
We evaluate the shock speed and location $\chi = \chi_s(\tau)$ for times $\tau > \tau_b$ using the Rankine--Hugoniot condition in \eqref{eq:shock_location}, as shown by the green curve in figure \ref{fig:am}(\textit{c}). 
The solid curves for $\tau = 0.17$, $0.21$, $0.25$ and $0.29$ in figure \ref{fig:am}(\textit{a}) are composed of the solution found using the method of characteristics combined with the above results for $\chi = \chi_s(\tau)$.
For $\tau > 0.25$, the minimum in the drag reduction starts to increase owing to nonlinear interaction with the SHS.

Alternatively, we can solve the advection--diffusion equation in \eqref{eq:advection_1} numerically using the method in Appendix \ref{app:0}. 
Diffusion over the length of the channel regularises the flow in the vicinity of the shock.
This methodology allows us to compute the distribution of $DR_0$ for $\tau \geq \tau_b$, as shown by the dotted curves for $\tau \geq 0.17$ in figure \ref{fig:am}(\textit{a}). 
The solutions, inclusive of a small amount of diffusion (dotted lines), remain fairly close to the solutions without diffusion (solid lines). 
However, diffusion marginally reduces the maximum (minimum) amplitude of the surfactant-flux (drag-reduction) distribution and widens the surfactant-flux (drag-reduction) distribution as time progresses. 

\subsubsection{Flow and surfactant transport at the scale of the periodic cell}

As the bolus of surfactant flux passes over a given plastron at the AM boundary, the proportion of the liquid--gas interface that is shear-free at the upstream end and no-slip at the downstream end varies; as seen from the concentration field in figure \ref{fig:am}(\textit{d}) and the streamwise slip velocity in figure \ref{fig:am}(\textit{e}). 
Using the asymptotic solution (Appendix \ref{subsec:a_strong_diff_strong_surfactant_strength}), 
\begin{equation}  \label{eq:am_limited_leading_results}
c_{0}(x; \, K_0) \approx \Gamma_0(x; \, K_0) \approx \begin{cases}
\st{0 \hspace{2.945cm} \text{for} \ \ -\phi_x\leq x\leq\phi_x-\gamma K_0/\beta,} \\ \st{\beta(x - \phi_x)/\gamma + K_0\quad \text{for} \quad \phi_x-\gamma K_0/\beta \leq x \leq \phi_x,}
\end{cases}
\end{equation}
the length of the shear-free region increases as $K_0$ decreases, $\beta$ increases and $\gamma$ decreases, as less surfactant is held at higher concentrations at the plastron's downstream stagnation point, $x = \phi_x = 0.5$.
The maximum concentration at the downstream end of the interface increases with $K_0$ and $\Delta c_0 = K_0$. 
Therefore, ${DR}_0$ attains a minimum when $K_0$ is at its maximum, where the amplitude ($K_0$) and length ($\gamma K_0/\beta$) of the stagnant cap are greatest. 
At the leading edge of the bolus of surfactant flux, large stagnant caps slowly transition into small ones over a long range in $\chi$, and at the trailing edge of the bolus of surfactant flux, small stagnant caps rapidly transition into large ones over a shorter range in $\chi$.

\subsection{Diffusion--Marangoni boundary}
\label{subsec:Diffusion--Marangoni boundary}

We derive asymptotic solutions at the diffusion--Marangoni (DM) boundary in Appendix \ref{subsec:a_strong_diff_and_surf}, where both diffusion and Marangoni effects dominate over interfacial advection (see the DM boundary in figure \ref{fig:dr}\textit{a}). 
The asymptotic solution for $a_{\text{DM}}$ has a complex dependence on $\alpha$, $\beta$, $\delta$, $\theta$, $\zeta$ and $K_0$ that is different to $a_{\text{AM}}$. Nevertheless, we find that $a_{\text{DM}}$ exhibits similar trends as $a_{\text{AM}}$ with respect to $\alpha$, $\beta$, $\delta$, $\theta$, $\zeta$ and $K_0$; furthermore, because of its nonlinear dependence on $K_0$, wave-steepening effects can lead to shock formation in the surfactant-flux and drag-reduction distribution. 
When bulk--surface exchange is weak, e.g. at the DM$_\text{E}$ boundary depicted in figure \ref{fig:dr}(\textit{b}), we find that $a_{\text{AM}} = a_{\text{DM}} \approx 1/(\theta+\zeta \phi_x)$, such that the propagation speed is the same for all $\alpha$, $\beta$, $\gamma$ and $\delta$ and there is no wave-steepening at leading order.

\section{Discussion} \label{sec:discussion}

Superhydrophobic surfaces (SHSs) can be contaminated by trace amounts of surfactant, which may reduce their drag-reducing potential for applications in microchannels and marine hydrodynamics \citep{peaudecerf2017traces, tomlinson2023model}. 
\jl{In order to provide some fundamental insight into the effect of these local spatio--temporal variations in surfactant concentration, we have considered a scenario that is simplified compared to real applications, which take place in complex natural, industrial or laboratory environments.
We have derived an asymptotic theory to model the unsteady transport of soluble surfactant in a laminar pressure-driven channel flow bounded between two SHSs.} 
The SHSs are textured with periodic grooves in the streamwise and spanwise directions.
Exploiting the multiple length scales in the problem, we have derived and solved a quasi-steady advection--diffusion equation for surfactant concentration transport over moderate length scales, which is coupled to an unsteady advection--diffusion equation for surfactant flux transport over large length scales.
The governing partial differential equations for surfactant flux transport can be rewritten in terms of the slip length or drag reduction, key quantities of interest for practical applications.
When there is a disturbance to the surfactant flux that varies over a large number of periods \cite[similar to those measured in the ocean by][]{frossard2019properties}, our model allows us to make predictions about the propagation speed, the shape of the disturbance, and the evolving distribution of slip length and drag reduction. 
Furthermore, in certain regions of parameter space, higher surfactant concentrations can lead to surface immobilisation and therefore slower surfactant flux transport, leading to wave-steepening and shock formation in the distribution of surfactant flux, slip length and drag reduction.

We have investigated the transport of the surfactant flux and the corresponding reduction in drag along the channel length in distinct asymptotic regimes (figure \ref{fig:dr}), defined by the relative strength of bulk diffusion ($\alpha$), the partition coefficient ($\beta$), surfactant strength ($\gamma$), surface diffusion ($\delta$), exchange strength ($\nu$) and background surfactant flux ($K_b$).
Extending the results of \citet{tomlinson2023laminar}, we also identify the bulk capacitance ($\theta$) and surface capacitance ($\zeta$) as key parameters, which quantify the bulk and surface response to time-dependent changes in the surfactant flux, respectively. 
Distinct flow patterns emerge across parameter space.
If a bolus of surfactant flux is injected into the channel, the propagation speed is constant in the Marangoni- (M), advection- (A) and diffusion-dominated (D) regimes, and the shape of the bolus of surfactant flux does not change appreciably along the length of the channel (figure \ref{fig:m}).
In region M, the interfacial concentration profile along each plastron is linear, the liquid--gas interface is immobilised and there is negligible drag reduction (${DR}_0 \ll 1$); in region A, the interfacial concentration is uniform and then increases in a boundary layer at each plastron's downstream contact line, the liquid--gas interface is \jl{largely} shear-free and there is substantial drag reduction ($1-{DR}_0 \ll 1$); and in region D, the interfacial concentration profile in each plastron is uniform, the liquid--gas interface is shear-free and there is substantial drag reduction ($1-{DR}_0 \ll 1$).

The speed of propagation of disturbances to the surfactant flux across different asymptotic regimes is summarised in tables \ref{tab:1} and \ref{tab:2}.
\st{All of the propagation speeds presented in table \ref{tab:1} decrease with $\zeta$ and $\theta$. 
When the surfactant is very soluble, $\zeta \ll \theta$, changes in the surfactant flux are advected at the cross-channel-averaged bulk propagation speed. 
As the surfactant becomes more insoluble (increasing $L_d$) and the area for adsorption increases (increasing $\phi_x\phi_z$), more surfactant is adsorbed and the propagation speed falls as the liquid--gas interface is more immobilised. 
Because the volume flux of fluid is fixed, changes in the cross-channel area of the channel through $P_z$ or $\phi_z P_z$ reduce the streamwise velocity and therefore the propagation speed.}
However, at the advection--Marangoni (AM) and diffusion--Marangoni (DM) boundaries, the values of $DR_0$ can span the whole range from 0 to 1 along the \jl{whole channel}, depending on the local surfactant flux.
Here, a bolus of surfactant flux steepens at its rear side as smaller concentrations of surfactant are advected faster than larger concentrations, because the liquid--gas interface is more mobile at lower concentrations, resulting in the formation of a shock (figure \ref{fig:am}). 
We anticipate that the results observed here for a Gaussian initial distribution in the surfactant flux would be applicable to other distribution profiles due to the dominant advective nature of the transport at the channel scale. 
\st{Hence, for other distribution profiles, the dimensionless parameters in table 3 can be chosen to move  into a favourable region of parameter space (second column of table 2) for drag-reduction applications and optimise the propagation speed (third column of table 2) to maximise the space- and time-average drag reduction.}
Wave-steepening effects arise only in the strong-exchange limit (large $\nu$). 
In contrast, when exchange is weak, the propagation speed given in table \ref{tab:1} is the same for all $\alpha$, $\beta$, $\gamma$, $\delta$, $\theta$, $\zeta$ and $\phi_x$. 
\begin{table}
\resizebox{\columnwidth}{!}{%
\centering
    \begin{tabular}{c c c c}
    \hline \\[-6pt]
    \multicolumn{3}{c}{\hspace{3cm} Strong exchange $\displaystyle \nu \gg \max(1,\,\alpha,\,\beta,\,\delta)$} \\[4pt]
    \hline
    Region & Parameter space & Propagation speed ($a$) & Regime \\[6pt]
    $\text{M}$ & $\displaystyle \gamma \gg \max\left(1,\,\alpha,\,\beta,\,\delta\right)$, $K_b > 2\phi_x\beta/\gamma$ & $\displaystyle\frac{1}{\theta + \zeta \phi_x}$ & ${DR}_0 \ll 1$ \\[10pt]
    $\text{A}$ & $\max(\alpha, \, \delta) \ll 1, \quad \gamma \ll \min(1,\,\beta)$ & $\displaystyle\frac{(\beta+1)}{\phi_{x} (\zeta +\theta )+ \theta (1 - \phi_{x})(\beta+1)}$  & $1-{DR}_0 \ll 1$\\[10pt]
    D & $\min(\alpha,\,\delta) \gg \max(1,\,\gamma)$ & $\displaystyle\frac{\alpha(1 + \beta \phi_x) + \delta(1-\phi_x)}{(\theta + \zeta \phi_x)(\alpha + \delta(1-\phi_x))}$ & $1-{DR}_0 \ll 1$\\
    AM & $\min(\beta, \, \gamma)\gg1\gg\max(\alpha, \, \delta)$, $K_b \leq 2\phi_x\beta/\gamma$ & $\displaystyle\frac{2\beta}{\gamma(\zeta+\theta) K_0 + 2\beta \theta(1-\phi_x)}$ & $0<{DR}_0 < 1$ \\
    \hline \\[-6pt]
    \multicolumn{3}{c}{\hspace{3cm} Weak exchange $\displaystyle \nu \ll \min(1,\,\alpha,\,\beta,\,\delta)$}\\[4pt]
    \hline
    Region & Parameter space & Propagation speed ($a$) & Regime \\[6pt]
    $\text{M}_\text{E}$ & $\gamma \gg \max(1,\,\alpha,\,\beta,\,\delta)$ & $\displaystyle \displaystyle\frac{1}{\theta + \zeta \phi_x}$ & ${DR}_0 \ll 1$ \\[10pt]
    $\text{A}_\text{E}$, $\text{D}_\text{E}$ & $\gamma \ll \min(1,\,\alpha,\,\beta,\,\delta)$ & $\displaystyle \displaystyle\frac{1}{\theta + \zeta \phi_x}$ & $1-{DR}_0 \ll 1$ \\
    \hline
    \end{tabular}%
    }
    \caption{Summary of the propagation speed and drag reduction regime in the main regions analysed in the strong-exchange problem: the Marangoni-dominated region (M), the advection-dominated region (A), the diffusion-dominated region (D) and the advection-Marangoni (AM) boundary; and their analogues in the weak-exchange problem: the $\text{M}_\text{E}$, $\text{A}_\text{E}$ and $\text{D}_\text{E}$ regions. 
    The propagation speed is expressed in terms of the transport coefficients $\alpha$, $\beta$, $\gamma$, $\delta$, $\nu$, $\theta$ and $\zeta$ given in \eqref{eq:coefficients}.
    }
    \label{tab:1}
\vspace{.25cm}
\resizebox{\columnwidth}{!}{%
    \centering
    \begin{tabular}{c c c c c c c c c c}
    \hline \\[-6pt]
    Quantity & $\displaystyle \alpha$ & $\displaystyle \beta$ & $\displaystyle \gamma$ & $\displaystyle \delta$ & $\displaystyle \nu$ & $K_b$ & $\displaystyle\theta$ & $\displaystyle\zeta$ \\[12pt]
    Prop. to & $\displaystyle \frac{\hat{D} \hat{P}_z \hat{H}}{\hat{P}_x \hat{Q}}$ & $\displaystyle \frac{\hat{K}_a}{\hat{H} \hat{K}_d}$ & $\displaystyle \frac{\hat{K}_a^2 \hat{A} \hat{H}^2 \max(\hat{K}(\hat{x}, \, 0))}{\hat{P}_x \hat{K}_d^2 \hat{\mu} \hat{Q}^2}$ & $\displaystyle \frac{\hat{D}_I \hat{K}_a \hat{P}_z }{\hat{P}_x \hat{K}_d \hat{Q}}$ & $\displaystyle \frac{\hat{K}_a \hat{P}_z \hat{H}^2}{\hat{P}_x \hat{Q}}$ & $\displaystyle \frac{\min(\hat{K}(\hat{x}, \, 0))}{\max(\hat{K}(\hat{x}, \, 0))}$ & $\displaystyle \frac{\hat{P}_z}{\hat{H}}$ &  $\displaystyle \frac{\hat{K}_a \hat{P}_z}{\hat{H}^2 \hat{K}_d }$ \\
    \hline
    \end{tabular}
    }
    \caption{Summary of the dimensionless parameters appearing in table \ref{tab:1} that affect the leading-order propagation speed, and their dependence on the dimensional quantities characterising the flow, surfactant properties and geometry (outlined in \S\ref{sec:Formulation}).}
    \label{tab:2}
\end{table}

As a practical illustration, we now evaluate the propagation speeds ($a$) presented in tables \ref{tab:1} and \ref{tab:2} using parameters characteristic of microchannel applications.
In the analysis that follows, the transport coefficients in \eqref{eq:coefficients} have been appropriately adjusted for the geometry employed in \citet{temprano2023single}, which is bounded by one SHS and solid wall rather than the two SHSs considered in \S\ref{sec:Formulation}.
In regions M, $\text{M}_\text{E}$, $\text{A}_\text{E}$ and $\text{D}_\text{E}$, $\theta \approx 4$, $\zeta \approx 20.8$, and therefore, the dimensionless propagation speed is predicted to be $a \approx 0.04$. 
Using $\epsilon \hat{U} = 2.4 \times 10^{-4}\SI{}{\meter/ \second}$ as the velocity scale, a surfactant-flux perturbation is advected out of the channel at approximately $9.6 \times 10^{-6}\SI{}{\meter/ \second}$ when Marangoni effects dominate, significantly slower than the fluid itself. 
For regions A and D, $\alpha \approx 0.4$, $\beta \approx 3.7$, $\delta \approx 1$, and therefore, $a \approx 0.19$.
\st{Thus, when advection or diffusion dominates the surfactant transport over each period, we find that the surfactant-flux disturbance is advected out of the channel at approximately $4.5 \times 10^{-5}\SI{}{\meter/ \second}$, approximately five times faster than the propagation speed in region M}.
The difference in propagation speed is because shear-free liquid--gas interfaces (regions A and D) that lack surfactant gradients give rise to greater streamwise velocities than immobilised liquid--gas interfaces (region M) with substantial surfactant gradients.
We can then vary these parameters to maximise the propagation speed in regions M, A and D, and therefore, if we suppose a bolus of surfactant enters and leaves the channel in the measurement time interval, we can minimise the space and time-averaged drag reduction for applications. 

Our theory rests on several assumptions, which we summarise below and suggest possible extensions to this study. 
\st{First, the asymptotic expansion requires $\epsilon = \hat{H} / \hat{P}_x \ll 1$ and $\mathcal{E} = \hat{P}_x / \hat{L}_x \ll 1$, which seems reasonable based on the microchannel configurations considered herein, e.g. $1\times 10^{-5}\SI{}{\meter} \lessapprox \hat{H} \lessapprox 1\times 10^{-2}\SI{}{\meter}$, $1 \times 10^{-3}\SI{}{\meter} \lessapprox \hat{P}_x \lessapprox 1 \times 10^{-1}\SI{}{\meter}$ and $1 \times 10^{-2}\SI{}{\meter} \lessapprox \hat{L}_x \lessapprox 1\SI{}{\meter}$ in \citet{ou2004laminar}, \citet{ou2005direct}, \citet{Daniello_etal_PF_2009}, \citet{bolognesi2014evidence}, \citet{song2018effect} and \citet{temprano2023single}. }
This may need to be revised in other applications, e.g. marine hydrodynamics, where the boundary layer grows over the surface of the vessel and $\hat{L}_x \gg 1\SI{}{\meter}$. 
\st{Furthermore, the homogenisation framework that we have developed here for laminar channel flow is a stepping stone towards the turbulent external flow problem, where transient values of the slip length and drag reduction are expected due to variations in the surfactant concentration \citep{frossard2019properties}. }
Second, we have only considered the case where diffusion is strong enough to eliminate cross-channel concentration gradients. 
The reader is referred to \cite{tomlinson2023laminar} for a discussion of the parameter regimes where cross-channel concentration gradients first become important.
\st{Third, several potential physical complications can arise when considering surfactant-contaminated superhydrophobic channels, such as liquid--gas interface deformation, the interaction of the interior flow with the external gas or liquid subphase \cite[as discussed in][]{lee2016superhydrophobic,sundin2022slip} or turbulence in the outer flow field \cite[as discussed in][]{tomlinson2023model}, which have been neglected in this study.} 
\st{Finally, with minor modifications, the theory outlined here could be used to analyse diabatic flows \citep{maynes2008thermal}, where thermocapillary stresses can affect the performance of SHSs in the thermal management of electronics \citep{kirk2020thermocapillary}.}  

To summarise, we have shown how a disturbance to the surfactant concentration field can undergo wave-steepening as it propagates under a laminar channel flow bounded by SHSs. 
This nonlinear evolution is shared by the distributions of the effective slip and drag reduction in microchannel applications, emphasising the importance of treating these as dynamic quantities in time-evolving flows. 
\jl{We hope our theoretical study will encourage further experimental work to study these effects and gain some practical insight for applications which take place in environments where surfactant concentrations can vary in time and space. 
In the first instance, our closed-form asymptotic solutions for the drag reduction in various parts of the parameter space can provide testable predictions for experimental studies in well-controlled laboratory settings. }

\section*{Acknowledgements}

We acknowledge support from CBET--EPSRC (EPSRC Ref. EP/T030739/1, NSF \#2054894), as well as partial support from ARO MURI W911NF-17-1-0306.
For the purpose of open access, the authors have applied a Creative Commons Attribution (CCBY) licence to any Author Accepted Manuscript version arising.
F. T-C. acknowledges support from a Distinguished Postdoctoral Fellowship from the Andlinger Center for Energy and the Environment.

\section*{Declaration of interests}

The authors report no conflict of interest.

\appendix

\section{Velocity functions and fluxes}\label{app:v}

\st{We define the streamwise flow due to the pressure gradient in $\mathcal{D}_1$,
\refstepcounter{equation}  \label{eq:tildeu_bvp}
\begin{multline} 
    \nabla^2_{\perp}\tilde{U} = 1, \quad \text{subject to} \quad \tilde{U}_{y_\perp}(0,\,z_s)=0, \quad \tilde{U}(0,\,z_{ns})=0, \quad \tilde{U}_{y_\perp}(2,\,z_s)=0, \\ \tilde{U}(2,\,z_{ns})=0, \quad \tilde{U}_{z_\perp}(y_\perp,\,-P_z)=0, \quad \tilde{U}_{z_\perp}(y_\perp,\,P_z)=0;
	\tag{\theequation\textit{a--g}}
\end{multline}
the streamwise flow due to the surfactant gradient in $\mathcal{D}_1$,
\refstepcounter{equation} \label{eq:baru_bvp}
\begin{multline} 
    \nabla^2_{\perp}\bar{U} = 0, \quad \text{subject to} \quad \bar{U}_{y_\perp}(0,\,z_s)= 1, \quad \bar{U}(0,\,z_{ns})= 0, \quad \bar{U}_{y_\perp}(2,\,z_s) = -1, \\ \bar{U}(2,\,z_{ns})= 0, \quad \bar{U}_{z_\perp}(y_\perp,\,-P_z)=0, \quad \bar{U}_{z_\perp}(y_\perp,\,P_z)=0;
	\tag{\theequation\textit{a--g}}
\end{multline}
and the streamwise flow due to the pressure gradient in  $\mathcal{D}_2$,
\refstepcounter{equation} \label{eq:hatu_bvp} 
\begin{equation} 
    \breve{U}_{y_\perp y_\perp} = 1, \quad \text{subject to} \quad \breve{U}(0)= 0, \ \quad \breve{U}(2)=0;
	\tag{\theequation\textit{a--c}}
\end{equation} 
with $z_s\equiv \{z_\perp\in [-\phi_z P_z, \,\phi_z P_z]\}$ and $z_{ns} \equiv \{z_\perp\in [-P_z,\, -\phi_z P_z]\}\cup\{z_\perp\in [\phi_z P_z, \,P_z]\}$.
We define the volume and surface fluxes
\refstepcounter{equation}  \label{eq:bulkfluxes}
\begin{equation}
    \tilde{Q} = \int_{\mathscr{A}_n} \tilde{U} \, \text{d} A, \quad \bar{Q} =  \int_{\mathscr{A}_n} \bar{U} \,  \text{d} A, \quad \breve{Q} = -\frac{4 P_z}{3},
	\tag{\theequation\textit{a--c}}
\end{equation}
and
\refstepcounter{equation}  
\label{eq:surfacefluxes}
\begin{equation}
q = \int_{\mathscr{I}_n} u_0 \, \text{d} z_\perp, \quad \tilde{q} = \int_{\mathscr{I}_n} \tilde{U} \, \text{d} z_\perp, \quad
\bar{q} = \int_{\mathscr{I}_n} \bar{U} \, \text{d} z_\perp . \tag{\theequation\textit{a--c}}
\end{equation}
}

\section{Asymptotic solutions}\label{app:A}

\subsection{Strong Marangoni effect: region M} \label{subsec:a_strong_surfactant_strength}

Assume that $\beta=O(1)$, $\gamma \gg \max(1,\,\alpha,\,\delta)$ and $\nu \gg \max(1,\,\alpha,\,\delta)$, so that exchange is strong, $c_0 = \Gamma_0$, and expand using $c_{0} = c_{00} + c_{01} / \gamma + ...$.
At $O(\gamma)$, Marangoni effects are comparable to bulk advection and diffusion, and \eqref{eq:strong_diff_2} reduces to
\refstepcounter{equation} \label{eq:marangoni_limited_leading_order}
\begin{multline}
c_{00} c_{00x} = 0 \quad \text{in} \quad \mathcal{D}_1, \quad c_{00} - c_{00x} = K_0 \quad \text{in} \quad \mathcal{D}_2, \\ 
\text{subject to} \quad c_{00}(\phi_x^{-}) = c_{00}(\phi_x^{+}), \quad c_{00}(-\phi_x) = c_{00}(2 - \phi_x),
    \tag{\theequation\textit{a--d}}
\end{multline}
which gives $c_{00}=K_0$ in $\mathcal{D}_1\cup\mathcal{D}_2$.
At $O(1)$, Marangoni effects are comparable to advection and bulk diffusion, and \eqref{eq:strong_diff_2} gives
\refstepcounter{equation} \label{eq:marangoni_limited_first_order}
\begin{multline}
\beta - c_{01x} = 0 \quad \text{in} \quad \mathcal{D}_1, \quad c_{01} - c_{01x} = 0 \quad \text{in} \quad \mathcal{D}_2, \\
\text{subject to} \quad
    c_{01}(\phi_x^{-}) = c_{01}(\phi_x^{+}), \quad c_{01}(-\phi_x) = c_{01}(2 - \phi_x),
    \tag{\theequation\textit{a--d}}
\end{multline}
such that $c_{01} = \beta (x - \phi_x (E + 1)/(E - 1))$ in $\mathcal{D}_1$ where $E \equiv \exp(2(1-\phi_x)/\alpha)$.
Similarly, at $O(1/\gamma)$, \eqref{eq:strong_diff_2} reduces to
\refstepcounter{equation} \label{eq:marangoni_limited_second_order}
\begin{multline}
K_0 c_{02x} = (\beta+1)c_{01} - c_{01}c_{01x} - (\alpha + \delta)c_{01x} \quad \text{in} \quad \mathcal{D}_1, \quad c_{02} - c_{02x} = 0 \quad \text{in} \quad \mathcal{D}_2, \\
\text{subject to} \quad
    c_{02}(\phi_x^{-}) = c_{02}(\phi_x^{+}), \quad c_{02}(-\phi_x) = c_{02}(2 - \phi_x),
    \tag{\theequation\textit{a--d}}
\end{multline}
which gives $c_{02} = \beta x^2/(2 K_0) + \beta x (\phi_x - \alpha - \delta - 2\phi_x E/(E - 1)))/K_0 + M_1$ where $M_1$ is an integration constant.
Hence, we have that $c_0$ and ${DR}_0$ are given by \eqref{eq:marangoni_limited_results} and (\ref{eq:eq_marangoni_results}\textit{b}) respectively in region M. 
Substituting $c_0$ into \eqref{eq:Rs}, at leading order we have that
\refstepcounter{equation} \label{eq:Rs_marangoni}
\begin{equation}
    C_0 \approx 2 (\theta + \zeta \phi_x) K_0, \quad A_0 \approx 2 (1 + \beta \phi_x)K_0, \quad  M_0 \approx 2  \beta \phi_x K_0, \quad D_0 \approx 0, \tag{\theequation\textit{a--d}}
\end{equation}
where $K_0 = K_0(\chi, \,\tau)$ and $D_0 = O(1/\gamma)$. 
Then \eqref{eq:advection} with $\lambda = 0$ has the solution
\begin{equation} \label{eq:eq_marangoni}
    K_0 = K_0 (\chi - a_{\text{M}} \tau, \, 0) \quad \text{where} \quad a_{\text{M}} \approx \frac{1}{\theta + \zeta \phi_x}.
\end{equation}
Note that when $\zeta \rightarrow 0$, $a_{\text{M}} \rightarrow 1/\theta$.

Next, assume that $\nu \ll \min(1,\,\alpha,\,\delta)$ so that exchange is weak, and expand using $c_{0} = c_{00} + c_{01} / \gamma + ...$ and $\Gamma_{0} = \Gamma_{00} + \Gamma_{01} / \gamma + ...$.
At $O(\gamma)$, Marangoni effects are comparable to bulk advection and bulk diffusion, and \eqref{eq:strong_diff_2} reduces to
\refstepcounter{equation} 
\label{eq:we_marangoni_limited_leading_order}
\begin{multline} 
    (c_{00} - \alpha c_{00x})_x = 0, \quad \Gamma_{00} \Gamma_{00x} = 0 \quad \text{in} \quad \mathcal{D}_1, \quad c_{00} - \alpha c_{00x} = K_0 \quad \text{in} \quad \mathcal{D}_2, \\ \text{subject to} \quad c_{00}(\phi_x^{-}) = c_{00}(\phi_x^{+}), \quad c_{00}(-\phi_x) = c_{00}(2 - \phi_x), \\
    c_{00}(\pm \phi_x) - \alpha c_{00x}(\pm \phi_x) = K_0, \quad  \Gamma_{00}(\pm \phi_x) \Gamma_{00x}(\pm \phi_x) = 0,
    \tag{\theequation\textit{a--g}}
\end{multline}
which gives $\Gamma_{00} = c_{00} = K_0$ as $\int_{x=-\phi_x}^{\phi_x} (\Gamma_{00} - c_{00}) \, \text{d}x$ = 0.
At $O(1)$, Marangoni effects are comparable to advection and bulk diffusion, and \eqref{eq:strong_diff_2} gives
\refstepcounter{equation} 
\label{eq:we_marangoni_limited_first_order}
\begin{multline} 
    (c_{01} - \alpha c_{01x})_x = 0, \quad \beta - \Gamma_{01x} = 0 \quad \text{in} \quad \mathcal{D}_1, \quad c_{01} - \alpha c_{01x} = 0 \quad \text{in} \quad \mathcal{D}_2, \\ \text{subject to} \quad c_{01}(\phi_x^{-}) = c_{01}(\phi_x^{+}), \quad c_{01}(-\phi_x) = c_{01}(2 - \phi_x), \\
    c_{01}(\pm \phi_x) - \alpha c_{01x}(\pm \phi_x) = 0, \quad \beta - \Gamma_{01x}(\pm \phi_x) = 0,
    \tag{\theequation\textit{a--g}}
\end{multline}
such that $\Gamma_{01} = \beta x$ and $c_{01} = 0$ as $\int_{x=-\phi_x}^{\phi_x} (\Gamma_{01} - c_{01}) \, \text{d}x = 0$.
Hence,
\refstepcounter{equation} \label{eq:marangoni_limited_we}
\begin{equation} 
    c_{0} = K_0 + ..., \quad \Gamma_{0} = K_0 + \beta x / \gamma..., \quad \Delta \Gamma_{0} = 2\beta\phi_x/\gamma + .... \tag{\theequation\textit{a,\,b}}
\end{equation}
Substituting \eqref{eq:marangoni_limited_we} into \eqref{eq:Rs} we recover \eqref{eq:Rs_marangoni} and the propagation speed in \eqref{eq:eq_marangoni}.

\subsection{Strong advection: region A} \label{subsec:a_strong_advection}

In the advection--dominated (A) region, assume that $\beta=O(1)$, $\gamma \ll 1$ and $\nu \gg\max(1,\,\alpha,\,\delta)$, such that $c_0 = \Gamma_0$ and expand using $c_{0} = c_{00} + \gamma c_{01} + ...$.
At $O(1)$, advection is comparable to diffusion, and \eqref{eq:strong_diff_2} reduces to
\refstepcounter{equation} \label{eq:advection_limited_leading_order}
\begin{equation}
(\beta+1) c_{00} - \left(\alpha + \delta\right) c_{00x} = K_0 \quad \text{in} \quad \mathcal{D}_1, \quad c_{00} -  \alpha c_{00x} = K_0 \quad \text{in} \quad \mathcal{D}_2,
    \tag{\theequation\textit{a,\,b}}
\end{equation}
subject to (\ref{eq:marangoni_limited_leading_order}\textit{c,\,d}), which gives $c_{00}=K_0/(\beta+1) + K_0 \beta \exp((\beta+1)(x-\phi_x)/(\alpha+\delta)) / (\beta+1)$ in $\mathcal{D}_1$ and $c_{00} = K_0$ in $\mathcal{D}_2$, for $\max(\alpha, \, \delta) \ll 1$.
Hence, we have that
\refstepcounter{equation} \label{eq:advection_limited}
\begin{equation} 
    c_{0} = \frac{K_0}{\beta+1}\left(1+ \beta \exp\left(\frac{(\beta+1)(x-\phi_x)}{\alpha+\delta}\right)\right)+ ..., \quad \Delta c_{0} = \frac{K_0 \beta}{\beta + 1} + .... \tag{\theequation\textit{a,\,b}}
\end{equation} 
From \eqref{eq:advection_limited}, $c_0$ is uniform over the upstream end of the liquid--gas interface and increases exponentially in a boundary layer close to the downstream stagnation point.
The surfactant gradient, size of the boundary layer and drag reduction increase with decreasing $K_0$, as the channel and liquid--gas interface becomes less contaminated with surfactant, or with increasing $\alpha$ and $\delta$, as diffusion eliminates the concentration gradient.
Substituting \eqref{eq:advection_limited} into \eqref{eq:Rs}, at leading order we have that
\refstepcounter{equation} \label{eq:Rs_advection}
\begin{equation}
    C_0 \approx \frac{2 \phi _{x} (\zeta +\theta) K_0 }{\beta +1} + 2 \theta (1 - \phi _{x} )K_0, \quad A_0 \approx 2 K_0 , \quad  M_0 \approx 0, \quad D_0 \approx 0, \tag{\theequation\textit{a--d}}
\end{equation}
where $K_0 = K_0(\chi, \,\tau)$ and $M_0 = D_0 = O(\gamma)$. 
Then \eqref{eq:advection} with $\lambda = 0$ has the solution 
\begin{equation} \label{eq:eq_advection}
     K_0 = K_0(x - a_{\text{A}} \tau, \, 0) \quad \text{where} \quad a_{\text{A}} \approx \frac{(\beta+1)}{\phi_{x} (\zeta +\theta )+ \theta (1 - \phi_{x})(\beta+1)}.
\end{equation}
Note that when $\beta \rightarrow 0$ and $\zeta \rightarrow 0$, $a_{\text{A}} \rightarrow 1/\theta$.

Next, assume that $\nu \ll \min(1,\,\alpha,\,\delta)$ and expand using $c_{0} = c_{00} + \gamma c_{01} + ...$ and $\Gamma_{0} = \Gamma_{00} + \gamma \Gamma_{01}+ ...$.
At $O(1)$, diffusion is comparable to advection, and \eqref{eq:strong_diff_2} reduces to
\refstepcounter{equation} 
\label{eq:we_marangoni_weak_leading_order}
\begin{multline} 
    (c_{00} - \alpha c_{00x})_x = 0, \quad \beta \Gamma_{00} - \delta \Gamma_{00x} + c_{00} - \alpha c_{00x} = K_0 \quad \text{in} \quad \mathcal{D}_1, \\ c_{00} - \alpha c_{00x} = K_0 \quad \text{in} \quad \mathcal{D}_2, \quad \text{subject to} \quad c_{00}(\phi_x^{-}) = c_{00}(\phi_x^{+}), \quad c_{00}(-\phi_x) = c_{00}(2 - \phi_x), \\
    c_{00}(\pm \phi_x) - \alpha c_{00x}(\pm \phi_x) = K_0, \quad \beta \Gamma_{00}(\pm \phi_x)  - \delta \Gamma_{00x}(\pm \phi_x) = 0.
    \tag{\theequation\textit{a--g}}
\end{multline}
which gives $c_{00} = K_0$ and $\Gamma_{00} = 2 \phi_x \beta K_0 \exp((\beta(\phi_x + x))/\delta)/(\delta(\exp((2\beta\phi_x)/\delta) - 1))$ as $\int_{x=-\phi_x}^{\phi_x} (\Gamma_{00} - c_{00}) \, \text{d}x = 0$.
Hence,
\refstepcounter{equation} \label{eq:d_limited_we}
\begin{equation} 
    c_{0} = K_0 + ..., \quad \Gamma_{0} = \frac{2 \phi_x \beta K_0 \exp\left(\frac{\beta \left(\phi_x +x\right)}{\delta }\right)}{\delta \left(\exp\left(\frac{2 \beta \phi_x }{\delta }\right)-1\right)}  + ..., \quad \Delta \Gamma_{0} = \frac{2 \phi_x \beta K_0}{\delta} + .... \tag{\theequation\textit{a,\,b}}
\end{equation}
Substituting \eqref{eq:d_limited_we} into \eqref{eq:Rs}, we have
\refstepcounter{equation} \label{eq:Rs_d_we}
\begin{equation}
    C_0 \approx 2 (\theta + \zeta \phi_x)K_{0} , \quad A_0 \approx 2  (1 + \beta \phi_x)K_{0}, \quad  M_0 \approx 0, \quad D_0 \approx 2 \phi_x \beta K_0, \tag{\theequation\textit{a--d}}
\end{equation}
and we recover \eqref{eq:eq_marangoni}. 

\subsection{Strong diffusion: region D} \label{subsec:a_strong_diffusion}

Assume that $\beta=O(1)$, $\min(\alpha,\,\delta)\gg\max(1,\,\gamma)$ and $\nu \gg\max(1,\,\alpha,\,\delta)$ such that $c_0 = \Gamma_0$. 
Let $\delta = d \alpha$ where $d = O(1)$ and expand using $c_{0} = c_{00} + c_{01} / \alpha + ...$.
At $O(\alpha)$, diffusion dominates and \eqref{eq:strong_diff_2} reduces to
\refstepcounter{equation} \label{eq:diffusion_limited_leading_order}
\begin{equation}
c_{00x} = 0 \quad \text{in} \quad \mathcal{D}_1, \quad c_{00x} = 0 \quad \text{in} \quad \mathcal{D}_2,
    \tag{\theequation\textit{a,\,b}}
\end{equation}
subject to (\ref{eq:marangoni_limited_leading_order}\textit{c,\,d}).
At $O(1)$, diffusion is comparable to advection, and \eqref{eq:strong_diff_2} gives
\refstepcounter{equation} \label{eq:diffusion_limited_first_order}
\begin{equation}
(\beta + 1) c_{00} - (1 + d) c_{01x} = K_0 \quad \text{in} \quad \mathcal{D}_1, \quad c_{00} - c_{01x} = K_0 \quad \text{in} \quad \mathcal{D}_2,
    \tag{\theequation\textit{a,\,b}}
\end{equation}
subject to (\ref{eq:marangoni_limited_first_order}\textit{c,\,d}). 
Integrating \eqref{eq:diffusion_limited_first_order} over $\mathcal{D}_1$ and $\mathcal{D}_2$ gives $c_{00} = (K_0(\alpha + \delta(1-\phi_x)))/(\alpha (\beta \phi_x + 1) + \delta (1 - \phi_x)).$
Hence, we have that
\refstepcounter{equation} \label{eq:diffusion_limited}
\begin{equation} 
    c_{0} = \frac{(\alpha + \delta(1-\phi_x))K_0}{(\alpha (\beta \phi_x + 1) + \delta (1 - \phi_x))} + ..., \quad \Delta c_{0} = \frac{2 \beta \phi_x (1-\phi_x)K_0}{\alpha (\beta \phi_x + 1) + \delta(1 - \phi_x)} + .... \tag{\theequation\textit{a--c}}
\end{equation}
From \eqref{eq:diffusion_limited}, the surfactant concentration and drag increase linearly as the flux of surfactant increases over the SHS, and the interface becomes completely shear-free as $K_0 \rightarrow 0$.
Substituting \eqref{eq:diffusion_limited} into \eqref{eq:Rs}, at leading order we have that
\refstepcounter{equation}  \label{eq:Rs_diffusion}
\begin{multline}
    C_0 \approx \frac{2 (\theta + \zeta \phi_x)(\alpha + \delta(1-\phi_x))K_0 }{\alpha (\beta \phi_x + 1) + \delta (1 - \phi_x)}, \quad A_0 \approx  \frac{2 (1 + \beta \phi_x)(\alpha + \delta(1-\phi_x))K_0 }{\alpha (\beta \phi_x + 1) + \delta (1 - \phi_x)}, \\  M_0 \approx 0, \quad D_0 \approx \frac{2  \beta \phi_x \delta(1-\phi_x)K_0}{\alpha (\beta \phi_x + 1) + \delta(1 - \phi_x)}, \tag{\theequation\textit{a--d}}
\end{multline}
where $K_0 = K_0(\chi, \, \tau)$ and $M_0 = O(1/\alpha)$. Then \eqref{eq:advection} with $\lambda = 0$ has the solution
\begin{equation} \label{eq:eq_diffusion}
     K_0 = K_0(x - a_{\text{D}} \tau, \, 0) \quad \text{where} \quad a_{\text{D}} \approx \frac{\alpha(1 + \beta \phi_x) + \delta(1-\phi_x)}{(\theta + \zeta \phi_x)(\alpha + \delta(1-\phi_x))}.
\end{equation}
Note that when $\beta \rightarrow 0$, $\delta \rightarrow 0$ and $\zeta \rightarrow 0$, $a_{\text{D}} \rightarrow 1/\theta$.

When $\nu \ll \min(1,\,\alpha,\,\delta)$, the expansion and solution are the same as in region A, such that we recover \eqref{eq:eq_marangoni}.

\subsection{Strong advection and strong Marangoni effect: the AM boundary} \label{subsec:a_strong_diff_strong_surfactant_strength}

At the advection--Marangoni boundary, assume that $\beta = O(\gamma)$, $\gamma \gg \max(\alpha, \, \delta)$, $\max(\alpha, \, \delta) \ll 1$ and $\nu \gg\max(1,\,\alpha,\,\delta)$, such that $c_0 = \Gamma_0$. 
Rescale $\alpha = a/\gamma$, $\beta = b\gamma$ and $\delta = d/\gamma$, where $a$, $b$ and $d$ are positive $O(1)$ constants.
Expand using $c_0 = c_{00} + c_{01}/\gamma + ...$.
At $O(\gamma)$, Marangoni effects are comparable to advection, and \eqref{eq:strong_diff_2} reduces to
\refstepcounter{equation} \label{eq:am_limited_leading_order}
\begin{equation}
c_{00}(b - c_{00x}) = 0 \quad \text{in} \quad \mathcal{D}_1, \quad c_{00} = K_0 \quad \text{in} \quad \mathcal{D}_2,
    \tag{\theequation\textit{a,\,b}}
\end{equation}
subject to (\ref{eq:marangoni_limited_leading_order}\textit{c,\,d}). 
For $K_0/b \leq 2\phi_x$, \eqref{eq:am_limited_leading_order} gives $c_0$ as \eqref{eq:am_limited_leading_results}. 
Hence, using \eqref{eq:dD_0ef}, we have that ${DR}_0$ is given by (\ref{eq:eq_AM_results}\textit{b}) at the AM boundary.
At $O(1)$, Marangoni effects are comparable to advection, and \eqref{eq:strong_diff_2} gives
\refstepcounter{equation} \label{eq:am_limited_first_order}
\begin{equation}
b c_{01} + c_{00} - c_{00}c_{01x} - c_{01}c_{00x}  = K_0 \quad \text{in} \quad \mathcal{D}_1, \quad c_{01} = 0 \quad \text{in} \quad \mathcal{D}_2,
    \tag{\theequation\textit{a,\,b}}
\end{equation}
subject to (\ref{eq:marangoni_limited_first_order}\textit{c,\,d}). 
For $K_0/b \leq 2\phi_x$, \eqref{eq:am_limited_first_order} gives
\begin{subequations}  \label{eq:am_limited_first}
\begin{align}
c_{01} &= K_0/b \quad \text{for all} \quad -\phi_x\leq x\leq\phi_x-K_0/b, \\ c_{01} &= x - \phi_x - K_0 \log(b(x-\phi_x)/K_0 + 1)/b \quad \text{for all} \quad \phi_x-K_0/b \leq x \leq \phi_x.
\end{align}
\end{subequations}
The solution for $K_0/b > 2\phi_x$ is outlined in \citet{tomlinson2023laminar} and is not included here as it gives a similar propagation speed to region M.
Substituting (\ref{eq:am_limited_leading_results},\ref{eq:am_limited_first}) into \eqref{eq:Rs}, we have
\refstepcounter{equation} \label{eq:Rs_am}
\begin{multline}
    C_0 \approx \frac{(\zeta + \theta) K_0^2}{2 b} + \theta (2 - 2\phi_x)K_0 , \quad A_0 \approx \frac{\gamma K_0^2}{2} + 2K_0, \\  M_0 \approx \frac{\gamma K_0^2}{2} , \quad D_0 \approx 0, \tag{\theequation\textit{a--d}}
\end{multline}
where $K_0 = K_0(\chi, \, \tau)$. 
Then \eqref{eq:advection} with $\lambda = 0$ has the solution
\begin{equation} \label{eq:eq_am}
     K_0 = K_0(\chi - a_{\text{AM}} \tau, \, 0) \quad \text{where} \quad a_{\text{AM}} \approx \frac{2b}{K_0(\zeta+\theta)+ 2 b \theta(1 - \phi_x)}.
\end{equation}
As $b \rightarrow K_0 / (2\phi_x)$, the bulk concentration $c_{00} \rightarrow K_0$ and we recover \eqref{eq:Rs_marangoni}.
As $b \rightarrow \infty$, the bulk concentration $c_{00} \rightarrow 0$ everywhere except at $x \approx \phi$ where $c_{00} = K_0$ and we recover \eqref{eq:Rs_advection}.
Furthermore, as $K_0 \rightarrow 2\phi_x b$, the propagation speed $a_{\text{AM}}\rightarrow a_{\text{M}}$, and as $K_0 \rightarrow 0$, the propagation speed $a_{\text{AM}} \rightarrow a_{\text{A}}$ for $\beta \gg 1$.

Next, assume that $\nu / \epsilon^2 \ll \min(1, \, \alpha, \, \delta)$ and expand using $c_{0} = c_{00} + \gamma c_{01} + ...$ and $\Gamma_{0} = \Gamma_{00} + \gamma \Gamma_{01}+ ...$.
At $O(\gamma)$, Marangoni effects are comparable to advection, and \eqref{eq:strong_diff_2} reduces to
\refstepcounter{equation} 
\label{eq:we_marangoni_limited_leading_order_weak_diff}
\begin{multline} 
    c_{00x} = 0, \quad b \Gamma_{00} - \Gamma_{00} \Gamma_{00x} = 0 \quad \text{in} \quad \mathcal{D}_1, \quad c_{00} = K_0 \quad \text{in} \quad \mathcal{D}_2, \\ \text{subject to} \quad c_{00}(\phi_x^{-}) = c_{00}(\phi_x^{+}), \quad c_{00}(-\phi_x) = c_{00}(2 - \phi_x), \\
    c_{00}(\pm \phi_x) = K_0, \quad  b\Gamma_{00}(\pm \phi_x) - \Gamma_{00}(\pm \phi_x) \Gamma_{00x}(\pm \phi_x) = 0.
    \tag{\theequation\textit{a--g}}
\end{multline}
For $K_0 \gamma \leq \beta \phi_x$, \eqref{eq:we_marangoni_limited_leading_order_weak_diff} gives 
\begin{subequations}  \label{eq:am_limited_leading_we}
\begin{align}
\Gamma_{00} &= 0 \quad \text{for all} \quad -\phi_x\leq x\leq\phi_x-2(\phi_x K_0/b)^{1/2}, \\ \Gamma_{00} &= b(x - \phi_x) + 2(b \phi_x K_0)^{1/2} \quad \text{for all} \quad \phi_x-2(\phi_x K_0/b)^{1/2} \leq x \leq \phi_x,
\end{align}
\end{subequations}
as $\int_{x = -\phi_x}^{\phi_x} (\Gamma_{00} - c_{00}) \, \text{d} x = 0$ and $c_{00}=K_0$. 
At $O(1)$, Marangoni effects are comparable to advection, and \eqref{eq:strong_diff_2} gives
\refstepcounter{equation} 
\label{eq:we_marangoni_limited_leading_order_weak_diff_we}
\begin{multline} 
    c_{01x} = 0, \quad b \Gamma_{01} - \Gamma_{00} \Gamma_{01x} - \Gamma_{01} \Gamma_{00x} = 0 \quad \text{in} \quad \mathcal{D}_1, \quad c_{01} = 0 \quad \text{in} \quad \mathcal{D}_2, \\ \text{subject to} \quad c_{01}(\phi_x^{-}) = c_{01}(\phi_x^{+}), \quad c_{01}(-\phi_x) = c_{01}(2 - \phi_x), \quad
    c_{01}(\pm \phi_x) = 0, \\  b \Gamma_{01}(\pm \phi_x) - \Gamma_{00}(\pm \phi_x) \Gamma_{01x}(\pm \phi_x) - \Gamma_{01}(\pm \phi_x) \Gamma_{00x}(\pm \phi_x) = 0.
    \tag{\theequation\textit{a--g}}
\end{multline}
For $K_0 \gamma \leq \beta\phi_x$, \eqref{eq:we_marangoni_limited_leading_order_weak_diff} gives $\Gamma_{01} = 0$ for all $-\phi_x\leq x\leq\phi_x$ as $\int_{x = -\phi_x}^{\phi_x} (\Gamma_{01} - c_{01}) \, \text{d} x = 0$ and $c_{01}=0$. 
Substituting \eqref{eq:we_marangoni_limited_leading_order_weak_diff_we} into \eqref{eq:Rs}, we have
\refstepcounter{equation} \label{eq:Rs_ame}
\begin{equation}
    C_0 \approx 2 ( \theta + \zeta \phi_x)K_0, \quad A_0 \approx 2(1 + \beta \phi_x)K_0, \quad  M_0 \approx 2  \beta \phi_x K_0, \quad D_0 \approx 0, \tag{\theequation\textit{a--d}}
\end{equation}
and we recover \eqref{eq:eq_marangoni}.

\subsection{Strong diffusion and strong Marangoni effect: the DM boundary} \label{subsec:a_strong_diff_and_surf}

Assume that $\beta = O(1)$, $\gamma \gg 1$ and $\nu \gg \max(1, \, \alpha, \, \delta)$, so that $c_0 = \Gamma_0$. 
Rescale $\alpha = \mathcal{A} \gamma$ and $\delta = d \gamma$, where $\mathcal{A}$ and $d$ are positive $O(1)$ constants. 
Expand using $c_0 = c_{00} + c_{01}/\gamma + ...$.
At $O(\gamma)$, Marangoni effects are comparable to diffusion, and \eqref{eq:strong_diff_2} reduces to
\refstepcounter{equation} \label{eq:dm_limited_leading_order}
\begin{equation}
c_{00} c_{00x} + (\mathcal{A}+d) c_{00x} = 0 \quad \text{in} \quad \mathcal{D}_1, \quad c_{00x} = 0 \quad \text{in} \quad \mathcal{D}_2,
    \tag{\theequation\textit{a,\,b}}
\end{equation}
subject to (\ref{eq:marangoni_limited_leading_order}\textit{c,\,d}), which means that $c_{00}$ is uniform along the periodic cell. 
At $O(1)$, Marangoni effects are comparable to advection and diffusion, and \eqref{eq:strong_diff_2} gives
\refstepcounter{equation} \label{eq:dm_limited_first_order}
\begin{equation}
(\beta+1)c_{00} - c_{00}c_{01x} - (\mathcal{A}+d)c_{01x} = K_0 \quad \text{in} \quad \mathcal{D}_1, \quad c_{00} - \mathcal{A} c_{01x} = K_0 \quad \text{in} \quad \mathcal{D}_2,
    \tag{\theequation\textit{a,\,b}}
\end{equation}
subject to (\ref{eq:marangoni_limited_first_order}\textit{c,\,d}). 
Integrating \eqref{eq:dm_limited_leading_order} over $\mathcal{D}_1$ and $\mathcal{D}_2$ gives $c_{00} = c_{00}(K_0)$ as the solution to the quadratic equation
\begin{equation} \label{combined_dm_concentration_gradient}
    (c_{00} - K_0)(\phi_x - 1)(\mathcal{A} + d + c_{00}) = \mathcal{A}\phi_x(c_{00}(\beta + 1) - K_0).
\end{equation}
We can then substitute $c_{00}$ back into \eqref{eq:dm_limited_first_order}, which gives $c_{01} = (x((\beta+1)c_{00} - K_0))/(\mathcal{A} + d + c_{00}) + D_1$ in $\mathcal{D}_1$, where $D_1$ is an integration constant.
Hence,
\refstepcounter{equation} \label{eq:dm_limited}
\begin{equation} 
    c_{0} = c_{00}(K_0) + ..., \quad \Delta c_{0} = \frac{2 \phi_x ((\beta+1)c_{00} - K_0))}{\mathcal{A} + d + c_{00}} + .... \tag{\theequation\textit{a,\,b}}
\end{equation}
At the DM boundary, the shear stress at the liquid-gas interface is not sufficient to completely immobilise the interface and there is partial drag reduction, as seen in \eqref{eq:dm_limited}. 
The concentration field is approximately uniform with a shallow gradient in $\mathcal{D}_1$ and $\mathcal{D}_2$.
Substituting \eqref{eq:dm_limited} into \eqref{eq:Rs} gives
\refstepcounter{equation} \label{eq:Rs_dm}
\begin{multline} 
    C_0 \approx 2 (\theta + \zeta \phi_x)c_{00}, \quad \quad A_0 \approx 2 (1 + \beta \phi_x) c_{00}, \\  M_0 \approx \frac{4 \phi_x c_{00} ((\beta+1) c_{00} - K_0)}{\mathcal{A} + d + c_{00}} , \quad D_0 \approx \frac{2 \phi_x d ((\beta+1) c_{00} - K_0))}{\mathcal{A} + d + c_{00}}, \tag{\theequation\textit{a--f}}
\end{multline}
where $c_{00} = c_{00}(K_0)$. 
Then \eqref{eq:advection} with $\lambda = 0$ has the solution
\begin{equation} \label{eq:eq_dm}
    K_0 = K_0(\chi - a_{\text{DM}} \tau, \, 0) \quad \text{where} \quad a_{\text{DM}}(K_0) \approx \frac{A'_{0} - M'_{0} - D'_{0}}{C'_{0}},
\end{equation}
where primes denote derivatives of the functions defined in \eqref{eq:Rs_dm}.
As $\mathcal{A} \rightarrow 0$ and $d \rightarrow 0$, the bulk concentration $c_{00} \rightarrow K_0$ and we recover \eqref{eq:Rs_marangoni}.
As $\mathcal{A} \rightarrow \infty$ and $d \rightarrow \infty$, the bulk concentration $c_{00} \rightarrow (\alpha + \delta (1 - \phi_x))K_0/(\alpha (\beta\phi_x+1) + \delta(1 - \phi_x))$ and we recover \eqref{eq:Rs_diffusion}.
Furthermore, as $K_0 \rightarrow \infty$, we need $c_{00} = K_0$ to satisfy \eqref{combined_dm_concentration_gradient} and the propagation speed $a_{\text{DM}} \rightarrow a_{\text{M}}$, and as $K_0 \rightarrow 0$, we linearise \eqref{combined_dm_concentration_gradient} (neglecting terms $O(c_{00}^2, \, c_{00} K_0)$) and the propagation speed $a_{\text{DM}} \rightarrow a_{\text{D}}$.  

Next, assume that $\nu \ll \min(1,\,\alpha,\,\delta)$ and expand using $c_{0} = c_{00} + c_{01} / \gamma + ...$ and $\Gamma_{0} = \Gamma_{00} + \Gamma_{01} / \gamma + ...$.
At $O(\gamma)$, Marangoni effects are comparable to diffusion, and \eqref{eq:strong_diff_2} reduces to
\refstepcounter{equation} 
\label{eq:we_marangoni_limited_leading_order_strong_diff}
\begin{multline} 
    c_{00xx} = 0, \quad \mathcal{A} c_{00x} + \Gamma_{00} \Gamma_{00x} + d \Gamma_{00x} = 0 \quad \text{in} \quad \mathcal{D}_1, \quad c_{00x} = 0 \quad \text{in} \quad \mathcal{D}_2, \\ \text{subject to} \quad c_{00}(\phi_x^{-}) = c_{00}(\phi_x^{+}), \quad c_{00}(-\phi_x) = c_{00}(2 - \phi_x), \\
    c_{00x}(\pm \phi_x) = 0, \quad  \Gamma_{00}(\pm \phi_x) \Gamma_{00x}(\pm \phi_x) + d \Gamma_{00x}(\pm \phi_x) = 0,
    \tag{\theequation\textit{a--g}}
\end{multline}
which implies that $\Gamma_{00}$ and $c_{00}$ are uniform along the periodic cell.
At $O(1)$, Marangoni effects are comparable to surface advection and diffusion, and \eqref{eq:strong_diff_2} gives
\refstepcounter{equation} 
\label{eq:we_marangoni_limited_first_order_strong_diff}
\begin{multline} 
    c_{01xx} = 0, \quad c_{00} - \mathcal{A} c_{01x} + \beta\Gamma_{00} - (\Gamma_{00}+d) \Gamma_{01x} = K_0
    \quad \text{in} \quad \mathcal{D}_1, \quad c_{01x} = 0 \quad \text{in} \quad \mathcal{D}_2, \\ \text{subject to} \quad c_{01}(\phi_x^{-}) = c_{01}(\phi_x^{+}), \quad c_{01}(-\phi_x) = c_{01}(2 - \phi_x), \\
    c_{00} - \mathcal{A} c_{01x}(\pm \phi_x) = K_0, \quad \beta\Gamma_{00} - (\Gamma_{00}+d) \Gamma_{01x}(\pm \phi_x) = 0.
    \tag{\theequation\textit{a--g}}
\end{multline}
such that $c_{00} = \Gamma_{00} = K_0$, $\Gamma_{01} = (\beta x K_0)/(d+K_0) + \mathcal{C}$ and $c_{01} = \mathcal{C}$, as $\int_{x=-\phi_x}^{\phi_x} (\Gamma_{00} - c_{00}) \, \text{d}x = 0$ and $\int_{x=-\phi_x}^{\phi_x} (\Gamma_{01} - c_{01}) \, \text{d}x = 0$, where $\mathcal{C}$ is an integration constant.
Hence,
\refstepcounter{equation} \label{eq:dm_limited_we}
\begin{equation} 
    c_{0} = K_0 + ..., \quad \Gamma_{0} = K_0 + ..., \quad \Delta \Gamma_{0} = \frac{2 \phi_x \beta}{\gamma(d+K_0)} + .... \tag{\theequation\textit{a,\,b}}
\end{equation}
Substituting \eqref{eq:dm_limited_we} into \eqref{eq:Rs} we have
\refstepcounter{equation}  \label{eq:Rs_dm_we}
\begin{equation}
    C_0 \approx 2 (\theta + \zeta \phi_x) K_{0}, \ \ A_0 \approx 2 (1 + \beta \phi_x) K_{0}, \ \ M_0 \approx \frac{2 \phi_x \beta K_0^2}{d + K_0}, \ \ D_0 \approx \frac{2 \phi_x \beta d K_0}{d+K_0}, \tag{\theequation\textit{a--d}}
\end{equation}
and we recover \eqref{eq:eq_marangoni}.

\section{Numerical solution to the advection--diffusion equation}\label{app:0}

In \S\ref{sec:results}, we solve \eqref{eq:advection} whilst retaining the $O(\epsilon^2)$ secondary-diffusion operator, partly for numerical convenience and partly to provide a rational regularisation of shock-like structures that may arise. 
The unsteady advection--diffusion equation is solved numerically using the method of lines and a backwards-in-time and centered-in-space (BTCS) scheme.  
At each timestep, we iterate $C_0$, $A_0$, $M_0$, $D_0$ and $D_1$, using $c_0$, $\Gamma_0$ and $K_0$ at the current timestep \cite[which are solved using the Chebyshev collection technique described in Appendix A of][]{tomlinson2023laminar}. 
Discretising space such that $\chi_i = i \Delta \chi$ for $i = 0$, $1$, ..., $N_\chi = \st{2  \Delta \chi}$, where $\st{2}$ is length of the channel, we write \eqref{eq:advection} at each $\chi_i$ for $i = 1$, $2$, ..., $N-1$:
\begin{multline} \label{eq:blah1}
    \left(\frac{\partial C_0}{\partial K_0}\right)_i\left(\frac{\text{d} K_{0}}{\text{d} t}\right)_i = \left(\frac{\partial A_0}{\partial K_0} - \frac{\partial M_0}{\partial K_0} - \frac{\partial D_0}{\partial K_0} - \lambda \epsilon^2 \frac{\partial^2 D_1}{\partial K_0^2}\frac{\partial K_0}{\partial \chi} \right)_i\left(\frac{K_{0,\, i+1} - K_{0, \, i-1}}{2\Delta\chi}\right) \\ - \lambda \epsilon^2 \left(\frac{\partial D_1}{\partial K_0}\right)_i\left(\frac{K_{0,\, i+1} -2K_{0, \, i} + K_{0, \, i-1}}{\Delta\chi^2}\right) + O(\Delta\chi^3),
\end{multline}
with periodic boundary conditions applied at interior ($i=1$, $N-1$), boundary interior ($i=0$, $N$) and ghost nodes ($i=-1$, $N+1$):
\begin{equation} \label{eq:blah2}
    K_{0,\, 0} = K_{0,\, N}, \quad \frac{K_{0,\, 1} - K_{0, \, -1}}{2\Delta\chi} = \frac{K_{0,\, N+1} - K_{0, \, N-1}}{2\Delta\chi} + O(\Delta\chi^3).
\end{equation}
Assembling \eqref{eq:blah1}--\eqref{eq:blah2} into a matrix problem, the advection--diffusion equation in \eqref{eq:advection} reduces to solving the system of ODEs
\begin{equation} \label{eq:blah3}
    \frac{\text{d} \boldsymbol{K}_{0}}{\text{d} t} = \mathsfbi{A}(\boldsymbol{K}_{0}) \boldsymbol{K}_{0}.
\end{equation}
We solve the problem in \eqref{eq:blah3} using an implicit Euler scheme. Hence, defining $\tau^n = n\Delta\tau$ for $n = 1$, $2$, ..., $N$, we have that
\begin{equation} \label{eq:blah4}
    (\mathsfbi{I} - \Delta\tau \mathsfbi{A}(\boldsymbol{K}_{0}^{n+1}))\boldsymbol{K}_{0}^{n+1} = \boldsymbol{K}_{0}^{n}.
\end{equation}
Note that \eqref{eq:blah4} is nonlinear, therefore we initialise \eqref{eq:blah4} with the solution at the previous step $\boldsymbol{K}_{0}^{n}$ such that $\mathsfbi{A} = \mathsfbi{A}(\boldsymbol{K}_{0}^{n})$, we then solve \eqref{eq:blah4} for $\boldsymbol{K}_{0}^{n+1}$ and substitute the new solution into $\mathsfbi{A} = \mathsfbi{A}(\boldsymbol{K}_{0}^{n+1})$, until $\boldsymbol{K}_{0}^{n+1}$ varies less than some specified tolerance. 

\bibliographystyle{jfm}
\bibliography{jfm-instructions}

\end{document}